\RequirePackage{fix-cm}
\documentclass[twocolumn,epjc3]{svjour3}  
\smartqed  
\RequirePackage{graphicx}
\RequirePackage{latexsym}
\RequirePackage[numbers,sort&compress]{natbib}
\RequirePackage[colorlinks,citecolor=blue,urlcolor=blue,linkcolor=blue]{hyperref}
\usepackage{mathtools}
\usepackage{amsmath}
\usepackage{amsfonts}
\usepackage{multirow}
\usepackage{tikz}
\usetikzlibrary{arrows,quotes,angles}
\usepackage{pgfplots}
\usepackage[colorinlistoftodos]{todonotes}
\pgfplotsset{width=10cm,compat=1.9}
\usepackage{physics}
\usepackage{caption}
\usepackage{yfonts}
\usepackage{euscript}
\usepackage{subcaption}
\usepackage{tablefootnote}
\usepackage{float}
\usepackage{graphicx}
\usepackage{adjustbox}

\usepackage[normalem]{ulem}

\newcommand{\unit}{1\!\!1}

\journalname{Eur. Phys. J. C}
\begin{document}

\title{Dynamics and the Emergence of Geometry in an Information Mesh}


\author{Philip Tee\thanksref{e1,addr1}}

\thankstext{e1}{e-mail: ptee2@asu.edu}


\institute{Beyond Center, Arizona State University, \\ Tempe, Arizona 85287  \label{addr1}}

\date{Received: date / Accepted: date}

\maketitle

\begin{abstract}
The idea of a graph theoretical approach to modeling the emergence of a quantized geometry and consequently spacetime, has been proposed previously, but not well studied.
In most approaches the focus has been upon how to generate a spacetime that possesses properties that would be desirable at the continuum limit, and the question of how to model matter and its dynamics has not been directly addressed.
Recent advances in network science have yielded new approaches to the mechanism by which spacetime can emerge as the ground state of a simple Hamiltonian, based upon a multi-dimensional Ising model with one dimensionless coupling constant.
Extensions to this model have been proposed that improve the ground state geometry, but they require additional coupling constants.
In this paper we conduct an extensive exploration of the graph properties of the ground states of these models, and a simplification requiring only one coupling constant.
We demonstrate that the simplification is effective at producing an acceptable ground state.
Moreover we propose a scheme for the inclusion of matter and dynamics as excitations above the ground state of the simplified Hamiltonian.
Intriguingly, enforcing locality has the consequence of reproducing the free non-relativistic dynamics of a quantum particle.
\end{abstract}

\section{Introduction}
\label{sec:introduction}
\subsection{Background}
Reconciling General Relativity (GR) with Quantum Mechanics (QM) has not yielded a consistent and finite theory \cite{Davies1982}.
In part, this is due to the fundamentally different role that spacetime, and its geometry, plays in the two theories.
In QM  spacetime exists external to the theory and its geometry is input, whereas GR is essentially a theory of the geometry of spacetime.
As such a quantum theory of gravity entails quantizing the geometry and a fully quantized theory of Gravity, therefore, would have to explain what it means to quantize spacetime.
This is not a simple task to undertake.

The original observation that quantum theory must inevitably lead to a  discrete spacetime was made by Matvei Bronstein \cite{Bronstein1936}, but the first concrete proposal for how this could be reconciled with Lorentz invariance  has its origin in the work of Hartland Snyder \cite{Snyder1947}. 
He proposed a framework to directly consider the implications of discretized space with a minimum length, originally as an attempt to rationalize the presence of ultra-violet cut-offs in Quantum Field Theory (QFT).
The existence of a global fundamental length would seem at odds with the principle of Lorentz invariance, as observers in inertial frames moving relative to each other would  disagree about the magnitude of this length due to relativistic length contraction.
The approach in this work was subsequently extended to prove that such inconsistencies are reconcilable \cite{Amelino-Camelia2000,Hossenfelder2012}, provided one admits additional dimensions  in which to accommodate such a universal minimum distance scale.
In these modifications, referred to as Doubly or Deformed Special Relativity (DSR),  a discrete spacetime can avoid the presence of problematic preferred observers, by the modification of the invariant interval $ds^2$.
We subscribe to the opinion that a discrete microscopic structure, at the scale of the Planck Length $l_p$, must form part of a consistent quantum theory of gravity and the properties of quantized spacetime are important to consider.
In fact it is the starting assumption of our model is that at the fundamental level spacetime is discrete.

A direct consequence of this assumption leads to the concept of a quantized spatial mesh (or more mathematically a graph), which can be used to model and explore the quantum nature of spacetime.
Most of the studies of such a quantized spacetime have taken place in the context of emergent theories of gravity and geometry.
The particular focus of this work is to build upon models that represent the emergent spacetime as a graph.
As outlined by Antonsen some years ago \cite{Antonsen1994} the inevitable reduction of the nature of geometry to points and relations leads naturally to the idea of a spacetime graph, capturing the spirit of Wheeler's {\sl``it from bit''} hypothesis \cite{Wheeler1989}.
A direct application of this concept, `Quantum Graphity' (QG) \cite{Konopka2006,Konopka2008}, provides an intriguing pathway to emergent geometry that  can be shown to naturally produce a four dimensional universe.
The analysis proposes a Hamiltonian, from which  the graph emerges as a ground state as the universe `cools' from a hypothetical `hot' early universe.
The argument however works backwards from desired properties of the emerged graph to the necessary form of the Hamiltonian which would produce that graph, rather than being derived from an underlying physical model.
In an elegant fashion Trugenberger \cite{Trugenberger2015} combined some recent advances in Network Science to propose such a mechanism, and outlined an approach he terms the Dynamical Graph Model (DGM).
In this seminal work he showed how the emergence of a stable graph can be explained as a phase transition occurring as the universe cooled.
Further the emerged spacetime graph has a convergence of certain measures of dimensionality that would indicate a  preference for a four dimensional universe.
However the universe `graph' that emerges from this model contains a significant amount of clustering (presence of triangles in the emerged spacetime graph), which has the unfortunate effect of introducing non-locality, whereby common conceptions of the neighborhood of a point in space are violated.
In further works \cite{Trugenberger2016,Trugenberger2017} the DGM model was extended to improve the locality of the emerged geometry with addition of two perturbation terms to the Hamiltonian, which we refer to as the perturbed DGM (PDGM).
This has the effect of improving the regularity of the emerged spacetime, but at the expense of the addition of two further coupling constants, which serve as the perturbation parameters.

In QG there is no concrete proposal for the inclusion of time, and in PDGM the role of time is treated as an emergent extra dimension beyond the three spatial dimensions of traditional geometry.
However in all of these models there is no scheme for the inclusion of matter, and therefore dynamics, in the emerged geometry.

Introducing matter and dynamics is the ultimate objective of the work presented in this paper, but we first conduct an extensive exploration of the ground states of the DGM and PDGM models, focusing on the claim that the models produce a `large world' geometry, possessing a low clustering coefficient and therefore  an emerged geometry with high degrees of locality.
This analysis is numerical, and we compute ground states for $3,840$ combinations of the coupling constants for PDGM, and across a wide range of coupling constants for DGM, chosen to probe the range where the emergence of the `large world' occurs.
We also propose a model, extending PDGM using a single, common,  coupling constant.
Further, we propose a mechanism to introduce time, and this contribution should be viewed as an extension and simplification of the models of Trugenberger.
Our proposal for a simplification of PDGM, produces a ground state that is highly local and geometrically flat to the same, or even better extent as PDGM \cite{Trugenberger2016}, whilst requiring a single dimensionless coupling constant.
In the quantitative assessment of models, a more parsimonious scheme is often preferred \cite{Burnham2004}, as evidenced in the definition of information criteria for a model such as Akaike or Bayes Infornation criteria where the number of parameters is viewed as a penalty.
The simplified model utilizes only one versus three parameters, and as it produces the same desired zero clustering would automatically represent a better model, and would consequently be viewed as more efficient.
Indeed, using our numerical simulations we verify that for PDGM and the simplified model the ground states represent a Euclidean vacuum.
For our simplified model (although this extension would also work for the Trugenberger models), we propose a description of how matter can be modeled in the spacetime graph as excited state `defects' in the graph.
These defects are free to propagate in the graph, and by considering a minimal extension of the Hamiltonian that does not alter the original ground state, we are able to connect these excitations to non-relativistic quantum dynamics.
What is obtained is an intriguing mechanism that could explain how the quantum dynamics of matter emerges along with the geometry that evolves as the ground state of an informational graph.

Whilst well short of a firm proposal for a quantum theory of gravity, the presence of a graph is intriguing, as graphs possess measures of informational entropy and therefore a way to connect to thermodynamics.
Consistent with the analysis of (P)DGM presented in \cite{Trugenberger2015,Trugenberger2017}, we are able to show that at least one version of graph entropy leads to a dependence of the entropy of a graph on boundary terms, further strengthening the results reported in those treatments.
This entropy results from the encoding of information in the structure of the graph, i.e. the number of `bits' needed to precisely specify it.
The graph, in the case of our proposal, emerges directly from the entanglement of `spins' between adjacent vertices.
There is a deep and interesting correspondence between entanglement across causal boundaries and the emergence of spacetime curvature and hence GR, first detailed in the work of Bekenstein and later Jacobson \cite{Bekenstein1973,Jacobson1995}.
A striking result of Jacobson's work is the connection between the requirement for finite entropy and the need for a fundamental length scale, which he determines to be the Planck Length $l_p$.
Further, this observation directly leads to a derivation of the thermodynamic statement of GR established by Bekenstein and also an `area law' for entropy well understood from Hawking's laws of black hole thermodynamics \cite{Bardeen1973}.
More recently, rekindled interest in this approach has been outlined by Verlinde \cite{Verlinde2011,Verlinde2016} in his emergent gravity theories.

It seems likely that there is a fundamental connection between the quantized, discreet graph of spacetime, a fundamental length scale, and gravity.
To investigate this link it would seem natural to start with a concrete proposal for the structure of this discrete spacetime at the Planck Length, and this forms the basic motivation for the work presented in this paper.

\subsection{Outline of the paper}
The first contribution of this work is an extensive exploration of the ground states of the (P)DGM models introduced by Trugenberger, and also the simplification of PDGM we  propose.
As such we begin in Section \ref{sec:spacetime} with a brief introduction to the idea of a discrete spacetime and present an overview of  QG and (P)DGM, highlighting how they treat the emergence of the graphical structure of spacetime.
Building upon these two approaches, in Section \ref{sec:model} we describe on intermediate model that sits between DGM and PDGM, reducing the number of terms and requiring only one coupling parameter.
We  present in Section \ref{sec:simulation} and Appendix A numerical simulations that show that the obtained spacetime geometries have the same local nature as PDGM and QG, and preserve the attractive feature of possessing a low dimension where the difference between the extrinsic and intrinsic measures of dimensionality is minimized.
In Appendix A, for completeness, we present the full results of the ground states of PDGM across a wide range of parameters.
Given the computational constraints involved in the numerical simulations, these calculations are practically bounded to spacetime graphs up to $150$ nodes, and consequently the errors involved in computing dimensionality measures prevent a precise prediction of the `natural' dimension of the graph.
Indeed for the deep exploration of the PDGM model we are further constrained to graphs of $50$ nodes, and in further work it would be interesting to raise that limit.
We do not believe that larger graphs would significantly alter the results obtained, and we conclude it is  likely that the dimension of the ground state will lay between $3$ and $5$.
It is also important to note that these simulations are fully consistent with the analysis in \cite{Trugenberger2015,Trugenberger2016}, but represent an original direct computation of the ground state of the models described therein.

With an emerged geometry, it is natural to ask how matter and dynamics can be included.
The second contribution of this work is to consider this problem, beginning in Section \ref{sec:dynamics} by considering how matter may be modeled as excitations above the ground state.
These manifest themselves as defects or `holes' in the mesh, with a boundary of precisely dimension one less than that of the emerged mesh.
In Section \ref{ssec:meshdyn} we investigate what minimal extensions can be made to the Hamiltonian that can describe the dynamics of the matter defects in the mesh.
For the defects to describe particles in this model, we would ideally like at least approximate conservation of the number of excitations, and the dynamics to be governed by interactions that are local.
Using dimensional analysis we propose a Hamiltonian that satisfies these conditions and does not interfere with the emergence of the ground state.
The proposed dynamical Hamiltonian focuses on the propagation of the vertex spin defects, and we outline how this could be extended in a consistent way to the edge Hilbert space.
This last step provides the final Hamiltonian of the model.
In Section \ref{ssec:entropy} we investigate how these defects in the mesh alter the underlying graph entropy.
This analysis provides further evidence that the entropy of such an excited state is dictated by the size of the boundary.
We regard this as supporting the claim made in \cite{Trugenberger2017} that the entropy of the mesh obeys an `area' law for the excitations, as well as the whole of the emerged mesh.

The final contribution of this work is to consider the continuum limit of the mesh, and the dynamical Hamiltonian that we proposed.
Although not a rigorous proof, we sketch out in Section \ref{ssec:waveeqn} how, in this limit, the discrete dynamics  can lead to the familiar form of the wave equation.
As a reasonableness check on the model, it is reassuring that it reproduces the precise form of the  Schr\"{o}dinger equation for a free particle in a vacuum, and therefore non-relativistic quantum mechanics.

I conclude in Section \ref{sec:conclusion} with a brief discussion of possible  future directions.

\section{Spacetime as a Quantum Mesh}
\label{sec:spacetime}
\subsection{The Nature of Discrete Space}

Starting with the idea that spacetime is quantized, one is naturally led to the concept of a fundamental length, below which it is impossible to measure the separation of two points \cite{Snyder1947,Antonsen1994,Amelino-Camelia2000}.
Elementary considerations on the limit of localization of a quantum particle using the uncertainty principle, led to the proposal of a minimum distance called the `Planck Length', which using S.I. units is defined to be $l_p=\sqrt{ \frac{\hbar G}{c^3} }$, and has the value of $1.6 \times 10^{-35}m$.
This assumption can be justified from many different directions, either as a practical limitation of measurement, or as a consequence of the fundamental nature of spacetime at very small separations.
We refer the reader to the comprehensive review by Hossenfelder \cite{Hossenfelder2012}, but in this work we take the view that spacetime is fundamentally discrete.
Hartland Snyder originally proposed the concept of a quantized spacetime in 1947 \cite{Snyder1947} as a mechanism to explain UV cutoffs in QFT.
In much the same way that the requirement that all observers agree upon the velocity of light requires the mixing of space and time into the familiar Lorentz invariant Minkowski interval, forcing all observers to agree as well on a minimal length can be accomplished by the mixing of a fifth dimension into a new invariant interval.
In the `Snyder basis' he proposed a $5$ dimensional interval which is invariant under the Lorentz group $SO(4,1)$ using the coordinates $\eta_a$, with $a=0,1,2,3,4$, and a diagonal metric $g_{ab}=diag(+,-,-,-,-)$.
In this definition the interval is $ds^2=d\eta_0^2-d\eta_1^2-d\eta_1^2-d\eta_3^2-d\eta_4^2$.
To recover the Lorentz invariant subgroup $SO(3,1)$ he defined non-commuting physical coordinate operators $\hat{x}_{\alpha}$, with $\alpha=0,1,2,3$ in terms of the $\eta_a$ as 
\begin{align}
	\hat{x}_{\alpha} &= i l_p \Bigg ( \eta_4 \frac{\partial}{\partial \eta_{\alpha} } - \eta_{\alpha} \frac{ \partial }{ \partial \eta_4 } \Bigg) \mbox{, $\alpha=0,1,2,3$, } \\
	\hat{x}_{0} &=\frac{ i l_p }{c} \Bigg ( \eta_4 \frac{\partial}{\partial \eta_{0} } + \eta_{0} \frac{ \partial }{ \partial \eta_4 } \Bigg) \mbox{.}
\end{align}

Using this definition he is able to prove that Lorentz invariance of the $\hat{x}_{\alpha}$ is guaranteed and that as operators the spectra of the position coordinates are discretized in units of $l_p$.
The approach outlined by Snyder was significantly built upon in the framework of DSR in the work of Amelino-Camelia \cite{Amelino-Camelia2000}, which recast the additional coordinates as being those in a de Sitter momentum space where the Planck Length plays the role of the scalar curvature.
Fundamentally these attempts show that there is hope of reconciling the existence of a fundamental length with Lorentz invariance, but our goal is a physical model of how such a space could arise.
In the following sections we will survey two approaches upon which the proposed model builds.

\subsection{Graph Theoretical Preliminaries}

In this paper we will rely upon graph theoretical terminology, which may not be familiar to the reader, and we follow closely the terminology and definitions used in the standard text by Bollobas \cite{Bollobas1998}.
We will summarize the key concepts here.
A graph $G(V,E)$, is defined as a collection of $N$ vertices defined in a set $V$, and a collection of pairs of vertices that define the edges in the graph $E \subset V \times V$.
An individual edge $e_{ij}$ connects the vertices $v_i$ and $v_j$.
There are a number of special graphs, and in particular we will refer to the perfect or fully connected graph on $N$ vertices, $K_N$.
This graph $K_N$, has all possible $\frac{1}{2}N(N-1)$ edges present, that is each pair of vertices is connected by an edge and each vertex is therefore connected to every other vertex.
A graph is termed `simple' if it contains no edges beginning and terminating at the same vertex, and there is only one edge between any two vertices.
Further we say that it is undirected if $e_{ij}$ is indistinguishable from $e_{ji}$.
For the rest of this paper we will only concern ourselves with simple, undirected graphs.
We denote by $k_i$ the degree of a vertex $v_i$, defined as the number of edges that begin or terminate at an individual vertex, and we note the basic result that $\sum_i k_i = 2|E|$.

\begin{figure}
	\begin{center}
	\begin{tikzpicture}[node distance=1.0cm,
 			thick,main node/.style={circle,fill=blue!15,draw,font=\sffamily\small\bfseries}]
	 
  			\node[main node,scale=0.75] (1) {$v_1$};
  			\node[main node, scale=0.75] (2) [below right of=1] {$v_2$};
			\node[draw=none,fill=none]  (10) [below of=2] {Open Triple};
			\node[main node, scale=0.75] (3) [above right of=2] {$v_3$};
			
			\node[main node,scale=0.75] (4)  [right =2cm of 3] {$v_1$};
  			\node[main node, scale=0.75] (5) [below right of=4] {$v_2$};
			\node[main node, scale=0.75] (6) [above right of=5] {$v_3$};
			\node[draw=none,fill=none]  (11) [below of=5] {Closed Triple};

 			 \path[every node/.style={font=\sffamily\small}]
    					(1) edge node {} (2)
    					(2) edge node {} (3);
					
			 \path[every node/.style={font=\sffamily\small}]
    					(4) edge node {} (5)
					(4) edge node {} (6)
    					(5) edge node {} (6);
	\end{tikzpicture}
	\end{center}
	\caption{Open and closed triples used to define the degree of clustering of a graph.}
	\label{fig:clustering}
\end{figure}
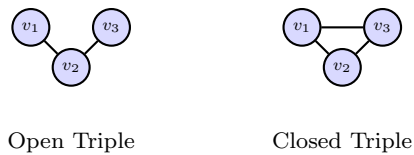
An important property of graphs is the degree of clustering present, referred to as the clustering coefficient.
This property of a graph measures how fully connected it is, and is directly dependent on how many of the possible edges in the graph are present.
Fundamental to this notion is the presence of closed triples, or triangles, in the graph.
In Fig. \ref{fig:clustering}. we denote the two possible ways in which three nodes of a graph can be connected by two or more edges.
An open triple has two edges, and a closed triple, or triangle, has three.
The ratio of open to closed triples is used to compute the clustering coefficient of the graph (see for example Bollobas \cite{Bollobas1998}, or Barabasi \cite{Barabasi2016}).
If $\tau_G$ is the number of closed triples, and $\lambda_G$ the number of open triples, with strictly $\tau_G < \lambda_G$, then the clustering coefficient $C_G$ is defined as
\begin{equation}\label{eqn:clustering}
	C_G=\frac{\tau_G}{\lambda_G} \mbox{.}
\end{equation}

It is possible to construct graphs where every node has the same degree in many ways.
In general these graphs are referred to as $k$-regular, and in particular a $k$-regular graph with little to no clustering we term a `mesh' and with significant clustering a lattice.

We can also associate a number of important matrices with a graph.
Firstly we can define the adjacency matrix $A_{ij}$ such that an element $A_{ij}=1$ if there is an edge between the vertices $v_i$ and $v_j$, zero if there is no edge, and by convention $A_{ii}=0$.
This matrix has a number of important and useful properties.
In particular when raised to the $n^{th}$ power, the value of $A_{ij}$ indicates the number of paths, defined as the non-unique sequence of traversed edges, that exist between vertices $v_i$ and $v_j$. 
It is important to note that edges in a path may be traversed more than once, so a path is not necessarily the shortest path, in terms of number of hops or edge traversals, between two vertices.
Associated with the adjacency matrix is the degree matrix $\Delta_{ij}$, which is defined as $\Delta_{ij}=k_i \delta_{ij}$, that is the diagonal values are set to the degree of the corresponding node.
Finally, constructed from the degree and adjacency matrices is the Laplacian matrix, defined as $L_{ij} = \Delta_{ij} - A_{ij}$.
The origin of the name of this matrix is that for certain dynamical processes on a graph (say for example heat diffusion) the Laplacian matrix plays the same role as that of the negative Laplacian operator in the differential equations of the continuous model analog of the same physical process \cite{Chung1997}.

\subsection{Quantum `Graphity'}

In a series of papers by Konopka {\sl et al} \cite{Konopka2006,Konopka2008} the authors proposed a graph theoretical origin for emergent spacetime.
The model for QG proceeds by defining a structure where the `atoms' of spacetime are quantum bits, whose permissible values of $0$ and $1$ define an `on' and an `off' state for each edge.
These states define a point and edge structure, {\sl i.e.} a graph, with a finite number of vertices.
The presence of an edge implies the nodes that it connects are `local', and, from this construct the traditional notions of geometry arise.
The model starts with a perfect graph $K_N$, of fixed size $N$, and  a Hamiltonian which maps the precise graph structure to an energy computed from the edges, vertices, loops and paths present in the graph.
The precise configuration of edges against a fixed set of vertices, is obtained as the ground state of a proposed Hamiltonian, and generates a very large, but finite, undirected graph of size $|V| = N \sim 10^{100-1000}$ nodes. 
The very large size of the graph is postulated to reflect the spatial extent of the universe, and also to encompass the estimated information content of the Universe.
Evidently the particular configuration of the produced spacetime graph is entirely dependent upon the Hamiltonian, and the authors propose a multi-term Hamiltonian to yield a $k$-regular mesh with minimal clustering.
This latter point regarding the absence of clustering is important to guarantee locality, which we discuss in detail in Section \ref{ssec:comparison}.
Each term in the Hamiltonian is crafted to favor a particular feature of the graph (regularity, locality and so on), and each term requires a new dimensionless coupling constant.
The quantum nature of the graph is described by the Hilbert space of the graph defined as the tensor product space of the individual vertex and edge Hilbert spaces.
It is proposed that the vertices are separated  by the Planck Length, and the graph is postulated to represent the actual structure of spacetime at the smallest scales.
In this way the total Hilbert space for the whole graph is defined as the tensor product of the edge and vertex spaces in the following way
\begin{equation}
	\mathcal{H}_{total} = \bigotimes_{e_{ij} \in E} \mathcal{H}_{edge} \bigotimes_{N} \mathcal{H}_{node}\mbox{.}
\end{equation}
In their model,  the $\bigotimes_N \mathcal{H}_{node}$ vertex Hilbert spaces are ignored, with the focus being upon the edge spaces.
The Hilbert space for each edge can be minimal, {\sl i.e.} $\mathcal{H}_{edge} = \mbox{span} \{ \ket{0}, \ket{1} \}$, with the $\ket{0}$ basis vector being associated with the absence of an edge, and $\ket{1}$ the presence of an edge.
An edge in their model is equated to the presence of a fermionic `particle', the nature of which is left unspecified.
Standard annihilation and creation operators are defined having the normal algebra $\hat{a}_{ij} \ket{0}=0$ and $\hat{a}^{\dagger}_{ij} \ket{1}=0$, and anti-commutation relations $\{\hat{a}_{ij}, \hat{a}^{\dagger}_{kl} \} = \delta_{ik} \delta_{jl}$, all other anti-commutators being zero.
One can extract a number operator $\hat{N}_{ij}=\hat{a}^{\dagger}_{ij}\hat{a}_{ij}$, which has the property of returning the number of particles in the edge `state', which can either be $1$ or $0$.
It is convenient to express the graph's adjacency matrix, in terms of these operators as follows, $A_{ij}=\hat{a}^{\dagger}_{ij}\hat{a}_{ij}$, or equivalently define the adjacency operator for an edge and its eigenvalue equation as $\hat{A}_{ij} \ket{ l } = a_{ij} \ket{ l } = \hat{a}^{\dagger}_{ij}\hat{a}_{ij}\ket{l} $, with $l=\{0,1\}$.

As stated the physics of the model is dictated by the choice of Hamiltonian for the graph.
There is no physical basis of this choice, other than to consciously construct one such that the ground state of the model has the desired properties.
The ground state of the QG model is a mesh, and importantly the pre-geometry, high temperature regime of the model is the initial perfect graph with no notion of locality, that is every point in space is local to every other point.
The authors propose that a phase transition occurs as the universe cools, whereby edges are deleted from the graph until locality emerges suddenly at the transition point.
It is an intriguing proposition but no rigorous proof of such a transition is given.

\subsection{The Trugenberger Model of Emergent Spacetime}
\label{ssec:trugenberger}
In two seminal papers \cite{Trugenberger2015,Trugenberger2016} Trugenberger proposed an alternative approach for the emergence of spacetime.
The basis of the approach utilizes competing pairs of Ising spin-spin interactions, a ferromagnetic interaction between spins on the vertices of a graph, and an anti-ferromagnetic interaction between spins on the edges of the graph.
In this way the ferromagnetic interaction seeks to create links in the graph, and the anti-ferromagnetic term acts to suppress links by introducing an energy penalty for triples of vertices sharing two or three edges.
The Hilbert space formulation is identical to basic QG, with the crucial distinction that a physical Hamiltonian is proposed, requiring only one dimensionless coupling constant. 
The basis for the formulation is that if two adjacent spins are aligned, there is a preference for an edge between them.
If we define the Hilbert space for the vertex $v_i$ as $\mathcal{H}_{i}=\mbox{span} \{ \ket{0},\ket{1} \}$, such that there is an operator $\hat{s}_{i} \ket{s}=s_i\ket{s}$, we can complete the Hilbert Space using the edge space of QG. 
This yields $\mathcal{H}_{total} = \bigotimes\limits_{e_{ij} \in E} \mathcal{H}_{edge} \bigotimes\limits_N \mathcal{H}_{i}$,  and the following Hamiltonian is proposed
\begin{equation}\label{eqn:ising}
	H_0=\frac{J}{2} \sum\limits_{j \neq i}^{j=N}  \sum\limits_{k \neq i,j}^{k=N}A_{ik}A_{kj} - \frac{1}{2} \sum\limits_{j \neq i}^{j=N} s_i A_{ij} s_j \mbox{, }
\end{equation}
where in the second term the Ising coupling has been absorbed into the dimensionless coupling constant $J$.
This can be converted into a quantum Hamiltonian operator by replacing the terms in $A_{ij},s_{i}$ by their appropriate operators on the total Hilbert space.
As noted, the two terms act in opposite directions, in that there is an energy reduction from the second term when edges are created between nodes of the same spin, and from the second term energy increases as a node acquires links.
This first term is the  `link frustration' term, and when the Hamiltonian is minimized to extract the ground state, the result is a constraint to the node degree, and the emergence of locality.
The balance between the two terms in the ground state of Eq. \eqref{eqn:ising} leads to a well defined minimum for a value of $k>0$.
The action of the link frustration term also reduces the number of connected triples in the graph.
Given the dynamical nature of the emergence of the graph structure, he terms these universe graphs Dynamical Graphs, and we refer to this approach as the Dynamical Graph Model (DGM).

By analyzing the graphs produced as the ground state of Eq. \eqref{eqn:ising}, many attractive properties of this model emerge.
In particular, it is possible to prove that a valid ground state is a $k$-regular graph with $k=2d$, where $d$ is the (approximate) Euclidean dimension of the emerged regular lattice.
Indeed if the assumption is made that the ground state is $k$-regular, it is possible to compute that this minimum energy occurs at precisely 
\begin{equation}\label{eqn:trug_min}
	k=\frac{1}{2} + \frac{1}{2J}\mbox{.}
\end{equation}
For a lattice (or mesh) of spatial dimension $d$,  $k=2d$, and it is possible to rearrange Eq. \eqref{eqn:trug_min} to establish that for such a mesh this minimum occurs at a value of 
\begin{equation*}
	J=\frac{1}{4d-1}\mbox{.}
\end{equation*}
For details we refer the reader to the original work of Trugenberger \cite{Trugenberger2015}.

The link frustration term disfavors triples in the graph, and clustering is lowered although not entirely absent, or indeed very different in magnitude to random graphs obtained using the Erd{\"o}s-R{\`e}nyi process \cite{Barabasi2016}.
The non-locality of the ground state was significantly improved upon by the addition of  perturbation terms to the Hamiltonian \cite{Trugenberger2016}, which then forms the basis for the PDGM model.
The perturbation of the Hamiltonian is given by an additional term $H_1$, defined as

\begin{equation}\label{eqn:pdgm}
	H_1=\frac{\lambda_3}{3} \Tr(A^3_{ij}) - \frac{\lambda_4}{4}\Tr(A^4_{ij})\mbox{.}
\end{equation}

The total Hamiltonian for the PDGM model is defined by adding this to Eq. \eqref{eqn:ising} obtaining

\begin{equation}\label{eqn:pdgm_total}
	H=H_0+H_1\mbox{.}
\end{equation}

In the original paper on PDGM the constants were chosen to be $\lambda_3=0.01$, and $\lambda_4=0.001$, and we analyze extensively the ground states achieved with this choice in Section \ref{sec:simulation}.
However, there is no {\sl a priori} reason for this choice, and we perform a deep exploration of the ground states obtained for differing choices of $\lambda_3$ and $\lambda_4$, around the value of the coupling constant $J$ that produces the `large world',  to understand the effect on the clustering coefficient and average node degree of the ground state.
This is a significant addition to the discussion of PDGM undertaken in the original treatment, where it is argued that the terms will result in a triangle free regular mesh as its ground state.
We will demonstrate that this is indeed the case, but that the effect of $\lambda_4$ is more complex, reducing clustering less effectively than $\lambda_3$ and at the expense of reducing average degree.
Reduction in node degree significantly alters the topology, specifically the dimension, of the ground state and is an unwelcome effect.
We will demonstrate that reduced clustering can also be achieved using a simplification of the PDGM Hamiltonian without incurring a reduction in  the average node degree.

An intriguing feature of both DGM and PDGM is the transition to the regular lattice from a hot `disordered information soup'.
This transition is marked by a divergence in measures of graph dimensionality (see Section \ref{ssec:dimensionality}) at or around $d=4$.
Intriguingly the emerged graph can also exhibit stable deformations in the lattice,  referred to as topological black holes.
We will make use of this concept to model matter later in this work.

%
%
\subsection{Comparison and Limitations}
\label{ssec:comparison}

\begin{figure}
	\begin{center}
	\begin{tikzpicture}[node distance=2.8cm,
 			thick,main node/.style={circle,fill=blue!15,draw,font=\sffamily\small\bfseries}]
	 
  			\node[main node,scale=0.75]  (1) {$v_1$};
  			\node[main node, scale=0.75] (2) [below  of=1] {$v_2$};
			\node[main node, scale=0.75] (3) [right of=1] {$v_3$};
			\node[main node, scale=0.75] (4) [below of=3] {$v_4$};
			\node[main node, scale=0.75] (5) [right of=3] {$v_5$};
			\node[main node, scale=0.75] (6) [right of=4] {$v_6$};

			 \path[every node/.style={font=\sffamily\small}]
    					(1) edge node {} (2)
					(1) edge node {} (3)
					(2) edge[dashed] node [pos=0.5,sloped,above ] {} (5)
					(3) edge node [right] {} (4)
					(3) edge node [right] {} (5)
					(4) edge[dashed] node [right] {} (5)
					(4) edge node [below] {$l_p$} (6)
					(5) edge node [right] {$l_p$} (6)
    					(2) edge node [below] {$l_p$} (4);
		
	\end{tikzpicture}
	\end{center}
	\caption{A small patch of a $d=2$ emerged mesh. The presence of the links $(v_2,v_5)$ and  $(v_4,v_5)$ violate conventional locality. Without these links $v_5$ is not in the local neighborhood of $v_2$ or in the neighborhood of any of $v_2$'s neighbors. With the links established $v_2$ is now 'local' to a node that is distant to all of the points in the node's neighborhood.}
	\label{fig:locality}
\end{figure}
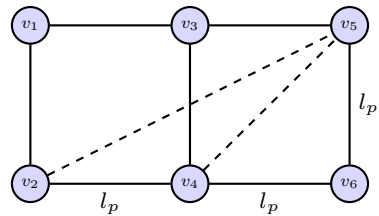

The two models both seek to propose a model of emergent geometry and locality, and (P)DGM  mostly differs from QG in that it proposes a physical basis for the graph Hamiltonian.
The QG model has a ground state that approximates a Euclidean vacuum, in that the graph produced is free of clustering and is locally flat.
It is worthwhile pointing out that the assertion of a flat geometry is not substantiated in the QG model.
The consequence of  clustering in a graph is problematic for the interpretation of the ground state as a Euclidean vacuum.
To illustrate this point, assume that all of the links in the mesh are assumed to be of equal length $l_p$.
The presence of a triangle in the graph can link nodes in violation of the normal conceptions of distance, and introduce non-locality.
In Fig. \ref{fig:locality}. we depict a subgraph of a $d=2$ emerged graph, with the presence of the triangle $\{v_2,v_4,v_5\}$.
It is clear that $v_5$ is not local to all of the points local to $v_2$, and that without the triangular links between $v_2,v_4$ and $v_5$ the shortest hop distance to $v_5$ would be $2$ hops from $v_4$, and 3 from $v_2$.
The inclusion of the triangle, and therefore clustering, brings $v_5$ into the neighborhood of $v_2$ and creates a `shortcut' through the graph compromising conventional ideas of distance and locality.

It is proven in both the DGM and QG model that $k$-regularity is a minimum of the Hamiltonian, but it is not the only possible minimum, and not necessarily guaranteed to be the global minimum.
This can include ground states with significant clustering.
The underlying reason for this is that the Hamiltonian does not have a term that introduces an additional energy penalty when a link completes a triangle.
It is for that reason that the PDGM extension to the original DGM model is more successful at eliminating clustering, as evidenced by the data from our simulations in Table \ref{tab:clustering}.
These numerical simulations constitute an independent verification of the results reported in \cite{Trugenberger2015}, as we are able to reproduce the results for clustering almost exactly, and certainly within the margin of error.
We investigate in detail the results from our simulations in Section \ref{sec:simulation}.
The QG model does have a triangle penalty term, but it is introduced in an arbitrary fashion with the requirement for an additional dimensionless coupling constant.
Further, and unlike the (P)DGM model, neither simulation or analysis is used to investigate the geometry of the emerged spacetime graph.

In both models there is not a well defined causal structure, and hence a rigorous framing of the concept of time and consequently temperature.
Temperature in the thermodynamic sense is problematic as the fundamental definition of thermodynamic temperature is a measure of equilibrium between bodies in thermal contact.
Equilibrium in turn depends upon `time' for such an equilibrium to be established, and time is not concretely defined.
In  Section \ref{ssec:time}, we attempt to place time on a firm footing as the labelling of discrete states of the spacetime graph as it evolves in a dynamical process, though it is not necessary for time itself to be discretized.
This evolution in time will proceed according to the graph Hamiltonian, with time being used as the iteration label used to control the convergence of the graph to its ground state.
In DGM, temperature is proposed to arise from the statistical fluctuations of the graph around the ground state.
In  broad terms, connectivity as measured by average node degree, decreases as temperature decreases, to a stable value, thereby yielding our current familiar spacetime geometry.
In technical terms, the coupling constant, as introduced in the Hamiltonian, runs with temperature such that the coupling decreases with temperature.

In this work we propose an intermediate model that sits between DGM and PDGM, with a simpler Hamiltonian.
We show that the graph obtained by minimizing this Hamiltonian approaches $k$-regularity, and is effectively triangle free {\sl without} having to introduce additional coupling constants.
It is this absence of additional constants that will be particularly advantageous, as we speculate how matter can be modeled  as persistent  defects in the mesh, and investigate the dynamics of such defects.
\section{Proposed Simplification of PDGM}
\label{sec:model}
\subsection{Simplifying the PDGM Hamiltonian}

Building upon PDGM, we propose a simplified Hamiltonian including a triangle suppressing term but using the same coupling constant, raised to the second power.
The triangle suppressing term is proportional to the third power of the adjacency matrix, which computes the number of paths of length $3$ between any two nodes.
Following the DGM model we consider a finite graph of $N$ nodes, is composed of $N$ spins, with each node $v_i$ associated with a Hilbert space $\mathcal{H}_{i} = \mbox{span} \{ \ket{i,0}, \ket{i,1} \}$, and a spin operator obeying $\hat{s}_{i} \ket{i,s}=s_i\ket{i,s}$ on this space.
If we were to more closely model a fermionic spin $1/2$ particle, the spin operator would have eigenvalues of $\pm \hbar/2$.
For simplicity we omit this detail and we bear in mind that in our model we choose to measure spin in the $z$ direction, that is $\hat{s}_i \equiv \hat{s}_z$.  
Associated with these states are the usual ladder operators and anti-commutators of spin $1/2$ fermions

\begin{align}\label{eqn:statistics}
	\{ \hat{s}^{+}_i, \hat{s}^{+}_j \},  \{ \hat{s}^{-}_i, \hat{s}^{-}_j \} = 0, \{ \hat{s}^{-}_i, \hat{s}^{+}_j \}&= \delta_{ij} \mbox{,} \\
	\hat{s}^{+}_i \ket{ i,1} = 0, \hat{s}^{+}_i \ket{ i,0} &= \ket{i,1} \mbox{,} \\
	\hat{s}^{-}_i \ket{ i,1} = \ket{i,0}, \hat{s}^{-}_i \ket{ i,0} &= 0 \mbox{.}
\end{align}
Between each of the pairs of spins, there are $\frac{1}{2}N(N-1)$ Hilbert spaces $\mathcal{H}_{ij} = \mbox{span} \{ \ket{i,j,0}, \ket{i,j,1} \}$, with the state $\ket{i,j,1}$ indicating the presence of an edge between the the vertices $v_i,v_j$, and $\ket{i,j,0}$ its absence.

For each edge, $e_{ij}$ we can define an adjacency operator $\hat{A}_{ij}$ with eigenvalue $a_{ij}$, such that the adjacency matrix $A_{ij}$,  is the collection of these eigenvalues.
One can then define the edge annihilation and creation operators as follows
\begin{align}
	\{ \hat{a}^{\dagger}_{ij}, \hat{a}^{\dagger}_{kl} \},  \{ \hat{a}_{ij}, \hat{a}_{kl} \} = 0, \{ \hat{a}_{ij}, \hat{a}^{\dagger}_{kl} \}&= \delta_{ik} \delta_{jl} \mbox{,}\\
	\hat{a}^{\dagger}_{ij} \ket{ i, j, 1} = 0, \hat{a}^{\dagger}_{ij} \ket{ i, j, 0} &= \ket{i, j, 1} \mbox{,}\\
	\hat{a}_{ij} \ket{ i, j, 1} = \ket{i,j, 0}, \hat{a}_{ij} \ket{ i, j, 0} &= 0 \mbox{.}
\end{align}

We can express the adjacency matrix in terms of these annihilation and creation operators, by imposing a normal ordering that places annihilation operators to the right of any creation operators.
This naturally defines an edge number operator $\hat{N}_{ij}$, and the adjacency matrix is defined as a collection of edge number operators as follows:
\begin{equation}
	A_{ij} = \hat{N}_{ij} = \hat{a}^{\dagger}_{ij} \hat{a}_{ij} \mbox{.}
\end{equation}
The value of the cube of the adjacency matrix at each vertex, $A^3_{ii}$, computes the number of triangles that involve the vertex $v_i$.
We can therefore propose a simplified  Hamiltonian, largely equivalent to the PDGM model when we set $\lambda_4=0$ and $g^2=J$ in Eq. \eqref{eqn:pdgm_total}.
We retain, and do not absorb the Ising coupling into the edge suppression term, as in DGM, as a matter of choice, but this does not alter the physics of the model.
We propose:
\begin{equation}\label{eqn:hamiltonian}
	H=\frac{g^2}{2} \Bigg ( \Tr(A^3_{ij})+  \sum\limits_{i \neq j}^N \sum\limits_{ k \neq i,j }^N A_{ik}A_{kj}  \Bigg ) - \frac{g}{2}  \sum\limits_{i,j} s_i A_{ij}  s_j \mbox{.}
\end{equation}
We will analyze this variant alongside DGM and PDGM, and refer to it as Quantum Mesh Dynamics (QMD), and demonstrate that the ground state is a `large world' cluster free mesh.

We can establish with simple calculations the relevant minima of Eq. \eqref{eqn:hamiltonian} by assuming that the graph is approximately a uniform $k$ regular mesh, with a small but non-zero clustering. 
In a strict $k$-regular graph it is possible for clustering to arise (in fact in the small world model of Watts and Strogatz \cite{Watts1998} they start with a high clustering $k$-regular graph), however such graphs are non-local in the sense that you can have paths that connect nodes that do not share any other neighbors (the so called small world property).
In terms of the paths that can be taken from two such nodes the `hop distance', that is the number of links traversed in a path from one node to the other, is on average much greater than the one hop that now connects these normally `distant' nodes.
As discussed in the previous section, as a model of spacetime this is highly undesirable as it would entail distances between points being dependent upon how they are measured.
For our emerged spacetime mesh to be a useful model, we require both $k$-regularity and extremely low or zero clustering, in short a mesh.
With zero clustering the triangle penalty term is zero, and the Hamiltonian reduces to the original DGM model Eq. \eqref{eqn:ising}.
Using the definition of clustering coefficient $C_G$ from Eq. \eqref{eqn:clustering} as the ratio of closed to open triples, it is possible to place some bounds on ground state of this graph, by considering the energy change upon addition and deletion of links, in the presence of clustering. 
The calculation relies upon the observation that the clustering coefficient also acts as a probability that any three connected nodes participate in a triangle versus being a connected triple \cite{Barabasi2016}, and so it is possible to compute the contribution from $\Tr(A^3_{ij})$ in Eq. \eqref{eqn:hamiltonian}.
Firstly we note that a typical node in a $k$-regular graph participates in $k(k-1)$ (non-unique) open triples for $k>2$, as any triple requires at least one node to have degree at least $2$.
As links connect spins of the same value,  the energy of a typical link from the Hamiltonian can be computed as $E=g^2(k-1)(C_G+1) - g/2$, and therefore the energy cost of removing a link is $\Delta E_{-}=g/2-g^2(k-1)(C_G+1)$. 
Using the same argument adding a link  contributes $\Delta E_{+} = g^2k(C_G+1)-g/2$.
At the ground state $\Delta E_{+/-} > 0$, that is the addition or removal of a link increases the energy.
After some manipulation can be expressed as the following limit on $k$
\begin{equation}\label{eqn:stability}
	\frac{1}{2g(C_G+1)} < k < \frac{1}{2g(C_G+1)} +1 \mbox{.}
\end{equation}
Finally we note that the total energy of the graph on $N$ nodes can be computed as 
\begin{equation}
H(N,k,g,C_G) = \frac{Ng^2}{2}k(k-1)(C_G+1) -\frac{gN}{2}\mbox{.}
\end{equation}
This is minimized for values of $k_{\min}$ obeying
\begin{equation}\label{eqn:ground_k}
	k_{\min}=\frac{1}{2g(C_G+1)} + \frac{1}{2} \mbox{,}
\end{equation}
satisfying Eq. \eqref{eqn:stability}.
All of these results reproduce the original DGM in the limit $C_G \rightarrow 0$, which underlines that our proposed Hamiltonian is a simplified version of PDGM.
%
%
\subsection{The role of time in the mesh}
\label{ssec:time}

In the extended PDGM model it is argued that one of the emerged dimensions can be interpreted as time, but there is no definitive prescription of how to select this dimension or interpret this emergent time.
We consider here  two alternatives to include a time like dimension to the mesh.
The first possibility is to choose, using some scheme, one of the dimensions of the graph, which would define the causal structure.
That is time is intrinsic to the emerged geometry.
In fact this method would mirror the approach taken in Causal Dynamical Triangulation (CDT) and Loop Quantum Gravity \cite{Nicolai2005,Markopoulou2006,Ambjorn2012}, and indeed in PDGM.
This requires the identification of one of the emerged dimensions as a time like coordinate, but there is no structure intrinsic to the QMD model which would identify such a dimension.
Further, the dimensionality of the emerged mesh is dependent upon the value of the coupling constant $g$, which we will see in Section \ref{sec:simulation} can be thought to `run' with temperature and increase as the universe cools.
To identify and preserve one of the dimensions as a time coordinate would need the mechanism to survive the running of this coupling constant, and would be an external and  arbitrary input to the model.

We will follow the second alternative approach, which considers time as defining the dynamic evolution of the mesh geometry. 
In this approach, time exists as a label for each of the sequence of configurations of the spacetime mesh.
It does not matter whether time is continuous or discrete, as there is no fundamental quantization of this degree of freedom of the mesh.

If we assume that the dynamical variables in the graph are the spins $s_i$ and the local correlations represented by the edges $A_{ij}$, we can represent the time evolution of the graph as being the operation of the Hamiltonian of the graph on the time independent states of these spins and edges at a particular instant in time, that is their quantum states are in the `Heisenberg picture'.
It is then possible to add in the time dependence of these states in the usual way for a Hamiltonian that is not time dependent, such as Eq. \eqref{eqn:hamiltonian}.
Denoting by $\ket{ A_{ij},s_i }$ as the state of the entire graph, the time-dependent state is given by

\begin{equation}\label{eqn:heisenberg}
	\ket{ A_{ij},s_i,t} = e^{- i\hat{H}t/\hbar }\ket{ A_{ij},s_i }\mbox{, }
	\end{equation}
where $\hat{H}$ is the operator equivalent of Eq. \eqref{eqn:hamiltonian}.
This relationship will be the starting point for our investigation of the dynamical behavior of the mesh analyzed in Section \ref{sec:dynamics}.

It should be emphasized that this approach does not represent a full causal and covariant description of the emerged mesh, and this remains an open question for future work.
\section{Numerical Simulations and Properties of the Emerged Mesh.}
\label{sec:simulation}

\subsection{Sampled Parameter Space, and Computational Technique}

The DGM and PDGM models have one and three free parameters, the coupling constant $J$, which henceforth we relabel $g$ for consistency, and the two `perturbation' parameters $\lambda_3$ and $\lambda_4$.
We are principally concerned with a deep exploration of this parameter space to investigate clustering and average node degree.
The focus on these two parameters is motivated by the simple observation that high clustering destroys the local `large world' geometry and that average node degree is intrinsic to the dimensionality of the emerged graph.

Alongside the PDGM and DGM models we also wish to investigate our simplified extension  to understand if the additional $\lambda_4$ term is actually necessary to produce a ground state with the desired properties.
We start with an extensive set of experiments on DGM and our intermediate model, which identifies a range of values of $g$ that yield a ground state of suitably low dimension and clustering.
We then use this range of $g$ to perform the deep exploration of the PDGM model.
The full range of results are reproduced in Appendix A, but we will discuss in detail in the text the results for three representative values of $g$.

To compute the ground states, we apply to our Hamiltonians, energy minimizing techniques taken from the field of neural networks \cite{Mehta2018,Mueller1995}.
The approach is stochastic, performing random spin and link manipulations, favoring those that reduce the total value of the Hamiltonian for the graph.
To compare the models we used the simulations to compute the ground states of the QMD Hamiltonian defined by Eq. \eqref{eqn:hamiltonian}, the DGM model defined by Eq. \eqref{eqn:ising}, and the extended model PDGM using Eq. \eqref{eqn:pdgm_total}.
Our approach to PDGM, however is twofold.
In the first instance we compute the ground state using the values of $\lambda_3=0.01, \lambda_4=0.001$, as proposed by Trugenberger \cite{Trugenberger2016}, and also with $\lambda_4=0.0$ to evaluate the effect of last term in Eq \eqref{eqn:pdgm}.
This is principally to compare the three models and establish the range of $g$ over which we obtain a condensed ground state with the desired dimensionality and clustering.
Beyond these simulations we then perform a deep exploration of the parameter space of PDGM, allowing the values of $\lambda_3$ and $\lambda_4$ to vary from $0$ to $0.15$, effectively spanning the full range of the values of $g$ investigated.

In the simulations we regard the dynamical variables as $A_{ij},s_i$ (for a fixed value of the coupling constant $g$), which we iteratively adjust to minimize $H$.
We describe that process in detail below.

We begin with a fully randomized graph, the spins of the vertices chosen with even probability to be either $\pm 1$.
Nodes are then  connected randomly, with each pair of nodes being linked with probability $0.5$.
The simulation proceeds at each unit time step to alter the values of $A_{ij}$ and $s_i$ in such a way as to minimize the Hamiltonian, effectively treating it as a `cost function'.
This algorithm  is a slightly modified variant of that used to solve Hopfield networks \cite{Mueller1995}; the presence of the triangle suppression term complicating the energy calculation of a single spin flip.
At each time step we perform the following steps until we achieve a stationary minimum of $H$, taken from Eq. \ref{eqn:ising}, \ref{eqn:hamiltonian} or \ref{eqn:pdgm_total}:

\begin{enumerate}
	\item \label{itr:start} Compute the value of graph Hamiltonian at time $t$ to be $H_{t}$.
	\item For each vertex, selected in a random sequence, compute
	 \begin{equation}\label{eqn:spin_itr}
	 h_i=\sum\limits_{j \in V } A_{ij} s_j \mbox{. }
	 \end{equation}
	 If $h_i \geq 0$ set $s_i=+1$, otherwise $s_i=-1$.
	\item \label{iter:cost} For every pair of vertices, with probability $0.5$ change the value of $A_{ij}$. The direction of the change is recorded in the value of $\delta$,  such that if initially $A_{ij} = 1$,  then set $A_{ij}=0, \delta=-1$; and conversely for $A_{ij}=0$ set $A_{ij}=1, \delta=+1$. We assert the symmetric nature of $A_{ij}$ by setting $A_{ji} =A_{ij}$. Then for each vertex pair compute
	\begin{equation}
	\begin{split}\label{eqn:energy_itr}
	 	h_{ij}&= \delta \frac{g^2}{2}  \Bigg[  (k_i + k_j -2 ) +  \sum\limits_{k=1}^N A_{ik}A_{kj} \Bigg ] \\
			&-\delta \frac{g}{2} s_i s_j  \text{, }
	 \end{split}
	 \end{equation}
	 representing the contribution to the Hamiltonian of the addition ($\delta=1$) or deletion of a link ($\delta=-1$). The last term computes the contribution from the addition, or deletion of a link that would complete or remove a triangle in the graph, as it calculates the number of open triples that the nodes $i,k$ participate in. For the DGM model the last term can be ignored. If the value of $h_{ij} \leq 0$ from Eq. \eqref{eqn:energy_itr} the random link change is preserved, otherwise the previous value of $A_{ij} \text{, and } A_{ji}$ are restored. For PDGM Eq. \eqref{eqn:energy_itr} is extended to add into the first term the terms in $\lambda_3$ and $\lambda_4$ with the signs adjusted to favor a reduction in overall mesh energy.
	 \item \label{itr:end}  After every link has been considered, the new value of the Hamiltonian at step $t+1$, $H_{t+1}$ is computed and if $H_{t+1} \geq H_{t}$, all changes are discarded.
	 \item Steps \ref{itr:start} to \ref{itr:end} are repeated until we reach a stationary value of $H_{t+n}$, that is $H_{t+n} = H_{t+n-1}$ for a configurable number of iterations, being no less than $4$. The stationarity condition was experimented with at values much higher than $4$, but once the minimum emerged it was observed to be stable.
\end{enumerate}

Due to the computational expense of solving for the minimum value of the Hamiltonian we are limited to ground states for graphs up to $N=150$, but in the case of PDGM and the deep parameter exploration we are further constrained to $N=75$ and $N=50$ respectively.
With all such minimization techniques, there is the risk that the obtained configuration is a local and not a global minimum.
To avert that we repeat the simulations  $10$ times and average all of the extracted results, each run starting from a different randomized configuration to mitigate the risk of a false or non-global minimum.

\subsection{Clustering and Degree}

\begin{figure*}[ht]
	\centering
	\begin{subfigure}[t]{0.45\textwidth}
		\centering
		\includegraphics[scale=0.42]{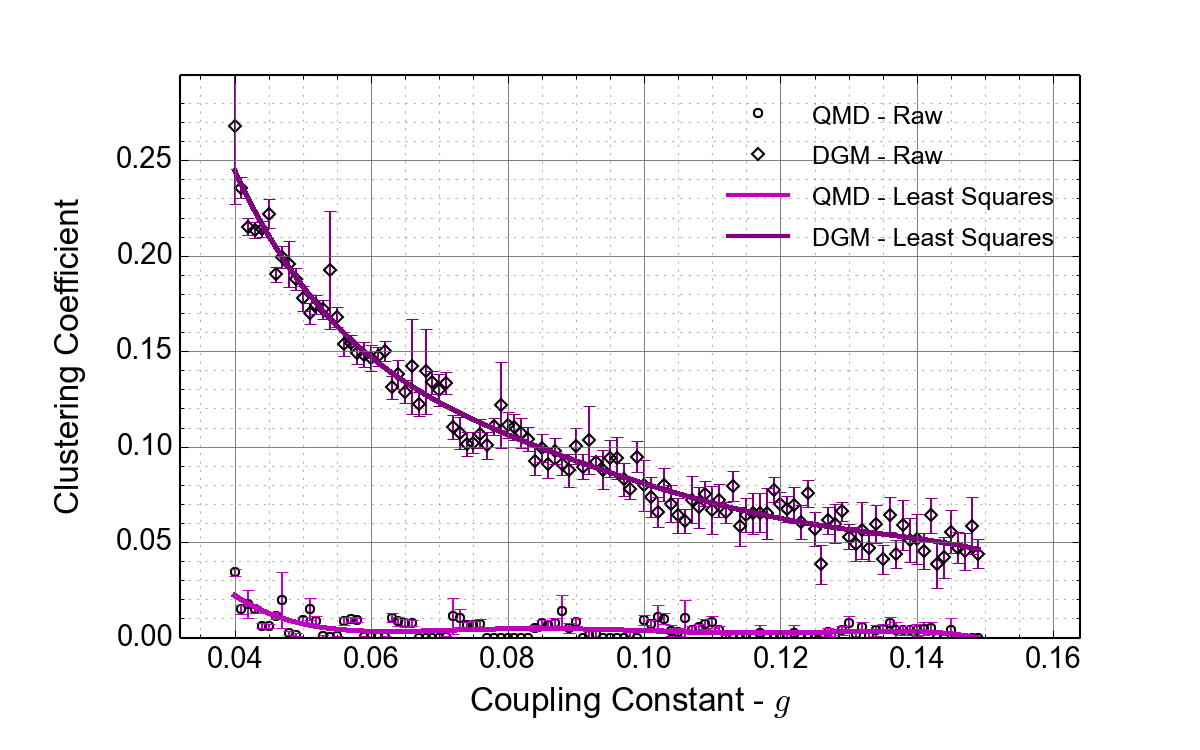}
		\caption{Averaged clustering coefficient, QMD and DGM, $N=50$.}
		\label{fig:CCN50}
	\end{subfigure}%
	~ 
	\begin{subfigure}[t]{0.45\textwidth}
		\centering
		\includegraphics[scale=0.42]{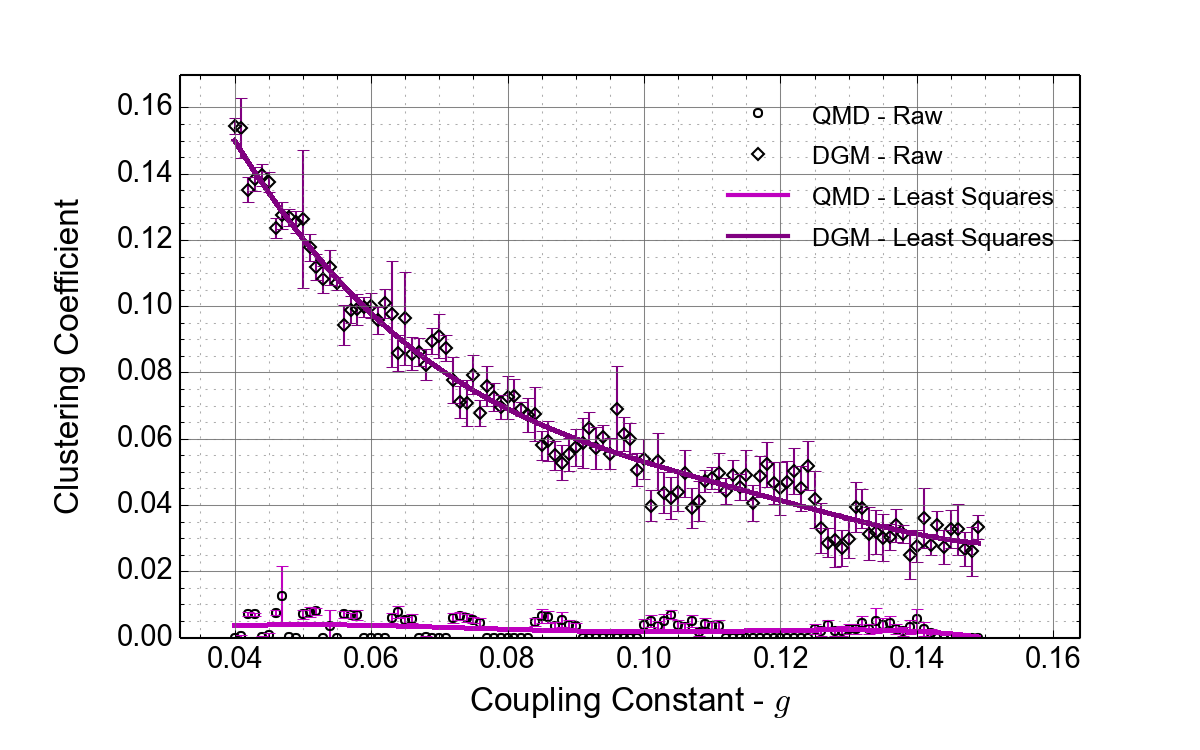}
		\caption{Averaged clustering coefficient, QMD and DGM, $N=75$.}
		\label{fig:CCN75}
	\end{subfigure}
	~ 
	\begin{subfigure}[t]{0.45\textwidth}
		\centering
		\includegraphics[scale=0.42]{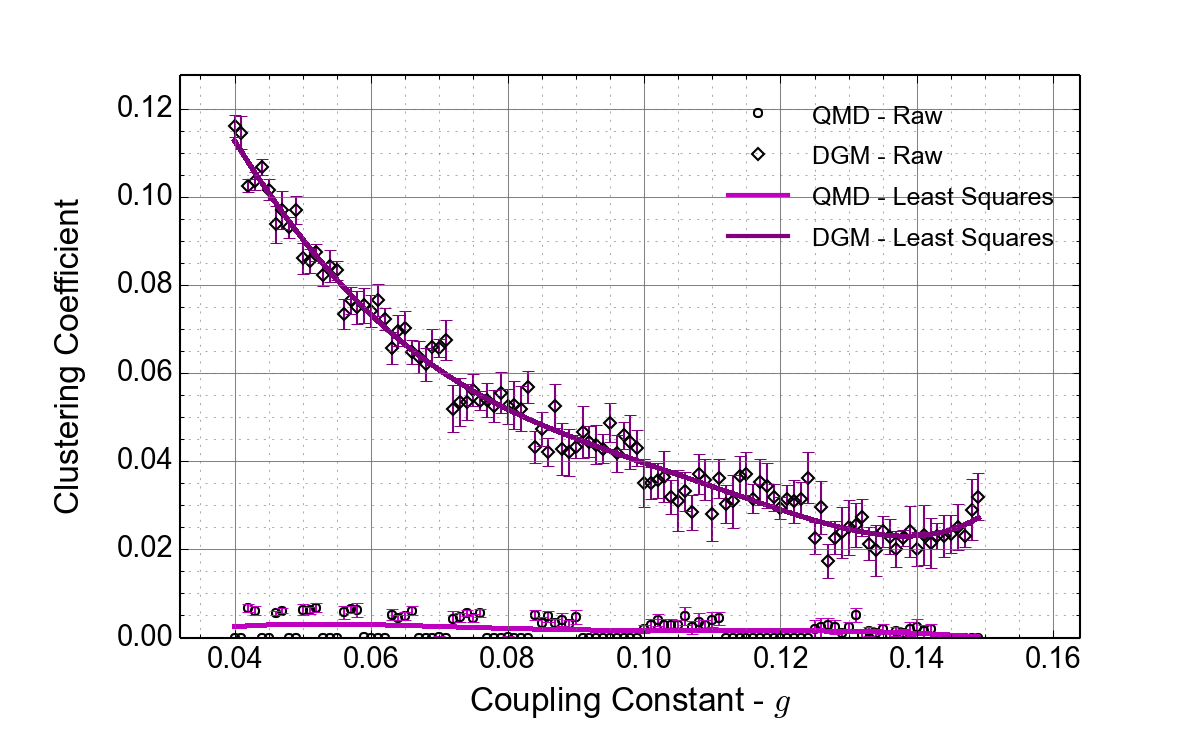}
		\caption{Averaged clustering coefficient, QMD and DGM, $N=100$.}
		\label{fig:CCN100}	
	\end{subfigure}%
	~ 
	\begin{subfigure}[t]{0.45\textwidth}
		\centering
		\includegraphics[scale=0.42]{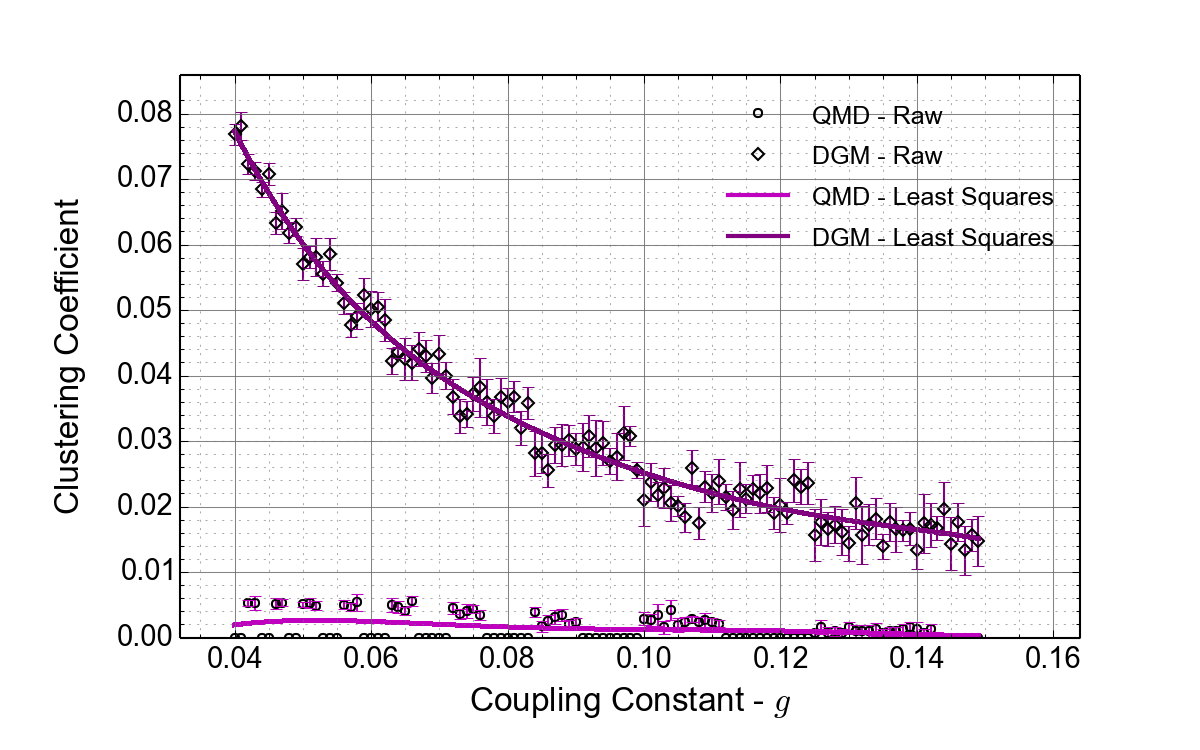}
		\caption{Averaged clustering coefficient, QMD and DGM, $N=150$.}
		\label{fig:CCN150}
	\end{subfigure}
	\caption{Comparison of both QMD and DGM clustering coefficients across a range of graph sizes. Each datapoint represents the average value across $10$ graph samples. Against each of the data points the error bars represent the $1^{st}$ standard deviation from the mean value. It is evident that the QMD ground state at each value of $N$ has a significantly lower degree of clustering.}
	\label{fig:clustering_coef}
\end{figure*}

\begin{figure*}[ht]
	\centering
	\begin{subfigure}[t]{0.45\textwidth}
		\centering
		\includegraphics[scale=0.42]{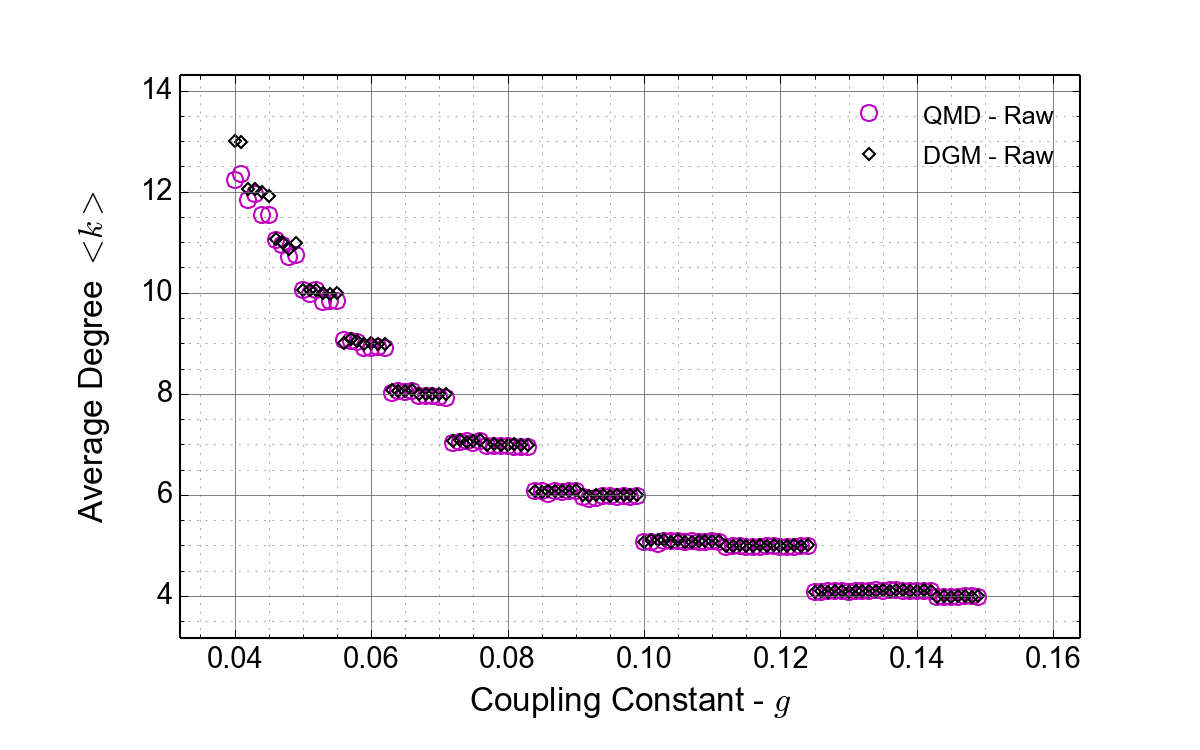}
		\caption{Averaged vertex degree, QMD and DGM, $N=50$.}
		\label{fig:KN50}
	\end{subfigure}%
	~ 
	\begin{subfigure}[t]{0.45\textwidth}
		\centering
		\includegraphics[scale=0.42]{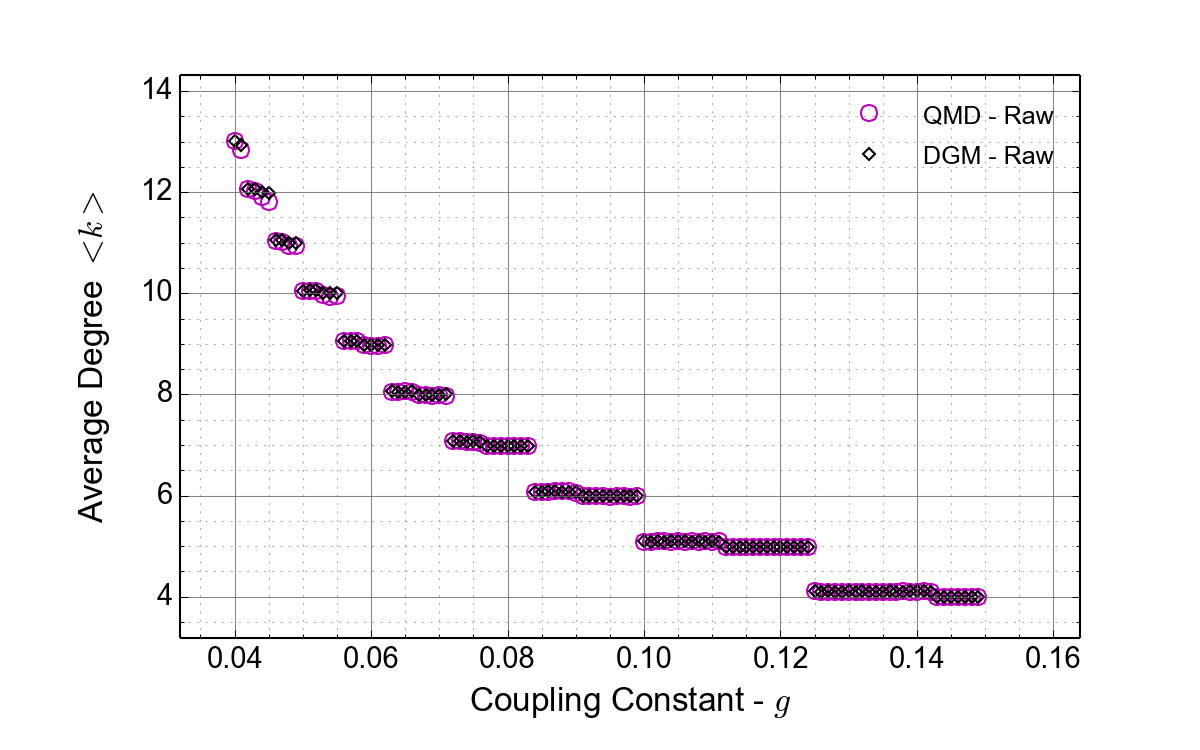}
		\caption{Averaged vertex degree QMD and DGM, $N=75$.}
		\label{fig:KN75}
	\end{subfigure}
	~ 
	\begin{subfigure}[t]{0.45\textwidth}
		\centering
		\includegraphics[scale=0.42]{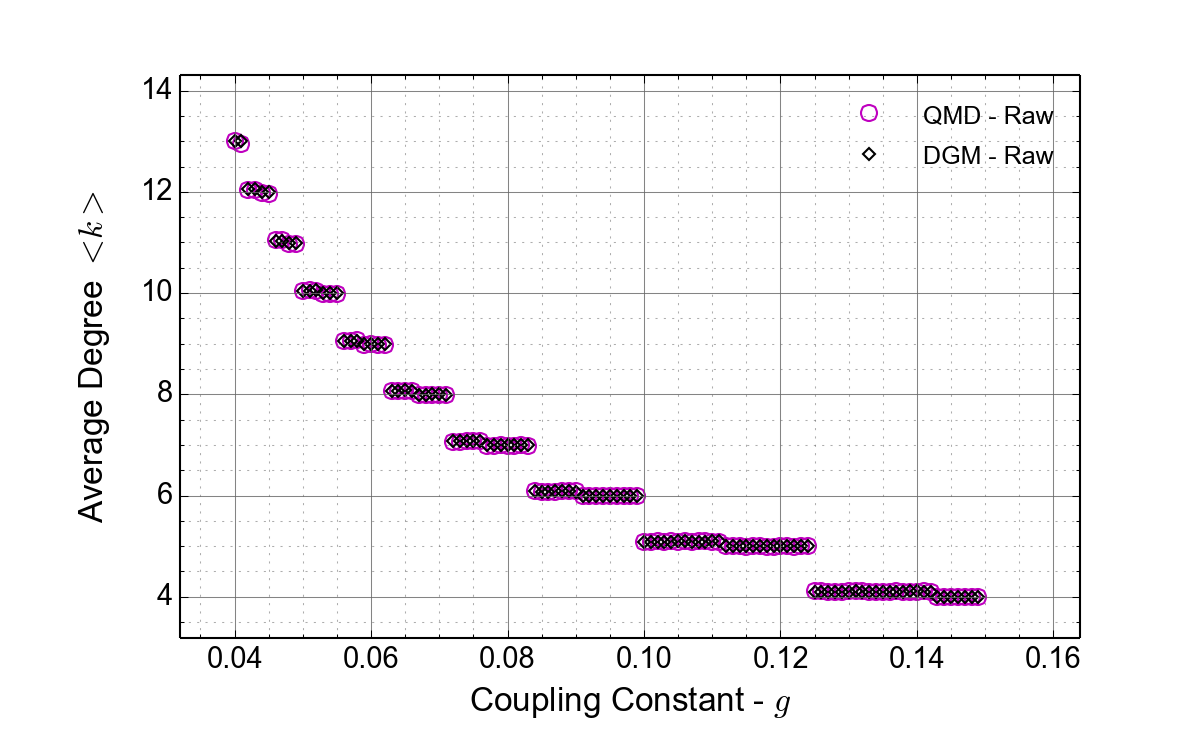}
		\caption{Averaged vertex degree, QMD and DGM, $N=100$.}
		\label{fig:KN100}	
	\end{subfigure}%
	~ 
	\begin{subfigure}[t]{0.45\textwidth}
		\centering
		\includegraphics[scale=0.42]{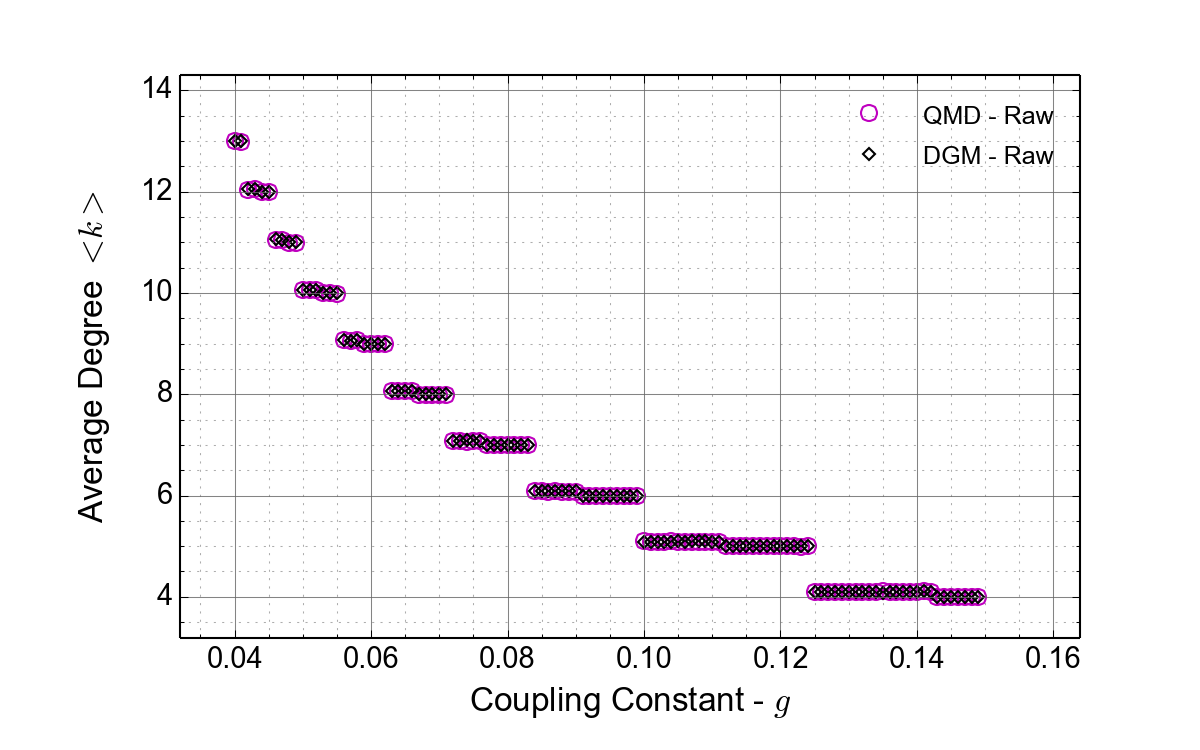}
		\caption{Averaged vertex degree, QMD and DGM, $N=150$.}
		\label{fig:KLN150}
	\end{subfigure}
	\caption{Comparison of both QMD and DGM  vertex degree across a range of graph sizes, with the valued averaged across $10$ graph samples. Against each of the data points the error bars represent the $1^{st}$ standard deviation from the mean value. Both models produce essentially identical values of $\langle k \rangle$.}
	\label{fig:node_degree}
\end{figure*}

We begin with an exploration of clustering coefficient and average node degree.
We plot in Figs. \ref{fig:clustering_coef}, \ref{fig:node_degree} and \ref{fig:comparative_pdgm}. clustering coefficient and average node degree results from the simulations for the DGM, PDGM and QMD Hamiltonian.
The effect of the additional terms in PDGM and QMD is very noticeable with considerably less clustering across a range of values of $N$ and $g$.
For the clustering coefficient values, it is also worth noting that the margin of error in the calculation indicates that the non-zero values may in fact not be significant.
For the PDGM model we compute the ground state both with and without the perturbation term in the fourth power of the adjacency matrix by setting $\lambda_4=0.001$ or $\lambda_4=0.0$ respectively, to determine the effect of the square favoring term.
Interestingly, without the square favoring term the model produces substantially the same outcome in terms of clustering as QMD, but for large values of $g$, QMD has substantially lower clustering than either variant of PDGM.
This is easy to understand as in QMD the benefit of the single coupling constant ensures that the energy penalty in forming a triangle is in proportion to the energy penalty of an open triple, maintaining the relative penalty of clustering, whereas in this initial simulation of PDGM  $\lambda_3$ is fixed at the value as originally proposed of $0.01$.  

%
%

In Fig. \ref{fig:node_degree}. and in Fig. \ref{fig:DegCompN100}. we plot the value of $\langle k \rangle$ as a function of the coupling constant $g$ for all models.
This is an important check to verify that the reduction in clustering in the QMD model is not simply a function of the ground state being significantly less connected.
Average clustering constant will reduce for smaller average node degree, as the number of edges in the graph is simply $\frac{N}{2} \langle k \rangle$.
What we can see in the plots is that the produced graphs have {\sl identical} average node degree, with the exception of the PDGM model with non zero $\lambda_4$.
At low values of $g$ the average node degree of the ground states graphs of this model is noticeably smaller than the values for DGM, QMD and PDGM with $\lambda_4=0$.

Using Eq. \eqref{eqn:ground_k} we can  define the degree dimension of the graph $d_k$ in terms of the minimum calculated degree of the graph $k_{\min}$ as $2d_k=k_{\min}$, reflecting the $k$-regularity of the emerged ground state. 
It is the most basic measure of dimensionality in a graph that is only exact when the $k$-regular mesh is lattice like.
We can however use this to check the accuracy of the simulation against the model by computing the ratio of $d_k$ to $\frac{1}{2}\langle k \rangle$.
For the QMD model this ratio is $0.997 \pm 0.039$, for DGM $0.999 \pm 0.041$, PDGM $1.017 \pm 0.053$, and for PDGM with $\lambda_4=0.0$ it is $0.993 \pm 0.019$.
These results indicate that all the simulations are reproducing the expected and computed minimum energy state with the anticipated average degree, with the exception of PDGM with $\lambda_4=0.001$ where the result is somewhat weaker.
This is a result of the noticeably lower average degree for low values of $g$ visible in Fig. \ref{fig:DegCompN100}.
In Table \ref{tab:clustering} we use this dimension, computed at different values of $g$ to compare the clustering coefficient extracted from the simulations for both models, and also the values published in the original paper by Trugenberger \cite{Trugenberger2015}.
We can see that our method for calculating the  ground state of the DGM Hamiltonian reproduces the published results within the margin of experimental error, providing validation of the algorithm outlined in this section.

\begingroup
\begin{table*}[h]
\centering
\begin{tabular}{|c|c|c|c|c|}
\hline
$d_k$ & $2$ & $3$ & $4$ & $5$ \\ \hline
QMD $NC_G$ & $0.187 \pm 0.23$ & $0.465 \pm 0.34$ & $0.610 \pm 0.12$ & $0.680 \pm 0.20$ \\ \hline
PDGM ($\lambda_4=0.001$) $NC_G$ & $1.30 \pm 0.12$ & $1.29 \pm 0.10$ & $0.859 \pm 0.08$ & $6.64 \pm 1.44$ \\ \hline
PDGM ($\lambda_4=0.0$) $NC_G$ & $0.188 \pm 0.23$ & $0.311 \pm 0.18$ & $0.111 \pm 0.05$ & $0.000 \pm 0.00$ \\ \hline
DGM $NC_G$ &  $2.30 \pm 0.60$ & $4.33 \pm 0.54$ & $6.48 \pm 0.55$ & $8.74 \pm 0.41$ \\ \hline
DGM $NC_G$ \cite{Trugenberger2015} &  $2.4 \pm 0.15$ & $4.52 \pm 0.09$  & $6.27 \pm 0.22$ & $8.39 \pm 0.17$ \\ \hline
\end{tabular}
\caption{Values of clustering coefficient, multiplied by network size for $N=100$ for QMD, DGM and both variants of PDGM. The significant reduction in clustering in the QMD and PDGM models are evident. Also reproduced from \cite{Trugenberger2015} are the published values for clustering, which are in line with our results.}
\label{tab:clustering}
\end{table*}
\endgroup

We also investigated the degree distributions  of the graphs produced as ground states of the model Hamiltonians, and for this experiment chose a fixed coupling constant of $g=0.07$, which is at the midpoint of the values simulated.
For a run of $10$ graphs of $100$ nodes, in all models close to $100\%$ of the nodes have degree exactly equal to $\langle k \rangle$.
For QMD, for a sample of $1000$ nodes all but $12$ have degree $\langle k \rangle$, the rest have degree $\langle k \rangle -1$.
In the case of DGM the values are $996$ and $4$ respectively.
In the case of PDGM, the results for $\lambda_4=0.0$ are identical to QMD, but in the case of $\lambda_4=0.001$ there is an interesting difference.
Although there is a similar peak at around $\langle k \rangle$, it is somewhat less pronounced, accounting for $924$ nodes.
A significant number, $67$, have degree $\langle k \rangle +1$ and $9$ nodes have degree $\langle k \rangle -1$.
These results are within the margin of error of being able to claim that all models produce $k$-regular graphs, but it would appear in the PDGM with $\lambda_4=0.001$, that claim is somewhat weaker.
It is tempting to speculate that the number of $\langle k \rangle -1$ degree nodes may indicate the presence of a lower dimensional boundary to the emerged geometry, as a closed $k$-regular graph would naturally have zero nodes of differing degree.
It is precisely this claim that is addressed in the discussion of PDGM in the works of Trugenberger \cite{Trugenberger2016,Trugenberger2017}.
It is however beyond the scope of this work to rigorously prove this, and it remains an open question.

\subsection{Dimensionality}
\label{ssec:dimensionality}
A critical feature of the emerged mesh is the dimensionality of the spacetime geometry that it represents.
The dimension of a graph has many definitions, categorized as intrinsic or extrinsic.
An intrinsic measure of dimensionality does not depend upon any embedding of the graph in a space of the same or higher dimensions, and corresponds to that which would be experienced by an observer in the mesh.
These measures of dimension are well understood and I direct the reader to the standard text by Ambj{\o}rn {\sl et al} \cite{Ambjorn1997}.

\begin{figure*}[hpt]
	\centering
	\begin{subfigure}[t]{0.45\textwidth}
		\centering
		\includegraphics[scale=0.43]{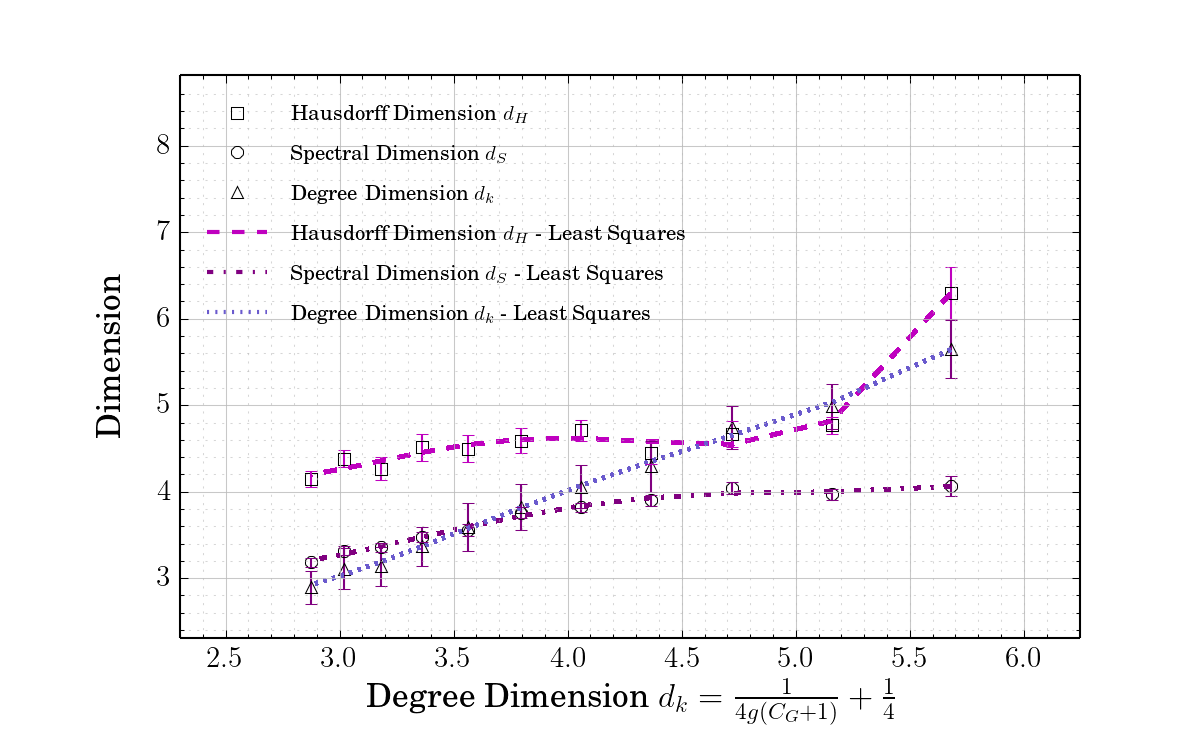}
		\caption{Extrinsic, Intrinsic and Graph dimension for QMD across a range of $d_k$, computed using Eq. \ref{eqn:ground_k}.}
		\label{fig:QMDPhase}
	\end{subfigure}%
	~ 
	\begin{subfigure}[t]{0.45\textwidth}
		\centering
		\includegraphics[scale=0.43]{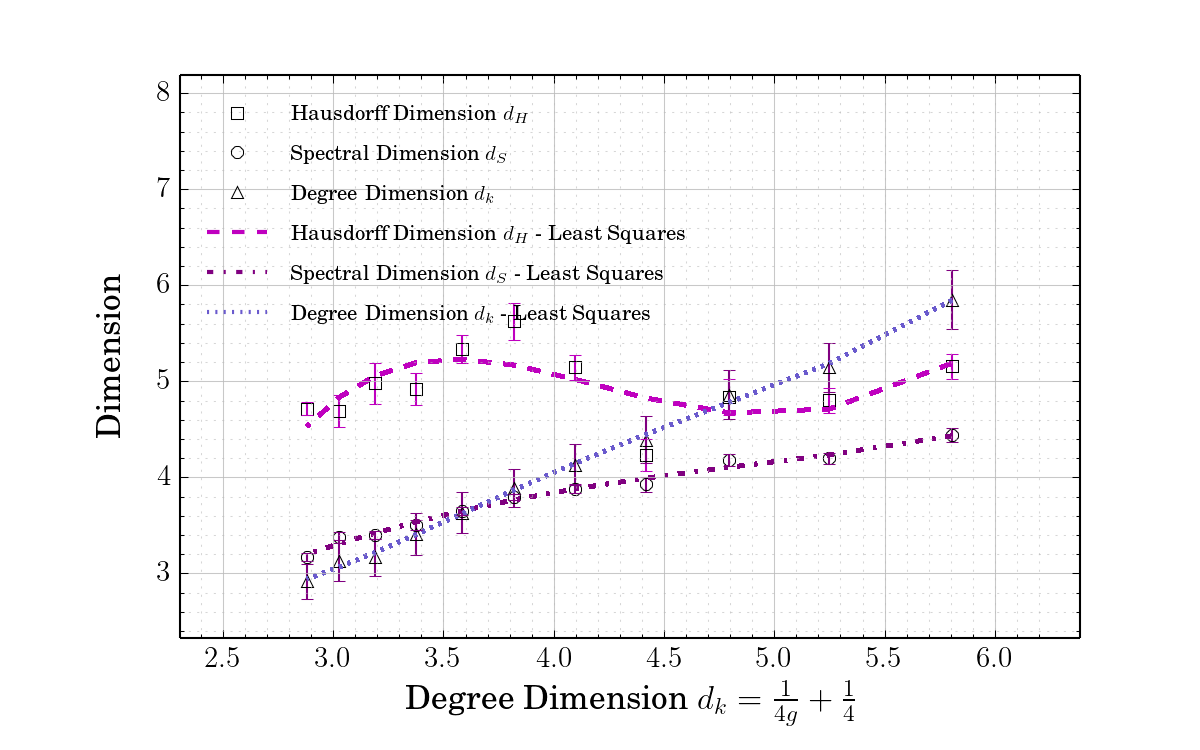}
		\caption{Extrinsic, Intrinsic and Graph dimension for DGM across a range of $d_k$, computed using Eq. \ref{eqn:trug_min}.}
		\label{fig:DGMPhase}
	\end{subfigure}
	~ 
	\begin{subfigure}[t]{0.45\textwidth}
		\centering
		\includegraphics[scale=0.43]{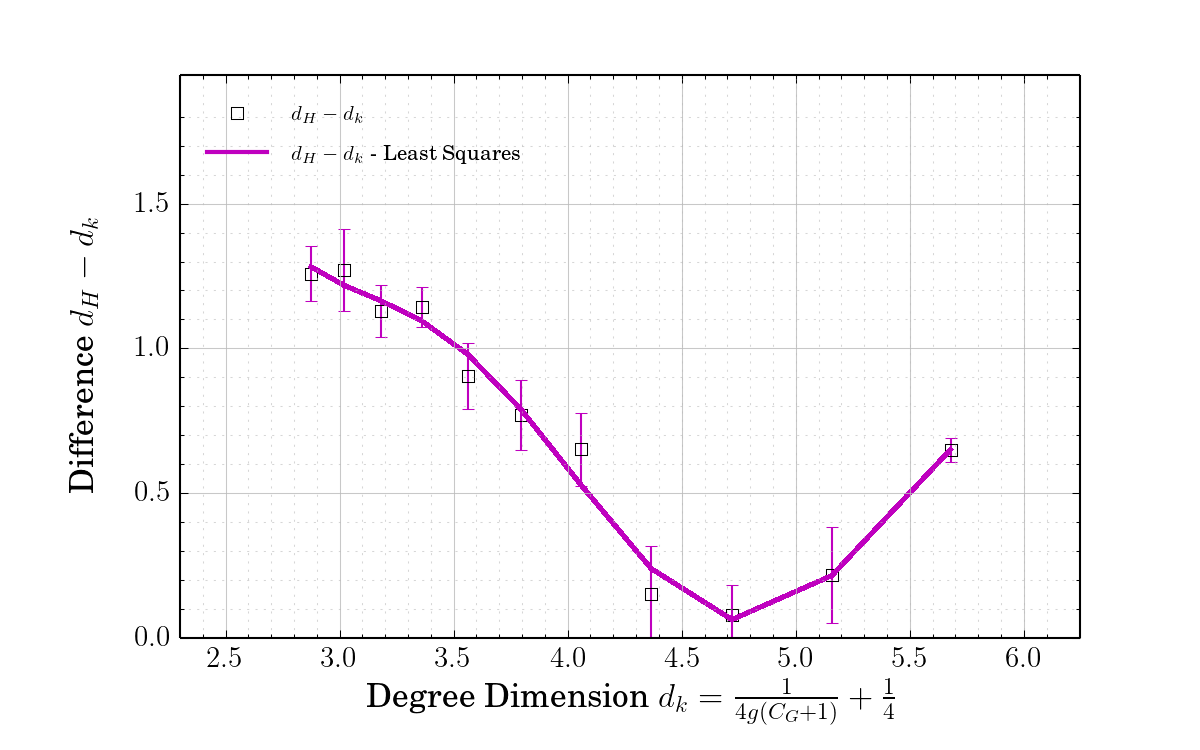}
		\caption{Difference between $d_H$ and $d_S$ for QMD across a range of $d_k$, computed using Eq. \ref{eqn:ground_k}.}
		\label{fig:QMDDiv}	
	\end{subfigure}%
	~ 
	\begin{subfigure}[t]{0.45\textwidth}
		\centering
		\includegraphics[scale=0.43]{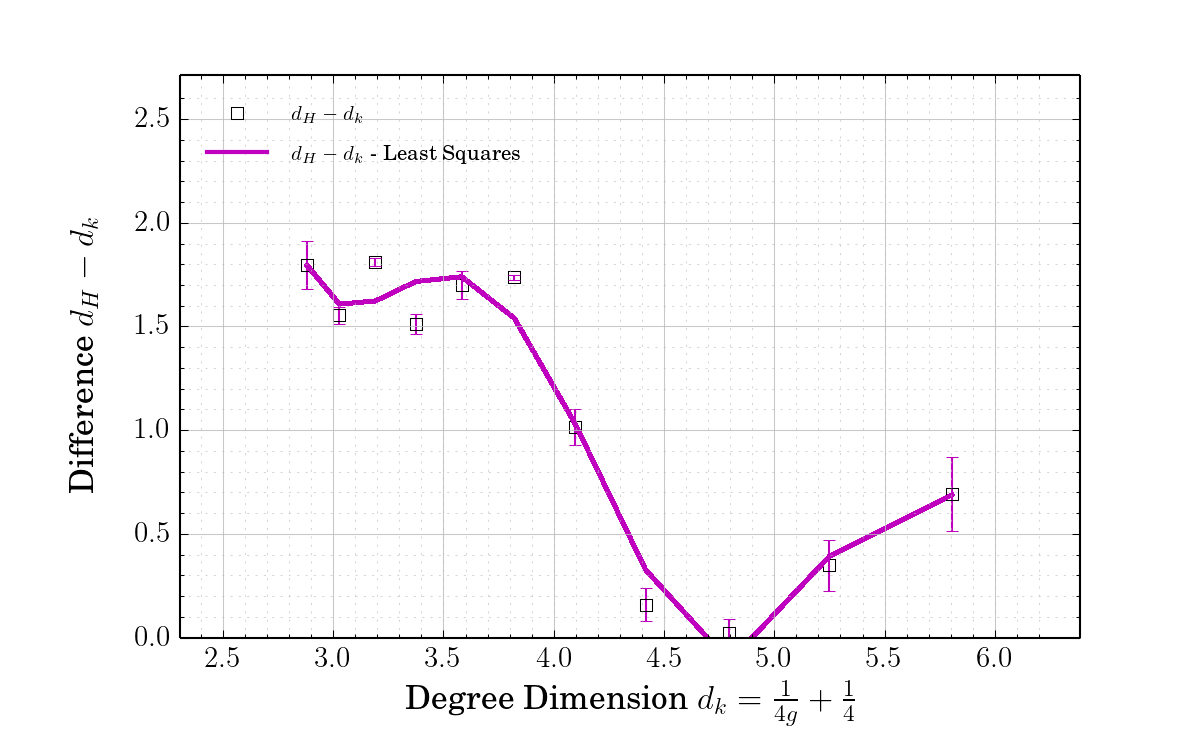}
		\caption{Difference between $d_H$ and $d_S$ for DGM across a range of $d_k$, computed using Eq. \ref{eqn:trug_min}.}
		\label{fig:DGMDiv}
	\end{subfigure}
	~ 
	\begin{subfigure}[t]{0.45\textwidth}
		\centering
		\includegraphics[scale=0.43]{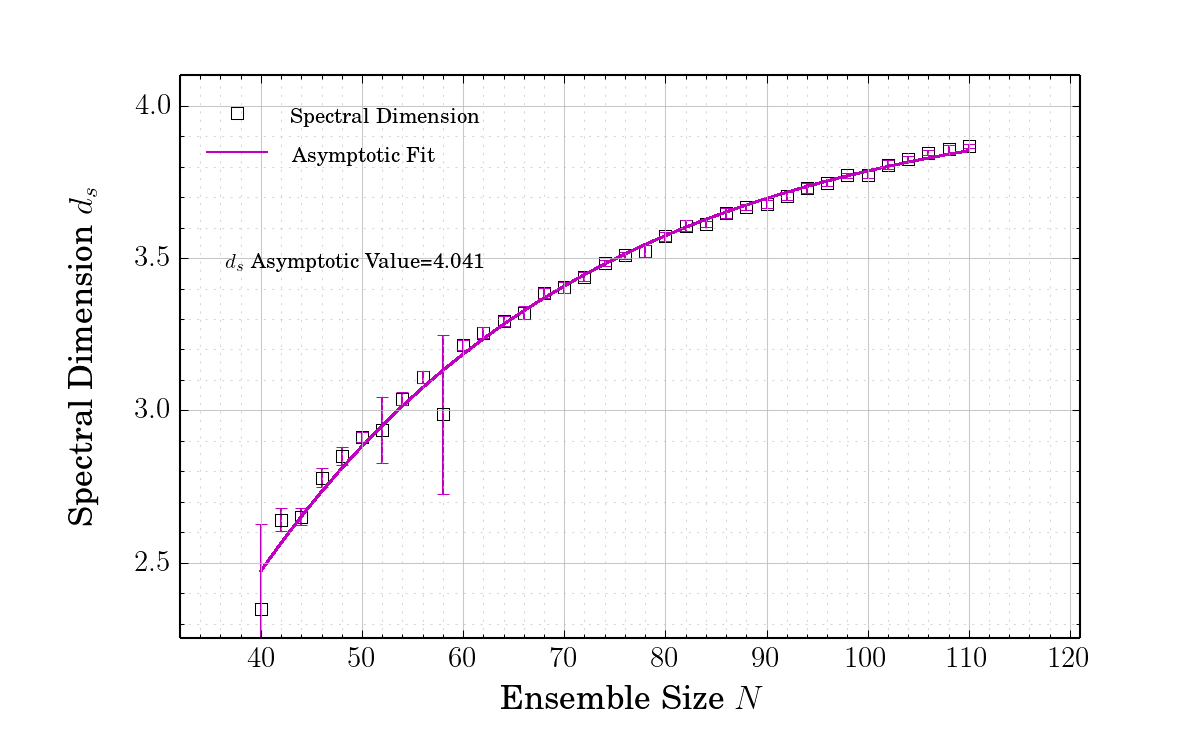}
		\caption{Evolution of spectral dimension $d_S$ with increasing network size $N$, for QMD, and $g=0.055$.}
		\label{fig:QMDSpec}	
	\end{subfigure}%
	~ 
	\begin{subfigure}[t]{0.45\textwidth}
		\centering
		\includegraphics[scale=0.43]{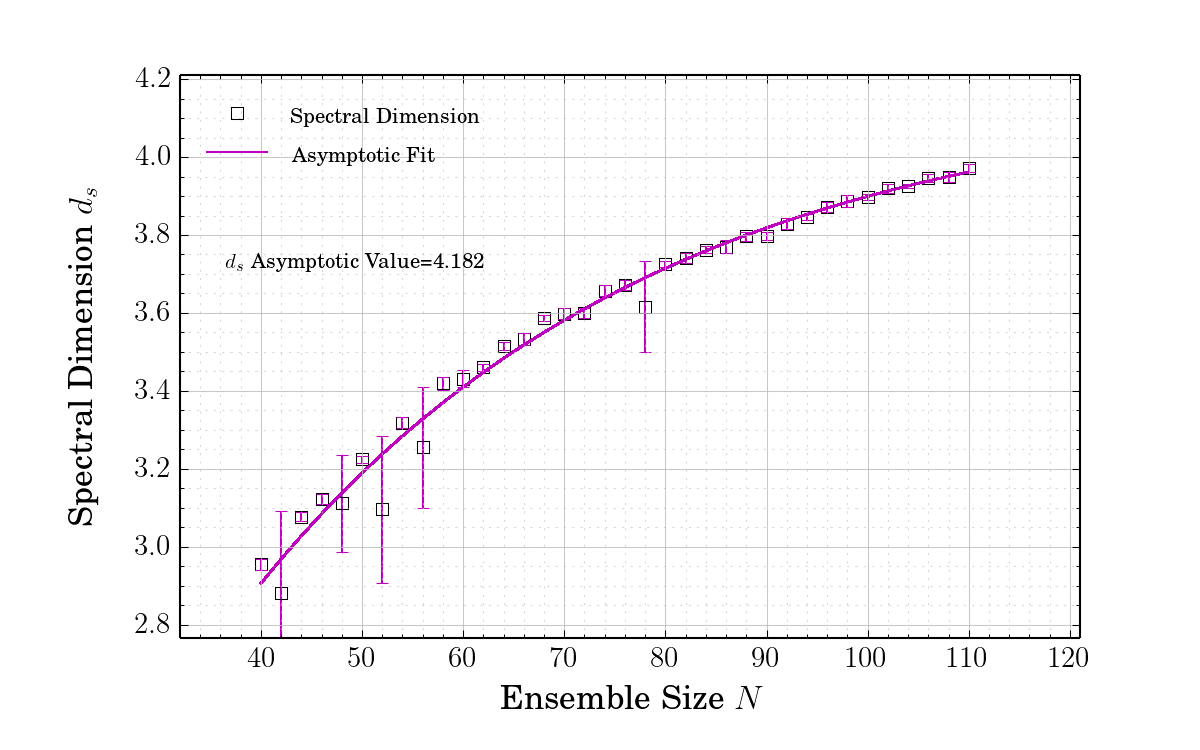}
		\caption{Evolution of spectral dimension $d_S$ with increasing network size $N$, for DGM, and $g=0.055$.}
		\label{fig:DGMSpec}
	\end{subfigure}
	\caption{Dimensional analysis of the graphs obtained as the ground state of QMD and DGM. Displayed are the three measures of dimensionality against graph dimension, the divergence of the extrinsic and graph measures and the evolution of spectral dimension with $N$.}
	\label{fig:dimension}
\end{figure*}

The most common of the intrinsic measures is the spectral dimension $d_S$, which measures the return probability of a random walk on the graph.
The walk is defined by starting at a random node, and at each time step moving to another connected vertex by randomly selecting an edge and traversing it.
As described in Ambj{\o}rn {\sl et al} \cite{Ambjorn1997}, for $t$ time steps, the probability of returning to the initial vertex $p(t)=t^{-d_S/2}$.
Unfortunately this is only well defined on an infinite graph, as the relation to the `spectral' dimension $d_S$ is only valid in the limit of $t \rightarrow \infty$, and for a finite graph  $p(\infty) = 1$.
It is possible to estimate the value of the spectral dimension by restricting the length of the random walk to time steps $t < N$, across a sample of graphs for different values of $N$.
When this is done, using a fit to the scaling law it is possible to extract $d_S$ for each value of $N$, and obtain the function for the dependence of $d_S$ upon $N$,  $d_S(N)$.
As $N \rightarrow \infty$ the value of $d_S(N)$ will tend to the final value.
Figures \ref{fig:DGMSpec} and \ref{fig:QMDSpec}  are presented as a representative sample of this scaling against increasing network size $N$, and also a fit of the curve to a test function.
To extract a value for the limit $N \rightarrow \infty$, a number of test functions were tested, and the best  least squares fit to the data was obtained using
\begin{equation}\label{eqn:test_fn}
	d_S(N) = d_S(1-be^{-aN})\mbox{.}
\end{equation}
With this scaling function it is possible to extrapolate to obtain the value of the spectral dimension. 
For a selection of graphs with $N$ ranging from $40$ to $110$, the Pearson correlation coefficient $\rho$ of the fit to Eq. \eqref{eqn:test_fn} was generally in the range of $0.96$ to $0.98$.

%
%
\begin{figure*}[ht]
	\centering
	\begin{subfigure}[t]{0.45\textwidth}
		\centering
		\includegraphics[scale=0.42]{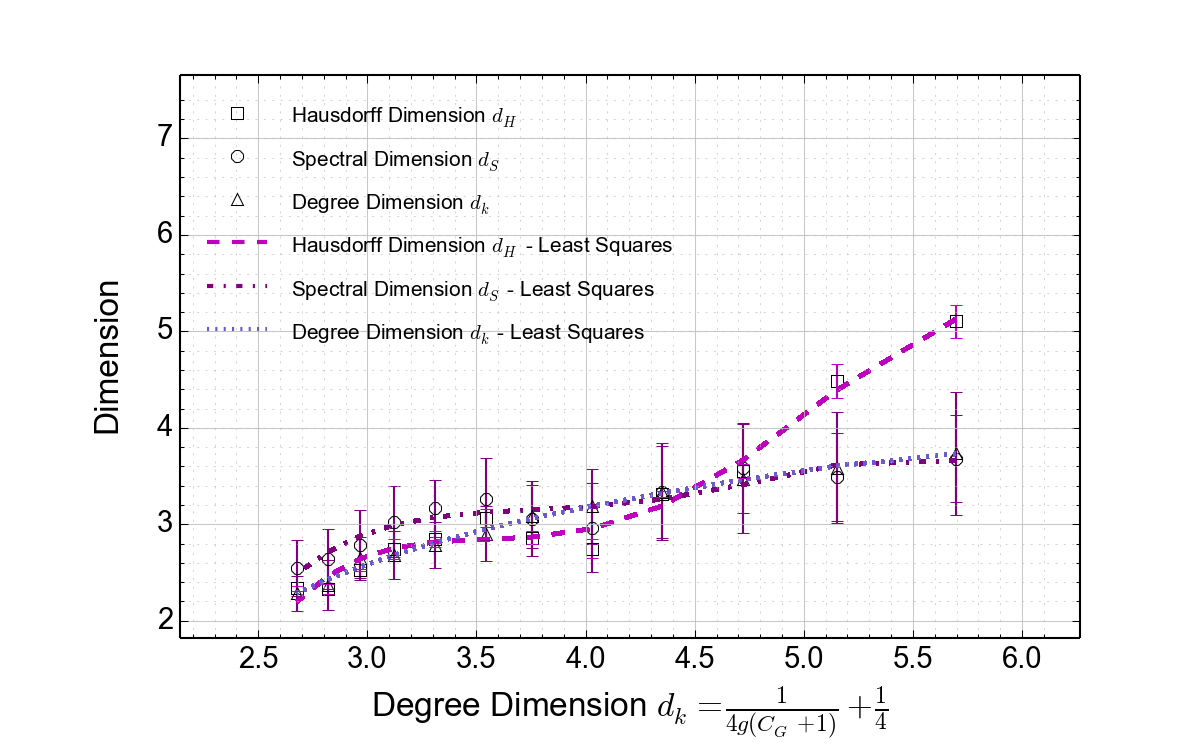}
		\caption{Extrinsic, Intrinsic and Graph dimension for PDGM with $\lambda_4=0.001$ across a range of $d_k$, computed using Eq. \ref{eqn:pdgm}.}
		\label{fig:PDGMLam4Zero}
	\end{subfigure}%
	~ 
	\begin{subfigure}[t]{0.45\textwidth}
		\centering
		\includegraphics[scale=0.42]{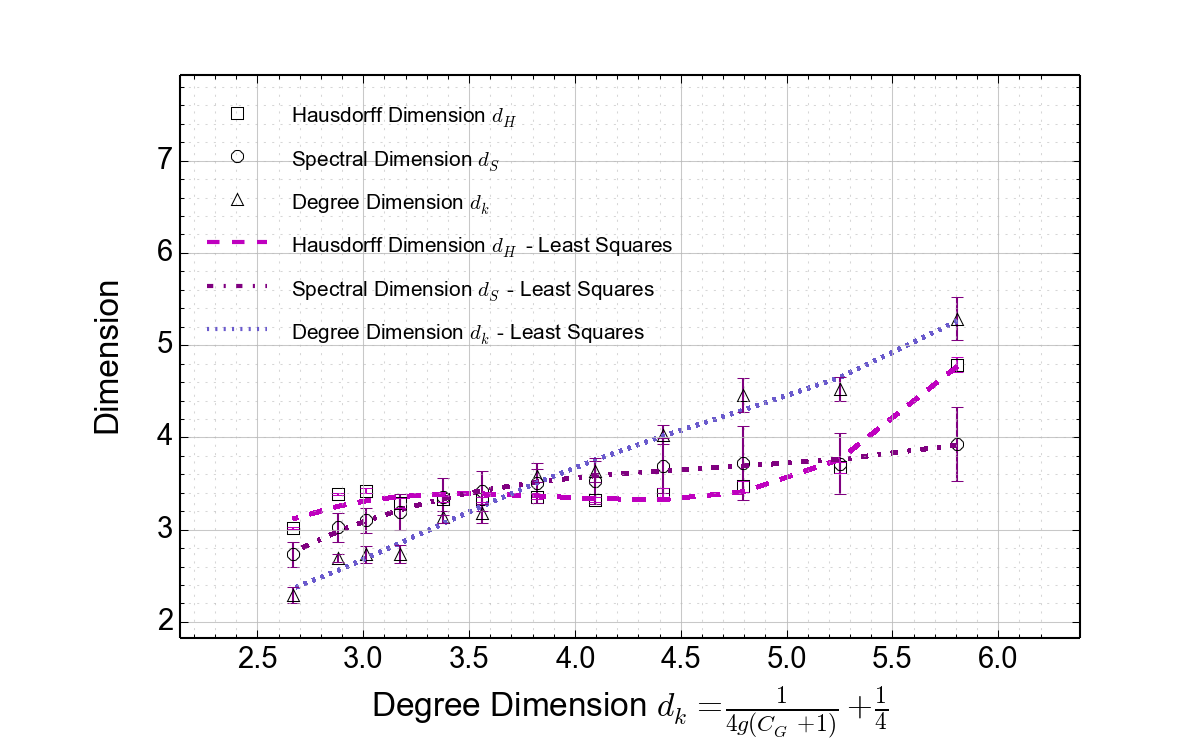}
		\caption{Extrinsic, Intrinsic and Graph dimension for PDGM with $\lambda_4=0.0$  across a range of $d_k$, computed using\\ Eq. \ref{eqn:pdgm}.}
		\label{fig:PDGMLam4Pos}
	\end{subfigure}
	~ 
	\begin{subfigure}[t]{0.45\textwidth}
		\centering
		\includegraphics[scale=0.42]{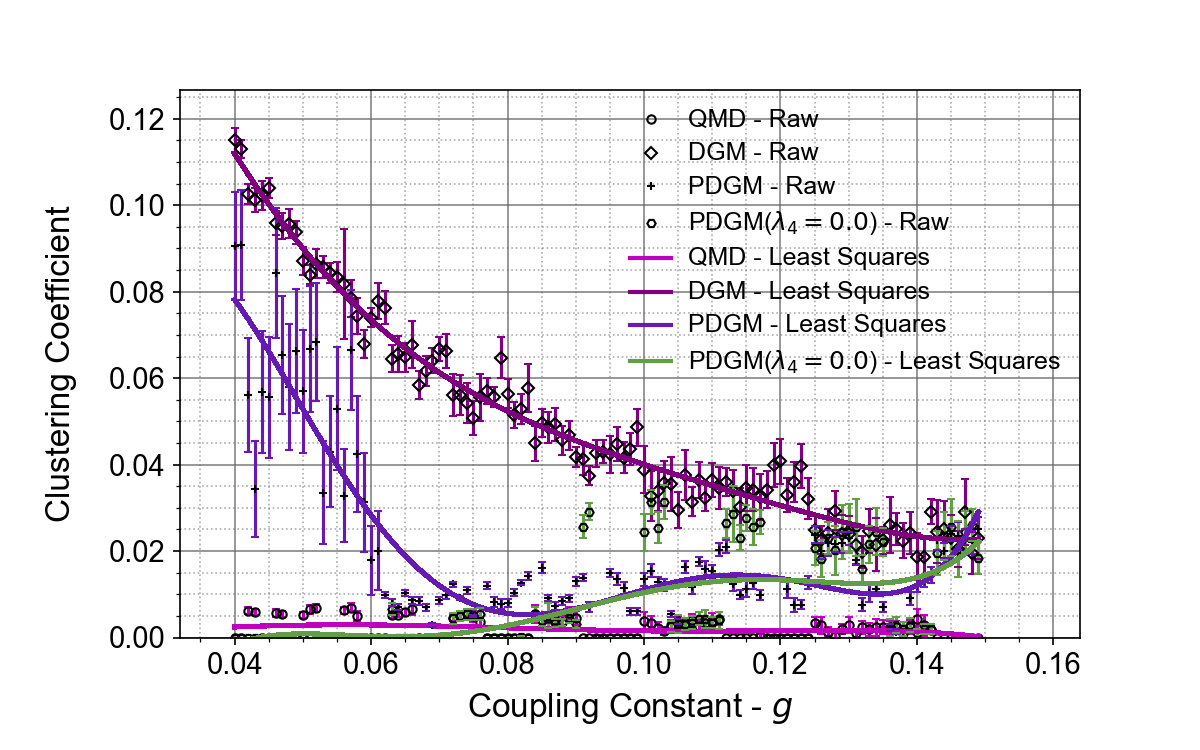}
		\caption{Averaged clustering coefficient, QMD, DGM, PDGM and PDGM with $\lambda_4=0.0$, and $N=75$.}
		\label{fig:CCompN100}
	\end{subfigure}%
	~ 
	\begin{subfigure}[t]{0.45\textwidth}
		\centering
		\includegraphics[scale=0.42]{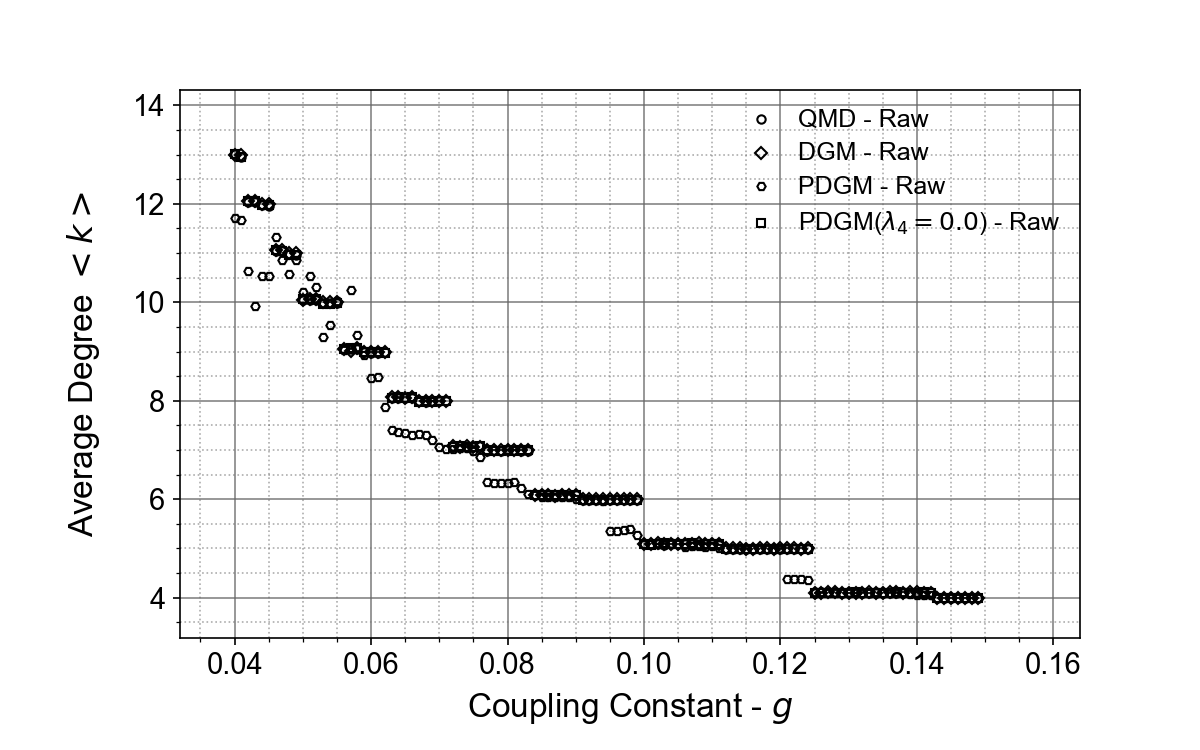}
		\caption{Averaged vertex degree, QMD, DGM, PDGM and PDGM with $\lambda_4=0.0$, and $N=75$}
		\label{fig:DegCompN100}
	\end{subfigure}
	\caption{Comparison of both QMD, DGM  and PDGM clustering coefficients and average degree across a range of graph sizes. Each datapoint represents the average value across $10$ graph samples. Against each of the data points the error bars represent the $1^{st}$ standard deviation from the mean value. It is evident that the QMD and PDGM models have a significantly lower degree of clustering, but that no significant advantage is gained from the $\lambda_4$ term in PDGM.}
	\label{fig:comparative_pdgm}
\end{figure*}

Turning to extrinsic measures, the standard measure, as described in Ambj{\o}rn {\sl et al} is the Hausdorff dimension, $d_H$.
This is defined on any sized graph, and so does not suffer from the scaling issues of the spectral dimension.
It measures how volume and area scale as the size of the graph increases.
Standard convention is that volume $V$ in this instance is the average distance between all pairs of nodes in the graph, which can be readily obtained from the adjacency matrix, or the use of a standard algorithm such as a breadth-first search \cite{Barabasi2016}.
Using this measure of volume, it is a standard result that this scales with the relation $V \propto N^{1/d_H}$ \cite{Ambjorn1997}, and by sampling the average distance from the same sample graphs used to calculate the spectral dimension it is possible to compute $d_H$.
In addition to the extrinsic measure,  the graph dimension $d_k$ is also computed using the Eq. \eqref{eqn:ground_k}.

In  Figs \ref{fig:PDGMLam4Zero}. \ref{fig:PDGMLam4Pos}. and Fig. \ref{fig:dimension}. we display the results of this analysis.
For PDGM computation constraints do not permit the same range of samples.
In the case of QMD and DGM,  Figs. \ref{fig:QMDPhase}. and \ref{fig:DGMPhase}. indicate that the graph and spectral dimensions converge at around $d_k=3.5$ to $4.0$ for both models.
In Figs. \ref{fig:QMDDiv}. and \ref{fig:DGMDiv}. we plot the differences between $d_H$ and $d_k$. 
For these measures the convergence is best for the range $d_k=4.0$ to $5.0$.
It is worthwhile remarking that in Fig. \ref{fig:DGMPhase}.  the obtained values of both the spectral and Hausdorff dimensions behave differently to QMD.
In particular $d_H$ is somewhat lower in DGM and $d_S$ is somewhat higher.
This can be explained by the presence of clustering in the graphs produced by DGM.
Essentially clustering introduces the `small world' property into the graphs \cite{Watts1998}, which in turn leads to average distances in the graph being smaller than a graph with no clustering (lower $d_H$), and a higher probability that a random walk will transport the walker to a distant location in the graph (higher $d_S$).
For QMD, and PDGM this `small world' property is replaced by an emerged `large world' \cite{Trugenberger2016}, which possesses a ground state with locality in line with a real world geometry.
Close examination of Figs \ref{fig:PDGMLam4Zero}. and  \ref{fig:PDGMLam4Pos}. underpins the observation that PDGM with $\lambda_4=0.0$ produces essentially the same results as QMD.
In particular, convergence of the dimensionality metrics occurs in the same range, but with less variance than is evident in the QMD model.
In the case of $\lambda_4=0.001$ the pattern is similar, but with a curious close correlation of spectral dimension with degree dimension in the low $g$ (high degree, high temperature) sector.
This could be as a result of the lower average degree at small values of $g$, and an effectively constant clustering suppressing both $d_k$ and $d_S$.
Although for our simulations the value of the preferred dimension, where the different measures of dimension converge, is not as precise as produced by Trugenberger, it is nevertheless supportive of a hypothesis that there is a preferred dimension, at or around $3.5$ to $5$.
Given the low sample size of the graphs, it is entirely possible that with a bigger sample set this result could be refined to something closer to the value of $3$ or $4$ familiar in classical spacetime.

\subsection{Probing the PDGM Parameter Space}

\begin{figure*}[ht]
	\centering
	\begin{subfigure}[t]{0.3\textwidth}
		\centering
		\includegraphics[scale=0.35]{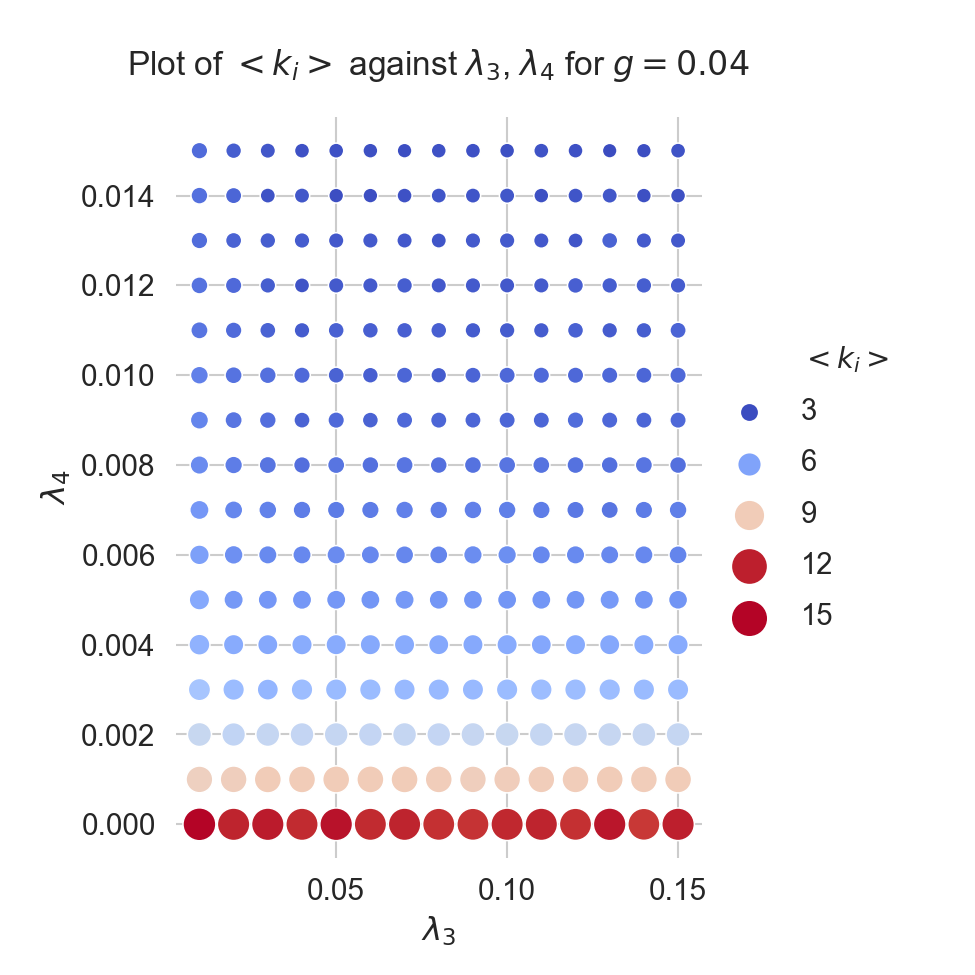}
	\end{subfigure}%
	~ 
	\begin{subfigure}[t]{0.3\textwidth}
		\centering
		\includegraphics[scale=0.35]{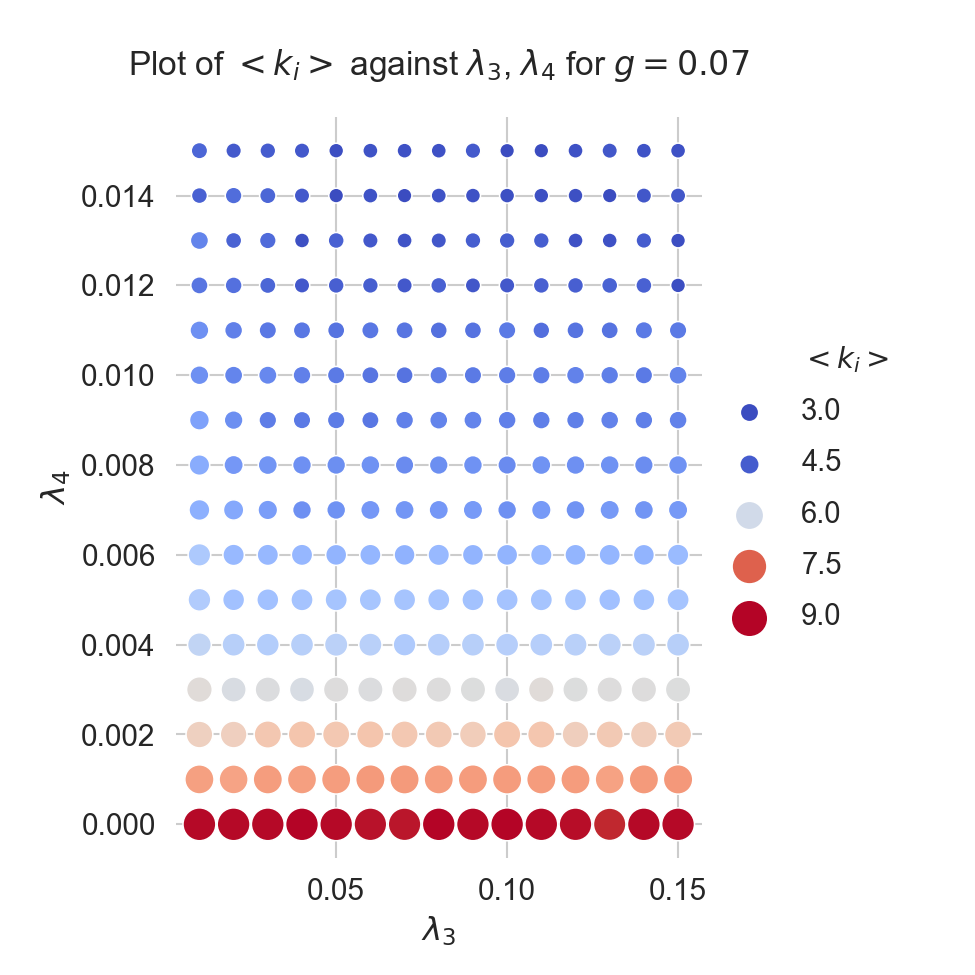}
	\end{subfigure}
	~ 
	\begin{subfigure}[t]{0.3\textwidth}
		\centering
		\includegraphics[scale=0.35]{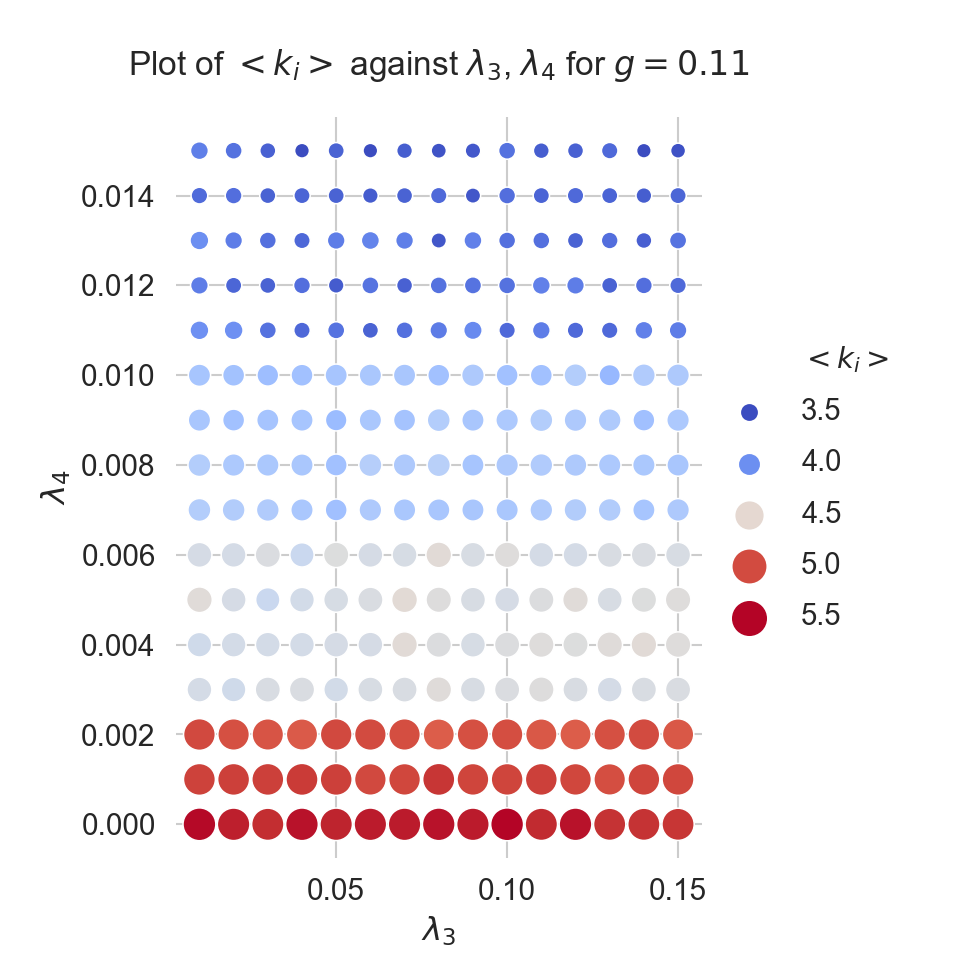}
	\end{subfigure}
	~
	\begin{subfigure}[t]{0.3\textwidth}
		\centering
		\includegraphics[scale=0.35]{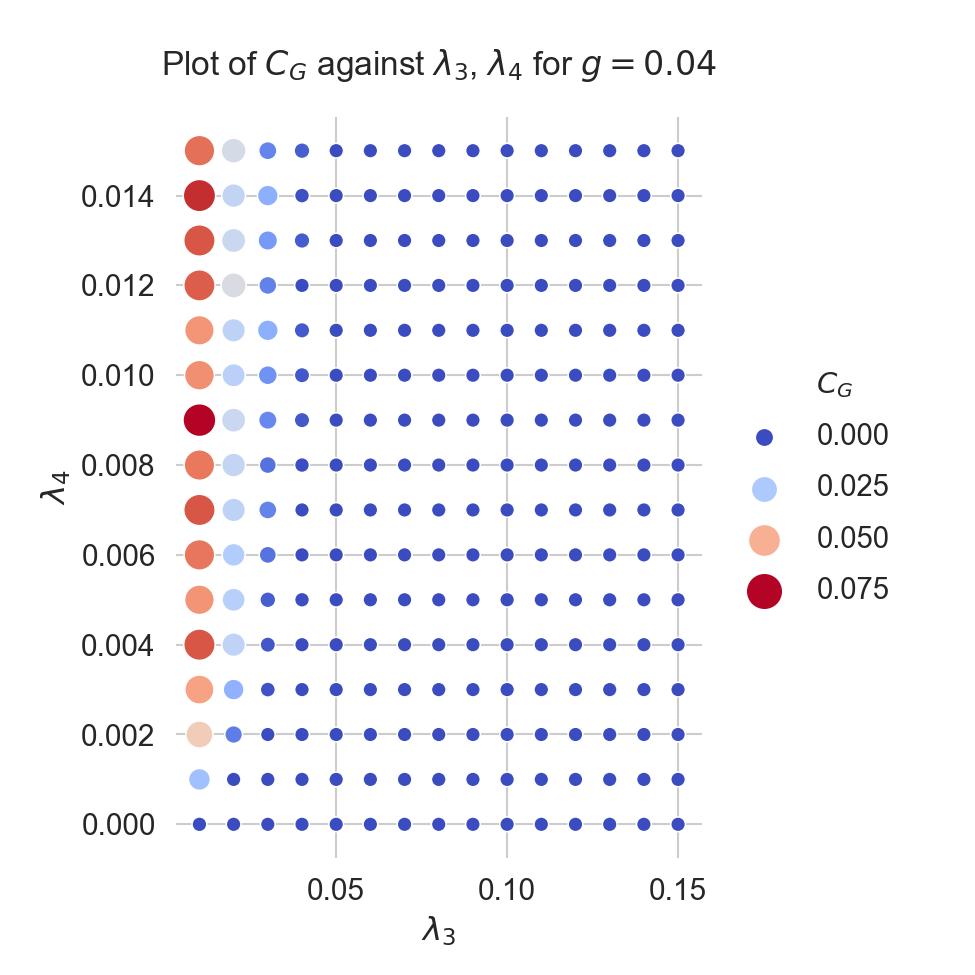}
	\end{subfigure}%
	~ 
	\begin{subfigure}[t]{0.3\textwidth}
		\centering
		\includegraphics[scale=0.35]{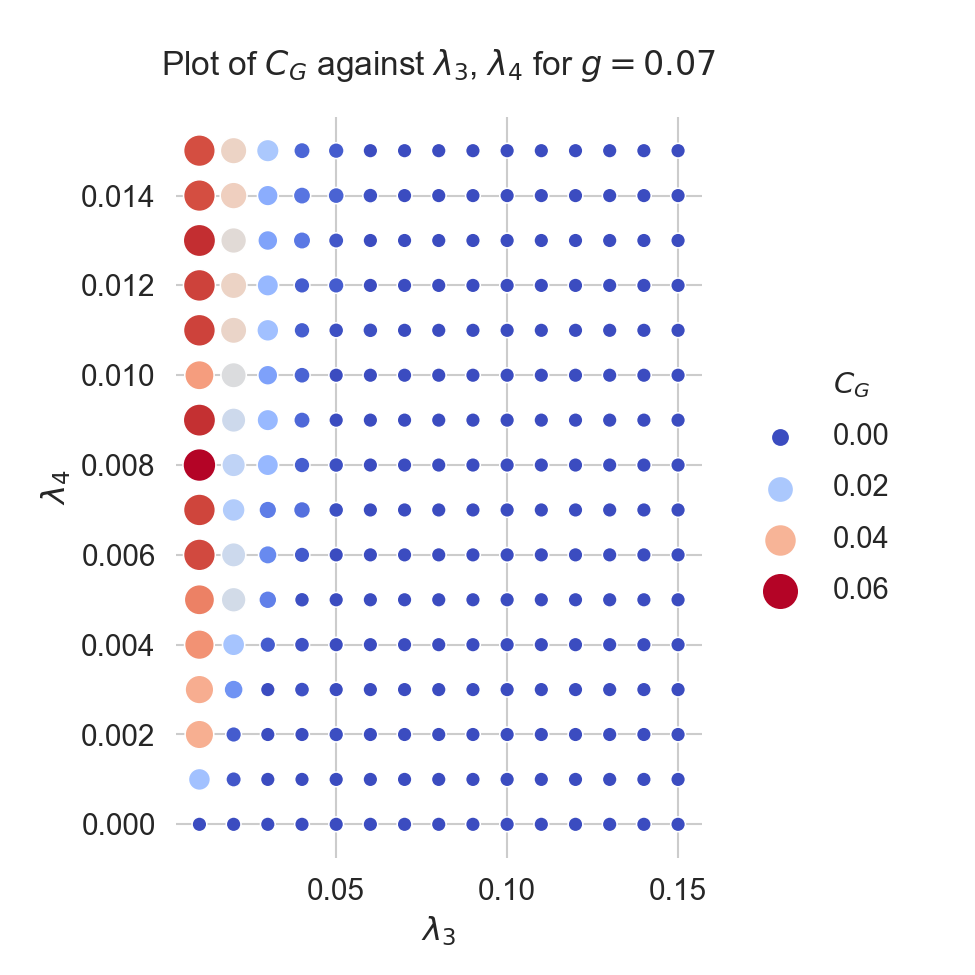}
	\end{subfigure}%
	~ 
	\begin{subfigure}[t]{0.3\textwidth}
		\centering
		\includegraphics[scale=0.35]{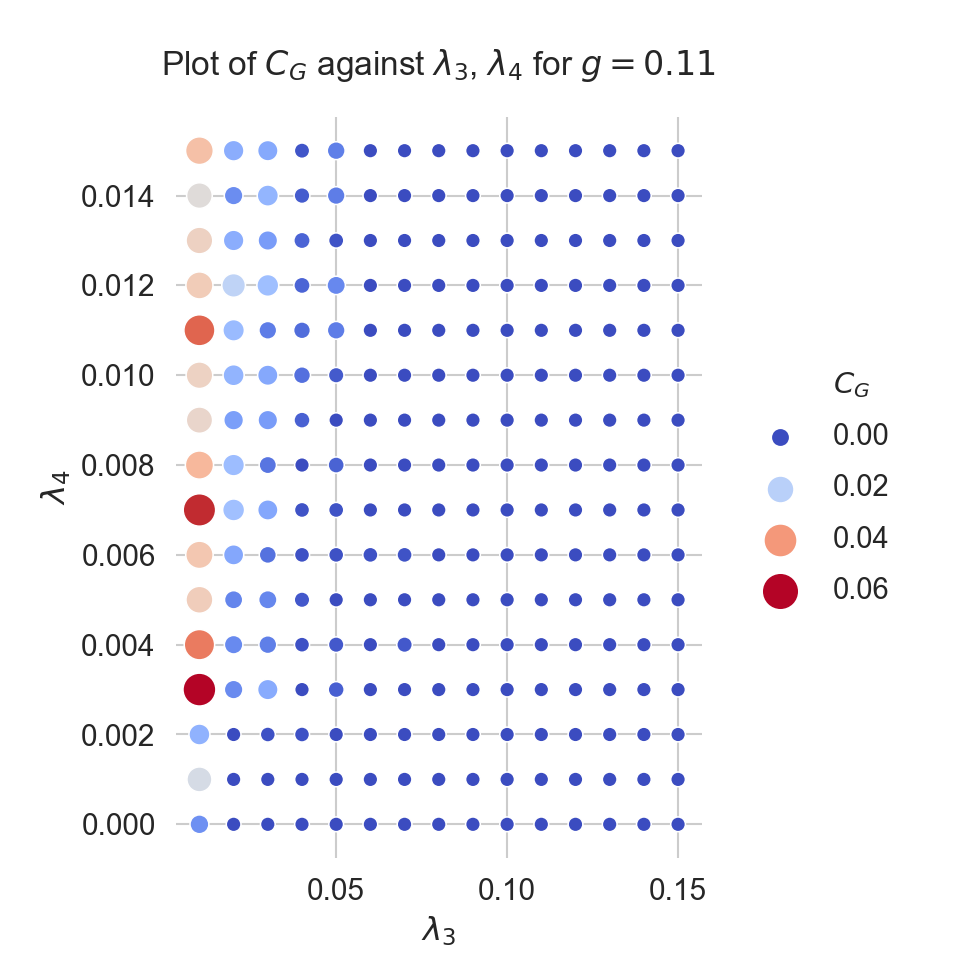}
	\end{subfigure}
	\caption{In a mesh of $N=50$ the ground state is computed of Eq. \eqref{eqn:pdgm_total}, across a range of values of $\lambda_3$,$\lambda_4$ from $0.0$ to $0.15$. We plot for values of $g=0.04,0.07,0.11$, both clustering coefficient and average node degree. }
	\label{fig:pdgm_params}
\end{figure*}

Having verified that the simulation reproduces the key results of the analysis in the original work of Trugenberger, we now seek to explore more deeply the full PDGM model with a view to understanding the effect of the additional terms in Eq. \eqref{eqn:pdgm_total}.
We reproduce in full the entire range of simulations in Appendix A, but we also present a sample of the graphs in Fig. \ref{fig:pdgm_params}.
Across the full range of values of $g$ the same pattern of ground state clustering coefficient is visible, in that for increasing $\lambda_3$ and $\lambda_4$, clustering decreases.
Closer inspection however reveals that there is {\sl no combination} of parameters where $\lambda_4 > 0$ that produces a ground state that is less clustered than when $\lambda_4 = 0$.
Further, inspecting the average degree graphs, it is clear that $\langle k_i \rangle$ decreases as $\lambda_4$ increases.

This could be as a result of the  $\Tr(A^4_{ij})$ term in Eq. \eqref{eqn:pdgm} interfering with both the triangle suppressing term $\Tr(A^3_{ij})$, and the first  `link frustration' term in Eq. \eqref{eqn:ising}, that disfavors any connected collection of three vertices.
To elaborate a little further, it is important to note that the diagonal terms in $\Tr(A^4_{ij})$ count the number of paths with length four that begin and end at a given node, as well as such paths that form a square.
For example, in a regular two dimensional mesh, out of the $36$ possible paths of length $4$, only $8$ involve square paths (remembering that a path traversed in each direction counts separately), and $4$ of the paths only involve one of the incident edges of a given vertex, traversed four times.
At low values of $g$, this term will compete with the original link frustration term in Eq. \eqref{eqn:ising}, favoring configurations of the graph where links are `absorbed' into squares.
This of course acts to reduce clustering, but also degree, a result born out in Fig. \ref{fig:pdgm_params}.

This investigation of the PDGM model parameter space would indicate that the optimal choice of parameters would be obtained by setting $\lambda_4=0$, which essentially reduces the PDGM model to the proposed  Hamiltonian of the QMD model.
We very much view the original model of PDGM as correct, producing an emerged ground state that is `large world' and low dimension, but the additional $\lambda_4$ parameter is not necessary.
With that in mind we will with due deference continue to use the simplified Hamiltonian as we investigate how we can incorporate dynamics and matter into the PDGM scheme.

%
%

\section{Dynamics}
\label{sec:dynamics}
\subsection{Modeling Matter in the Mesh, and the Temperature dependence of $g$}

In Section \ref{ssec:trugenberger} we noted that in the DGM treatment stable deformations of the mesh were able to form.
These correspond to an isolated spin in the mesh, that is a point at vertex $v_i$ with a value $s_{i} \neq s_{j}$, where $j$ ranges over all of the neighbors of $v_i$.
Unless  $A_{ij}=0$, that is there are no links between $v_i$ and its neighbors, this spin would contribute positively to the Hamiltonian, and so the links would be energetically disfavored.
Removing these links after a vertex spin is flipped would reduce the overall energy of the graph and result in a new stable minimum energy.
With the neighboring links removed the point is topologically isolated from the mesh.
In the DGM model these defects are proposed to be `topological black holes' and an argument is offered for their stability across a wide range of temperatures (using a statistical mechanical treatment of the ground state of the mesh).
The basis for the stability argument relies upon treating Equations \eqref{eqn:spin_itr} and \eqref{eqn:energy_itr} as energies in a Fermi-Dirac distribution.
It is then possible to postulate that there are temperature dependent fluctuations of both the vertex and link spins subject to a thermal probability of a new link forming being defined as
\begin{equation}
p(a_{ij}:  0 \rightarrow 1) = \frac{1}{1+ e^{-2h_{ij}/kT}} \text{,}
\end{equation}
with the factor of $2$ arising as links are undirected.
Without reproducing the details of the argument in \cite{Trugenberger2015}, it is possible to show that the average degree of the mesh $\langle k \rangle$ increases with increasing temperature, and its minimum value at $T=0$ is the ground state value that emerges from minimizing the Hamiltonian.
The additional triangle suppressing term in QMD does not affect the argument, as for the large part the resultant ground state graphs have very little clustering.
From Eq. \eqref{eqn:ground_k}, it is clear that $g  \propto 1/\langle k \rangle$, and assuming $k$-regularity $k_{min} \equiv k$.
Given that  $\langle k \rangle  \propto T$, it is possible to conclude $g  \propto 1/T$.

Consider a defect in the mesh, as described above.
To remove the defect it is necessary to flip one spin and connect $k$ links to the node.
From the earlier discussion it was established that the total energy cost of one new link to be $\Delta E=g^2/2(\langle k \rangle-1)(C_G+1) -g/2$, assuming that $C_G \approx 0$, stability is assured when $\Delta E > 0$. 
The ground state at $T=0$ has average degree $\langle k_0 \rangle = \frac{1}{2g} + 1/2$ and with some algebra, stability is guaranteed  for $\langle k \rangle > \langle k_0 \rangle + 1/2$.
We have already indicated that $\langle k \rangle$ is proportional to temperature and so this defect is stable until temperature drops close to $T=0$ where $\langle k \rangle = \langle k_0 \rangle$.
From Fig. \ref{fig:node_degree}. we note that average node degree jumps decreases discontinuously by an integer amount as $g$ increases and therefore $T$ decreases. 
It is also possible to conclude from the simulations that as $\langle k \rangle$ decreases its value remains constant for increasing ranges of $g$.
The stability equality will therefore hold until $\langle k \rangle = \langle k_0 \rangle + 1 $,  requiring a very low value of $T$. 

\subsection{Minimal Extensions to the Hamiltonian}
\label{ssec:meshdyn}

Having established the Hilbert Space associated with every edge and vertex in the graph, it is natural to ask how the state of the graph evolves with time.
For a vertex $v_i$ we can define in the standard way a general state vector $\ket{v_i}$ as an expansion in the basis states
\begin{equation}
	\ket{v_i}=c_0 \ket{0} + c_1 \ket{1}\mbox{,}
\end{equation}
where the $c_0,c_1$ are complex coefficients.
Clearly the same can be done for edges.
For the whole mesh we can of course construct the universe state vector $\ket{\Psi_G}$ as a an expansion in the $D=\frac{1}{2}N^2(N-1)$ dimensional tensor product basis vectors that span the total Hilbert space $\mathcal{H}_{total}$.
For simplicity of exposition we will only consider small segments of the whole graph, whilst remembering that wave functions and operators are properly defined in the whole tensor product Hilbert space. 

As time is not present in this definition we can consider the state vector to be in the Heisenberg, or time independent picture. To convert to the Schr{\"o}dinger picture we can write
\begin{equation}
	\ket{v_i, t}=e^{-i\hat{H}t/\hbar} \ket{v_i}\mbox{,}
\end{equation}
as the Hamiltonian $\hat{H}$, by definition, does not depend upon time.
If we now imagine the state of the mesh at some time $t_{in}$ evolving under the effect of this Hamiltonian a short time $t_{out}=t_{in} + \tau$, the evolution of the state vector can be  written as
\begin{equation}
	\ket{v_i,t_{out}} = e^{-i\hat{H}\tau/\hbar} \ket{v_i, t_{in}}\mbox{.}
\end{equation} 

As $\tau$ is very small we can expand the exponential, remembering that as $\ket{v_i,t}$ is an element of the entire tensor product space, and the operators are similarly defined to act  on elements of that space, we obtain
\begin{equation}\label{eqn:p_exp}
	\ket{v_i,t_{out}} = \hat{\unit} \ket{v_i, t_{in}} - \frac{i\tau}{\hbar} \hat{H} \ket{v_i, t_{in}} + O(\tau^2) \dots \mbox{,}
\end{equation}
where $\hat{\unit}$ is the identity operator on $\mathcal{H}_{total}$.

We first verify that the Hamiltonian for QMD cannot create dynamics for a defect introduced into the mesh, which we are using to model matter.
Let us first recast Eq. \eqref{eqn:hamiltonian} in an operator format, using the annihilation and creation operators for links and the spin operator, using the normal ordering introduced in Section \ref{sec:model}.
Taking each term by term we have three components of the Hamiltonian, which using our normal ordering convention are
\begin{align}
	\frac{g^2}{2} \Tr(A_{ij}^3) &= \sum\limits_{i,k,l \in [1,N]} \hat{a}^{\dagger}_{ik} \hat{a}^{\dagger}_{kl} \hat{a}^{\dagger}_{li}\  \hat{a}_{ik} \hat{a}_{kl} \hat{a}_{li} \mbox{,} \label{eqn:triangle}\\
	\frac{g^2}{2}  \sum\limits_{i \neq j} \sum\limits_{ k \neq i,j } A_{ik}A_{kj}  &=  \sum\limits_{i \neq j} \sum\limits_{ k \neq i,j } \hat{a}^{\dagger}_{ik} \hat{a}^{\dagger}_{kj} \hat{a}_{ik} \hat{a}_{kj} \mbox{,} \label{eqn:triplet}\\
	\frac{g}{2}  \sum\limits_{i,j} s_i A_{ij}  s_j  &= \sum\limits_{i,j} \hat{s}_i  \hat{a}^{\dagger}_{ij}\hat{a}_{ij } \hat{s}_j \mbox{,} \label{eqn:spins}
\end{align}
where the spin operators $\hat{s}_i$ are as defined in the definition of the Hilbert space for the vertices in Section \ref{sec:model}.

\begin{figure}[H]
\centering
\begin{tikzpicture}[node distance=0.5cm]
	\node [black] at (0,1) {\textbullet};
	\node [black] at (1,1) {\textbullet};
	\node [black] at (2,1) {\textbullet};
	\node [black] at (3,1) {\textbullet};
	\node [black] at (0,0) {\textbullet};
	\node [black] at (1,0) {\textbullet};
	\node [black] at (2,0) {\textbullet};
	\node [black] at (3,0) {\textbullet};
	\node [black] at (0,-1) {\textbullet};
	\node [black] at (1,-1) {\textbullet};
	\node [black] at (2,-1) {\textbullet};
	\node [black] at (3,-1) {\textbullet};

	\draw [dashed] (0,1) -- (0,2);
	\draw [dashed] (1,1) -- (-1,1);	
	\draw [-] (0,0) -- (0,1);
	\draw [dashed] (1,1) -- (1,2);
	\draw [-] (0,0) -- (0,-1);
	\draw [-] (0,1) -- (1,1);
	\draw [dashed] (2,1) -- (2,2);
	\draw [-] (2,1) -- (2,0);
	\draw [-] (2,0) -- (2,-1);
	\draw [-] (2,0) -- (3,0);
	\draw [-] (0,-1) -- (1,-1);
	\draw [-] (1,-1) -- (2,-1);
	\draw [-] (2,-1) -- (3,-1);
	\draw [-] (1,1) -- (2,1);
	\draw [-] (2,1) -- (3,1);
	\draw [dashed] (3,1) -- (3,2);
	\draw [dashed] (3,1) -- (4,1);
	\draw [-] (3,0) -- (3,1);
	\draw [-] (3,0) -- (3,-1);
	\draw [dashed] (3,0) -- (4,0);
	\draw [dashed] (3,-1) -- (4,-1);
	\draw [dashed] (3,-1) -- (3,-2);
	\draw [dashed] (2,-1) -- (2,-2);
	\draw [dashed] (1,-1) -- (1,-2);
	\draw [dashed] (0,-1) -- (0,-2);
	\draw [dashed] (0,-1) -- (-1,-1);
	\draw [dashed] (0,0) -- (-1,0);
	
	\node at (1,-0.25) {$v_i$};
	\node [fill=white, text=black] at (2,-0.25) {$v_1$};
	\node [fill=white, text=black] at (3,-0.25) {$v_2$};
	\node [fill=white, text=black] at (3,0.75) {$v_3$};
	\node [fill=white, text=black] at (3,-1.25) {$v_4$};
\end{tikzpicture}
\caption{A section of the emerged quantum mesh with $\langle k \rangle=2d=4$. Depicted is an isolated defect surrounding vertex $v_i$, with vertices $v_1,v_2,v_3,v_4$ labelled. For the purposes of discussion assume $s_i=\ket{i,0}$, all other spins $s_j=\ket{j,1}$.}
\label{fig:vertex}
\end{figure}
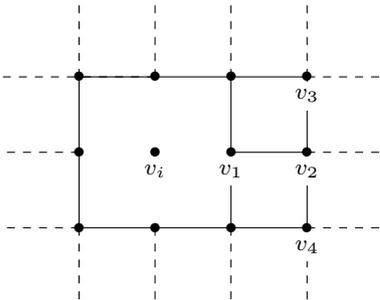

Fig. \ref{fig:vertex} depicts a single isolated defect in a $d_k=2$ dimensional emerged mesh.
It is possible to extend the following argument to higher dimensions, but for clarity the argument is outlined in $2$ dimensions.
We can consider the Hamiltonian as a sum over all possible edges $e \in E$ in the mesh.
For a given configuration not all of these edges are present, so let us denote by $E_G \subset E$ those edges that are present and the complement $E^c_G \subset E$ as the set of edges that are not present in a given state of the mesh.
Formally $E=E_G \bigoplus E^c_G$, and we can correspondingly write
\begin{equation}\label{eqn:comp_H}
	\hat{H}=\hat{H}_{E_G} + \hat{H}_{E^c_G} \mbox{.}
\end{equation}

By considering the action of Eq. \eqref{eqn:comp_H} on the the total state vector $\ket{\Psi_G}$ for the mesh we seek to show that the mesh with the defect is an eigenstate of $\hat{H}$ with an energy strictly above the ground state, and therefore that a defect is unchanged by the operation of $\hat{H}$ on $\ket{\Psi_G}$.
Let us first consider the action of the second term on $\ket{\Psi_G}$.
As this will apply only operator expressions from Equations \eqref{eqn:triangle},\eqref{eqn:triplet},\eqref{eqn:spins} that refer to edges not present in $\ket{\Psi_G}$, the annihilation operator $\hat{a}_{ij}$, which by our ordering convention will act first on $\ket{\Psi_G}$, will always return a zero state vector.
Our operator expression therefore only contains terms arising from $\hat{H}_{E_G}$, and as every edge annihilation operator is paired with a corresponding edge creation operator it must return a state vector proportional to $\ket{\Psi_G}$.
As such we arrive at the result
\begin{equation}
	\hat{H} \ket{\Psi_G} = E_d \ket{\Psi_G} \mbox{,}
\end{equation}
where $E_d$ is the energy eigenvalue for the mesh with one defect.
This calculation satisfies the assertion that the defect is preserved under the operation of $\hat{H}$, but we can go further and consider $E_d$ as compared to $E_0$ the eigenvalue of the mesh ground state with no defects present.
We already know that the removal of edges requires energy, so any operator that added back in an edge to a general state vector in the total Hilbert space, would contribute a negative value to the total energy eigenvalue.
As such we conclude that $E_d > E_0$, and conclude that the operator representation of Eq. \eqref{eqn:hamiltonian} as applied in the exponential expansion of Eq. \eqref{eqn:p_exp} leaves a defect static and unchanged if one is present in $\ket{v_i,t_{in}}$.

In order for the matter defect to propagate through the mesh, a new term in the Hamiltonian is needed that will result in the movement of the defect in the mesh.
What is required is a term that flips spins in the mesh in such a way that it favors spins at lattice positions that are disconnected but nearby, in other words the term must favor locality in the mesh.
At the same time, it would be desirable for this term not to require additional coupling constants, and be minimally constructed from the same operators used to construct the QMD Hamiltonian.
Further, the new term must not interfere with the emergence of the ground state.
This can be accomplished by a term of the form $-\hat{s}^{+}_i ( 1-A_{ij} ) \hat{s}^{-}_j$, with the minus sign inserted to guarantee that a dynamic interaction has a positive value.
For the purposes of the following analysis, noting that the eigenvalues of $\hat{s}^{\pm}_i$ are $\pm \hbar/2$, the convention of setting $c=h=1$ is temporarily dropped.
The middle term is zero for all nodes in the network that do not have a link to the node $v_i$ at position $i$, and as noted this is guaranteed to be energetically favorable.
It remains to prefer local over distant interactions, and this can be done by inserting an inverse proportionality to the square of the hop distance $l_{ij}^2$ as measured by the smallest number of links to a neighbor of $v_i$.
The effect of this term on the ground state is negligible as the interaction is specifically between nodes that are not adjacent, and the effect of the distance $1/l_{ij}^2$ will quickly reduce to zero the interaction between spins that are distant.
For a defect this requires some care in its definition as technically the vertex contained in the defect is disconnected from the graph, and we choose to define $l_{ij}$ as the distance between the two vertices {\sl as if} the defect vertex was connected into the mesh in the normal way. 
To define $l_{ij}$ in a way that is reducible to the fundamental operators we have, we require a discrete analog of the Dirac delta function defined for integers $n \in \mathbb{Z}$ as $\delta(n) = \delta_{n,0}$.

We are now able to define distance in terms of the adjacency matrix, itself reducible to edge annihilation and creation operators, using the well known property that when it is raised to the power $n$, the value at $A^n_{ij}$ counts the number of $n$ length paths between vertices $v_i$ and $v_j$, as:
\begin{equation}
	l_{ij} = \sum\limits_{n=0}^{\infty} \left \{ \delta( A^n_{ij} ) \times \delta \left(\sum\limits_{p=0}^{p=n} A^p_{ij} \right) \right \}\text{.}
\end{equation}
The choice of an inverse square dependency on $l_{ij}$ is of course arbitrary, but the choice is motivated by dimensional considerations described below. 
To propose the dynamical Hamiltonian, we use the Laplacian of the graph, noting that the equivalent of $(1-A_{ij})$ is $(1+L_{ij})$ as this is only non-zero  when no edge is present between vertices $v_i$ and $v_j$, or when $i=j$.
We can always express the Laplacian matrix in terms of edge operators as follows:
\begin{equation}
	L_{ij} = \sum\limits_{k,j=0}^{k,j=N} \delta_{i}^{j}\hat{a}^{\dagger}_{ik} \hat{a}_{kj} - \hat{a}^{\dagger}_{ij} \hat{a}_{ij} \text{,}
\end{equation}
which permits us to use it in our Hamiltonian, which remains defined independently of any particular state of the mesh.

In order for the dynamic Hamiltonian to use the same dimensionless coupling constant, we recall that each spin operator $\hat{s}_i^{\pm}$ has eigenvalues $\pm \hbar/2$.
To include the hop distance $l_{ij}$ and maintain the measurement of the Hamiltonian in units of energy, one can first introduce the energy to create or destroy a defect $\epsilon_m$, and identify it as the mass-energy represented by this defect.
Using these definitions the proposal for the dynamic term of the Hamiltonian is

\begin{equation}\label{eqn:dynamic}
	\hat{H}_d = -\frac{gc^2}{2\epsilon_m l_{ij}^2} \hat{s}^{+}_i ( 1+L_{ij}) \hat{s}^{-}_j \text{.}
\end{equation}

This extension will only deal with the realignment of vertex spins as a defect propagates between locations in the mesh.
For a full dynamic Hamiltonian we also need to include a mechanism that deals with the edges between the vertex and its neighbors.
The subsequent analysis of Eq. \eqref{eqn:dynamic}, however,  is not affected by the extensions necessary for the correct arrangement of the edges, but for completeness we will demonstrate how this may be done, whilst preserving the form of our proposed Hamiltonian.

We begin by defining a new operator using a combinatorial sum of all the possible subsets of vertices in the mesh that could be neighborhoods of a vertex.
A neighborhood of a given vertex $v_i$ is defined as the collection of vertices in the graph that share an edge with $v_i$, and essentially we need to pick out the appropriate neighborhood of each of the vertices $v_i$, $v_j$  and apply creation and annihilation operators to obtain the correct final arrangement of edges as a defect propagates.
On a general graph $G(V,E)$, we define a new operator $\hat{S}_i^{-}$, where $I$ is a subset of the set of vertices as:

\begin{equation}\label{eqn:S_minus}
	\hat{S}_i^{-} = \hat{s}_i^{-}  \sum\limits_{I \subset V} \prod\limits_{j \in I} \hat{a}_{ij} \text{.}
\end{equation}

The effect of the sum over products will create a large number of combinations of $\hat{s}^{-}_i$ with edge annihilation operators, but fortunately as $\hat{a}_{ij} \ket{0} = 0$, most of the terms in the sum will vanish.
We can define the conjugate of $\hat{S}_i^{-}$ by taking the complex conjugate of Eq. \eqref{eqn:S_minus} as follows:

\begin{equation}\label{eqn:S+plus}
	\hat{S}_i^{+} = \left( \hat{S}_i^{-} \right )^{\dagger} =  \hat{s}_i^{+}  \sum\limits_{I \subset V} \prod\limits_{j \in I} \hat{a}_{ij}^{\dagger} \text{.}
\end{equation}

Using these new operators we can redefine our dynamic Hamiltonian as:
\begin{equation}\label{eqn:H_mod}
	\hat{H}_d^{'} = -\frac{g}{2 \epsilon_m l^2_{ij} }\hat{S}_a^{+} ( 1 + L_{ij} ) \hat{S}_b^{-} \text{.}
\end{equation}
Defining the initial state of the defect present at $v_i$ and the normal connected vertex at $v_j$ as $\ket{i}$, and subsequently the translated defect as present at $v_j$ and $v_i$ restored to the vacuum configuration as $\ket{f}$, it is easy to verify that $\bra{f} \hat{H}_d \ket{i}$ only retains contributions from the correct initial and final edges states.

\subsection{Quantum Mechanics and the Continuum Limit}
\label{ssec:waveeqn}
As it is possible to assume that the other terms in the Hamiltonian will have no effect on a matter defect in the mesh, it is possible to use Eq. \eqref{eqn:dynamic} to illustrate how this formulation could be viewed as equivalent to the non-relativistic formulation of quantum mechanics for a defect of mass $m$.
Begin by noting that for a small (infinitesimal) increment in time $\tau$, the time evolution of the state vector $\ket{v_i, t}$ is
\begin{equation*}
	\ket{v_i, t+\tau} = e^{-i\hat{H}_d \tau/ \hbar} \ket{v_i,t} \mbox{,}
\end{equation*}
whilst expanding as a Taylor series in $\tau$ we have,
\begin{equation*}
	\ket{v_i, t+\tau} = \ket{v_i,t} + \tau \frac{\partial \ket{v_i,t} }{\partial t}  + O(\tau^2) \dots 
\end{equation*}
Expanding the exponential to $O(\tau)$, and gathering terms in the first power of $\tau$ yields
\begin{equation}
	-\frac{\hbar}{i} \frac{\partial \ket{v_i,t} }{\partial t}  =  \frac{gc^2\hbar^2}{8\epsilon_m l_{ij}^2} \ket{v_i,t} +  \frac{gc^2\hbar^2}{8\epsilon_m l_{ij}^2}L_{ij} \ket{v_i,t} \text{, }
\end{equation}
by noting that $L_{ij}$ is simply a number and therefore the state vector $\ket{v_i,t}$, is operated upon by the combination $\hat{s}^{+}_i \hat{s}^{-}_j$.
In the continuum limit, as the vertices $v_j$ and $v_i$ are neighboring points in the graph, they can be approximated as operating on the same vertex, and so each spin operator contributes the eigenvalue $\hbar/2$ to the terms on the right hand side.
Further, at the continuum limit, each vertex of the mesh becomes identified as a point in the $d$ dimensional space $\va*{x}$, where $d$ is the dimension of the mesh.
At this limit we propose the correspondence between the state vectors on the vertices with a position state vector be as follows $\ket{v_i,t} \rightarrow \ket{\va*{x},t}$.

It is well known \cite{Chung1997} that for functions defined upon the vertex of a graph (such as in our case the state vector), the continuum limit of the Laplacian matrix is $-\nabla^2$.
However as we shrink the edge length in our matrix to zero, then $l_{ij} \rightarrow 0$, which will cause infinities.
The freedom exists to claim that the coupling constant $g$ is a `bare' coupling constant, valid only when $l_p > 0$, and should not be expected to hold in the continuum limit.
Instead  the $l_{ij}^2$ factor can be absorbed into $g$, now denoted as the bare coupling constant $g_b$.
Additionally as $\epsilon_m=m c^2$, one can replace the defect energy with its mass $m$, by absorbing the $c^2$ in the numerators on the right hand side.
One can then redefine the physical coupling constant $g_p=g_b /4 l_{ij}^2$, and choose it to be numerically one with dimension length squared.

Bringing this together and making the substitutions  $\ket{v_i,t} \rightarrow \ket{\va*{x},t}$ and $L_{ij} \rightarrow -\nabla^2$ gives the following expression
\begin{equation}
	-\frac{\hbar}{i} \frac{\partial \ket{\va*{x},t} }{\partial t}  = \frac{g_p\hbar^2}{2m} \ket{\va*{x},t}  + \frac{g_p}{2m} \bigg( \frac{\hbar}{i} \nabla \bigg )^2\ket{\va*{x},t}
\end{equation}

The first term on the right hand side, $\frac{g_p\hbar^2}{2m} \ket{\va*{x},t}$, is a constant multiplier times the state vector, and one can write this as $V(\va*{x})  \ket{\va*{x},t} $, for a constant potential,  $V(\va*{x}) = \text{const}$, at all points in the space  $\va*{x}$.
Inserting this back generates the final result

\begin{equation}
	-\frac{\hbar}{i} \frac{\partial \ket{\va*{x},t} }{\partial t}  = V(x)  \ket{\va*{x},t} +  \frac{1}{2m} \bigg( \frac{\hbar}{i} \nabla \bigg )^2\ket{\va*{x},t} \text{.}
\end{equation}

This is of course the non-relativistic Schr{\"o}dinger equation for a free particle of mass $m$, moving in a constant potential.
The result confirms that the proposed Lagrangian for the dynamics of the defect $\hat{H}_d$, reproduces the behavior of a free quantum particle in the continuum limit $l_p \rightarrow 0$, in a self consistent way, and hints that it is possible to treat the energy of a defect,  the excitation needed to flip a spin, as a quantum `particle' of mass $m=\epsilon_m/c^2$.

\subsection{Further evidence for an Entropy `Area' law }
\label{ssec:entropy}
It was established in \cite{Trugenberger2017} that the ground state of a toy combinatorial quantum gravity theory based on the Hamiltonian defined by Eq. \eqref{eqn:pdgm_total} results in a boundary dependence of the mesh entropy.
This result builds upon the arguments outlined in \cite{Trugenberger2015,Trugenberger2017} where the DGM  and PDGM modes were originally introduced.
We can however take an alternative view of the entropy of the mesh which arises directly from its graph structure.
From considerations of information theory it is possible to quantify the information in `bits' required to describe the structure of the graph.
Mathematically this is well understood, originating in the work of Janos K{\"o}rner in 1973 \cite{Korner1986,Simonyi1995}.
These measures though are perhaps less well suited to the consideration of a quantum mesh.
A more appropriate measure, the `Von Neumann' (VN) entropy of a graph, has been proposed and explored by many, notably Passerini {\sl et al} \cite{Passerini2008,Du2010} and lately by Bianconi {\sl et al} \cite{Anand2011}.
In what follows, when we refer to the entropy of the graph we shall use the Von Neumann form.

One constructs the entropy by solving the eigenvalue problem for $L_{ij}$, obtaining the set of $N$ eigenvalues $\lambda_i$.
These are then used to define the dimensionless entropy

\begin{equation}\label{eqn:vonneuman}
	S_G=-\sum\limits_{i=1}^{N} \lambda_i \log_2 \lambda_i \mbox{.}
\end{equation}

In the paper by Passerini \cite{Passerini2008} it is proved that this quantity is maximized by both the complete graph and also $k$-regular graphs.
Therefore any defect in the graph, which causes a departure from $k$-regularity will cause a drop in the VN entropy of the mesh, but the amount will depend upon the precise configuration of the defect.
As we create matter defects in the mesh, by definition we create isolated vertices.
It is similarly well known that the multiplicity of the eigenvalue $\lambda_i=0$ yields the number of  disconnected subgraphs \cite{Bollobas1998,Chung1997,Bapat2014}.
As such the addition of a new defect will create an additional zero eigenvalue of the Laplacian.
By convention the contribution of $\lambda=0$ to Eq. \eqref{eqn:vonneuman} is zero, so the only change to the spacetime graph that can affect the spectrum and therefore the entropy of the graph are the nodes on the boundary of the defect.
For a $d$ dimensional graph this boundary is $d-1$ dimensional, and so in the mesh corresponding to our universe, the boundary has spatial dimension $2$ , that is it is an area.
At the boundary the vertices will lose a link, previously connected to the opposite spin node, and so the total number of links will reduce. 
From elementary linear algebra $\Tr(L_{ij}) = \sum\limits_i k_i = \sum\limits_i \lambda_i$, and by the convexity of Eq. \eqref{eqn:vonneuman}, the VN entropy increases as the sum of the eigenvalues decreases.
In practice this establishes that the entropy of the zero defect mesh (i.e. the vacuum) is lower than a mesh with defects, and that the change is proportional to the defect areas.
This is an intriguing result, providing a direct connection between the size of the boundary and the entropy of the contained matter defects.
\section{Conclusion and Future Directions}
\label{sec:conclusion}

Using extensive numerical simulations we have investigated the ground state graph properties of the emergent geometry first described by Trugenberger in \cite{Trugenberger2015} and extended in the  PDGM model \cite{Trugenberger2016}, including a deep exploration of the parameter space.
These models indicate how a workable discrete geometry for spacetime can emerge from a disordered and random collection of spins interacting using a similar mechanism to the Ising model of ferromagnetism.
Further we proposed a more parsimonious model as a simplification of PDGM that shares many of its desirable features.
This proposition is motivated and justified by the deep parameter space investigation of PDGM that indicates that the additional square favoring term does not produce noticeable benefit, and indeed could alter the topology of the emerged ground state.
We demonstrated that this  model performs  well and produces a triangle free, Euclidean flat ground state.
The simplification still possesses  the highly desirable feature of a low valued preferred dimension, approaching the three or four dimensions (if we include time) of classical physics.

For our proposed model, a scheme for the introduction of matter and dynamics is proposed using a minimal extension to the Hamiltonian defined in Eq. \eqref{eqn:hamiltonian}.
Intriguingly, as a direct result of the constraint that the dynamics term acts locally, it is possible in the continuum limit to recover the Schr{\"o}dinger equation for a free non-relativistic quantum particle.
The treatment of the emerged mesh as a graph also permits us to argue, that the intrinsic informational entropy of such a defect is proportional to its boundary and not its `bulk'.

The work presented is short of a concrete proposal for the quantum nature of spacetime in the extreme ultra violet regime, but it does represent a physical `toy' model of how such a spacetime could be manifested, and geometry could occur as an emergent phenomenon in a cooling universe.
The focus of future work will be to further refine the critical behavior of the graphs as the coupling constant is varied, and to consider in the presence of matter, how non-trivial geometrical features such as curvature could be represented.

\begin{acknowledgements}
The author would like to thank his colleagues at the Beyond Center at ASU for the many helpful discussions as this work was developed. In particular George Zahariade made numerous helpful suggestions as the model was developed. I also owe a thank you to Paul Davies for the many regular and lively discussions we had regarding the nature of the quantum mesh.
\end{acknowledgements}
\appendix
\label{sec:appendix}
\section{Results from the deep $\lambda_3$,$\lambda_4$ parameter space simulations.}

For completeness we present here the results from the full range of simulations conducted of the PDGM model.
This includes a full sweep of the $\lambda_3$ and $\lambda_4$ parameters for a range of core coupling constant $g$.
%
%
\begin{figure*}[ht]
	\centering
	\begin{subfigure}[t]{0.3\textwidth}
		\centering
		\includegraphics[height=0.2\textheight]{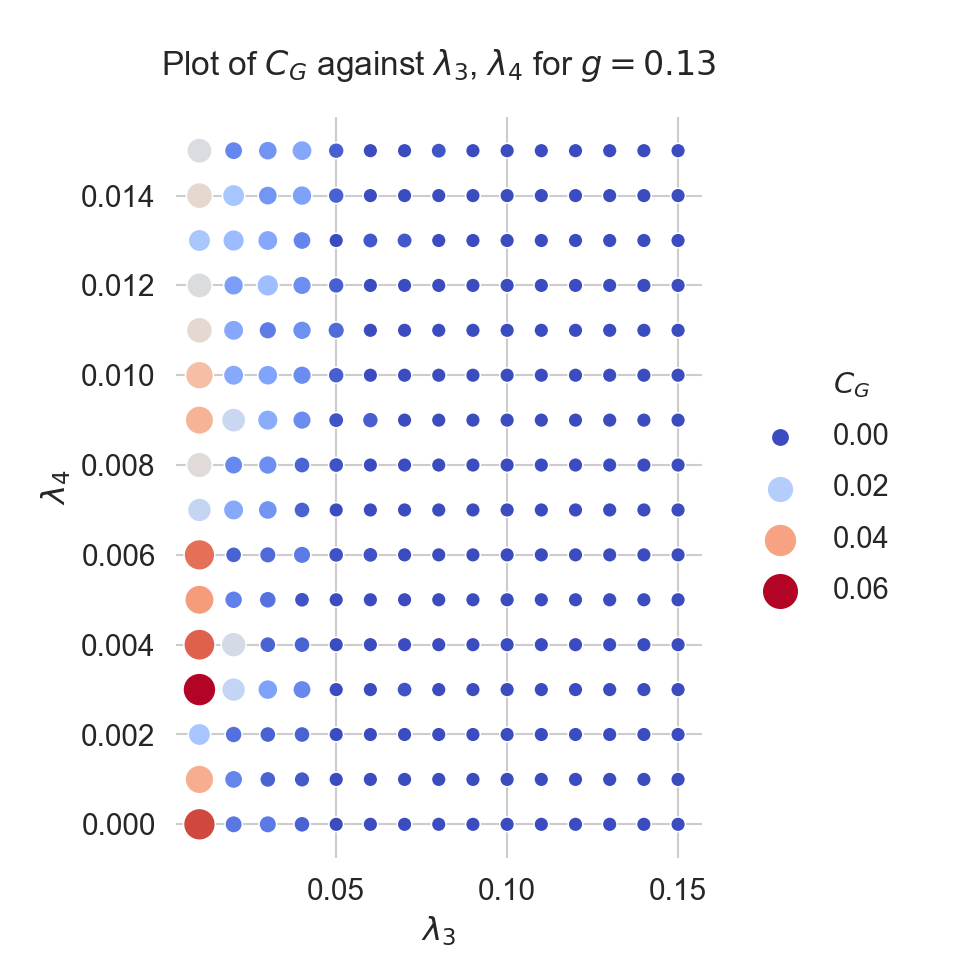}
		\label{fig:DCCg13}
	\end{subfigure}%
	~ 
	\begin{subfigure}[t]{0.3\textwidth}
		\centering
		\includegraphics[height=0.2\textheight]{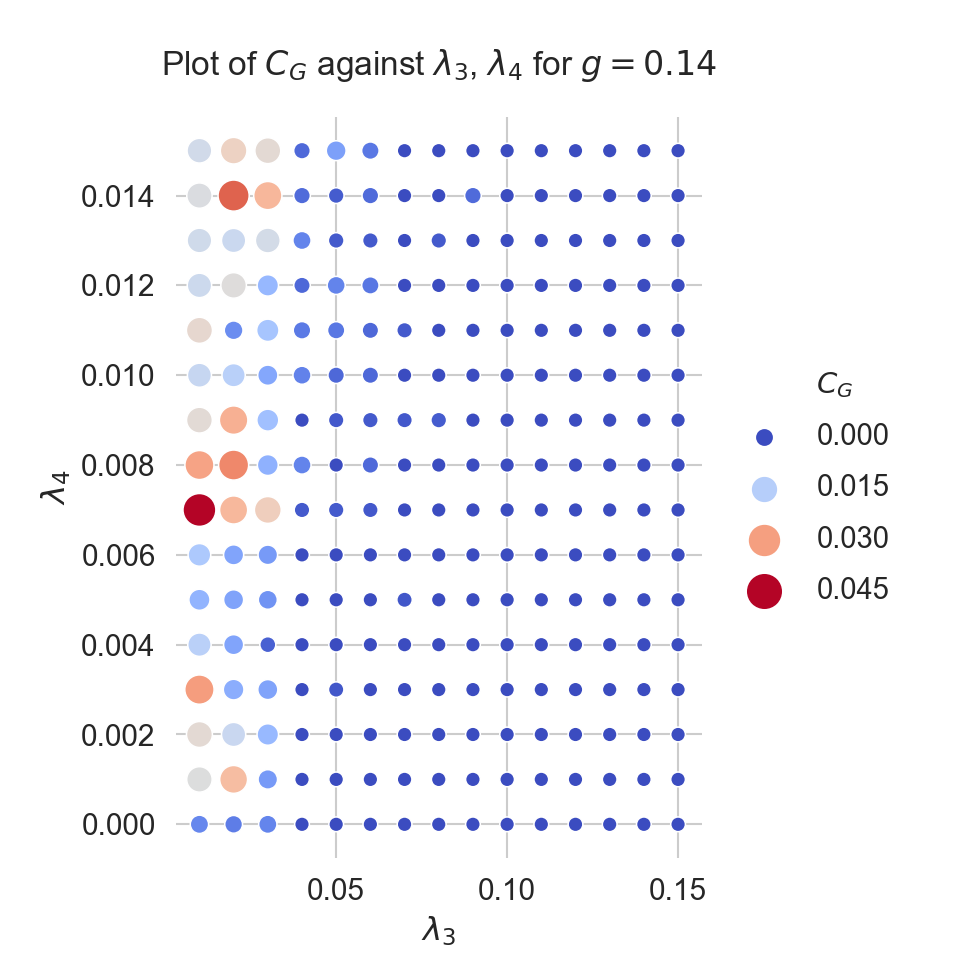}
		\label{fig:DCCg14}
	\end{subfigure}
	~ 
	\begin{subfigure}[t]{0.3\textwidth}
		\centering
		\includegraphics[height=0.2\textheight]{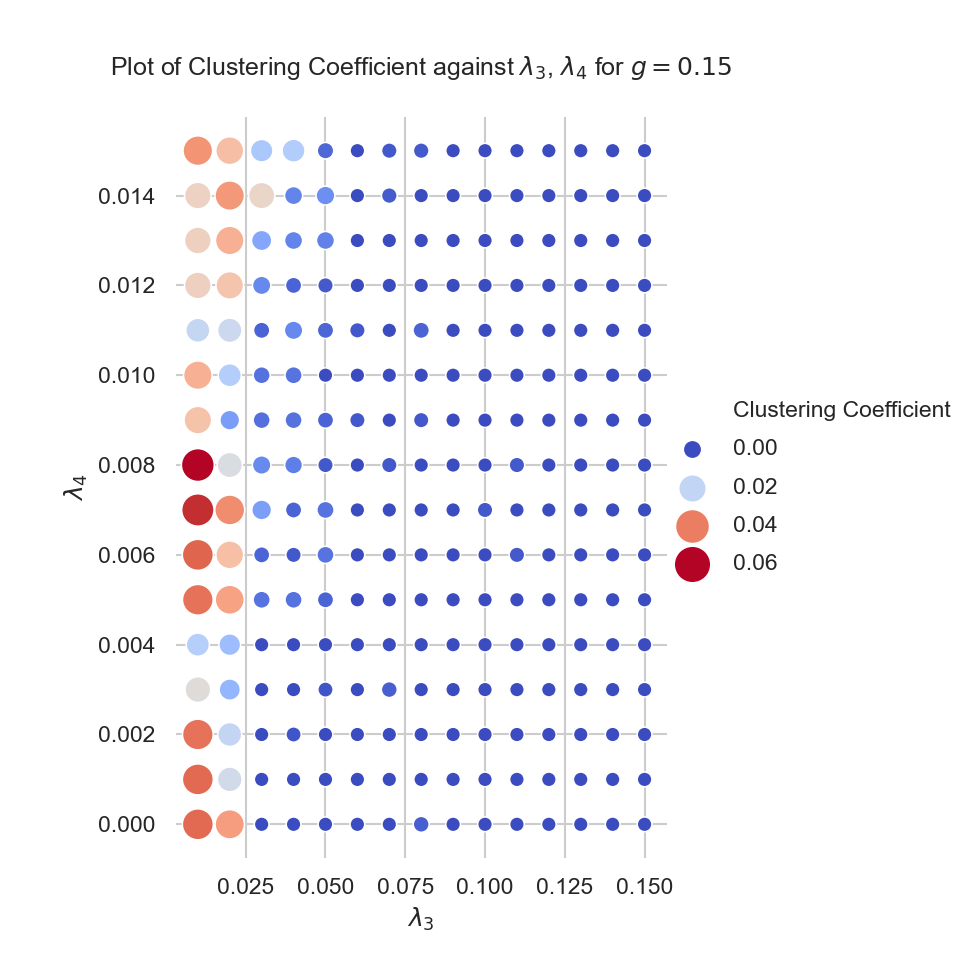}
		\label{fig:DCCg15}	
	\end{subfigure}
	~
	\begin{subfigure}[t]{0.3\textwidth}
		\centering
		\includegraphics[height=0.2\textheight]{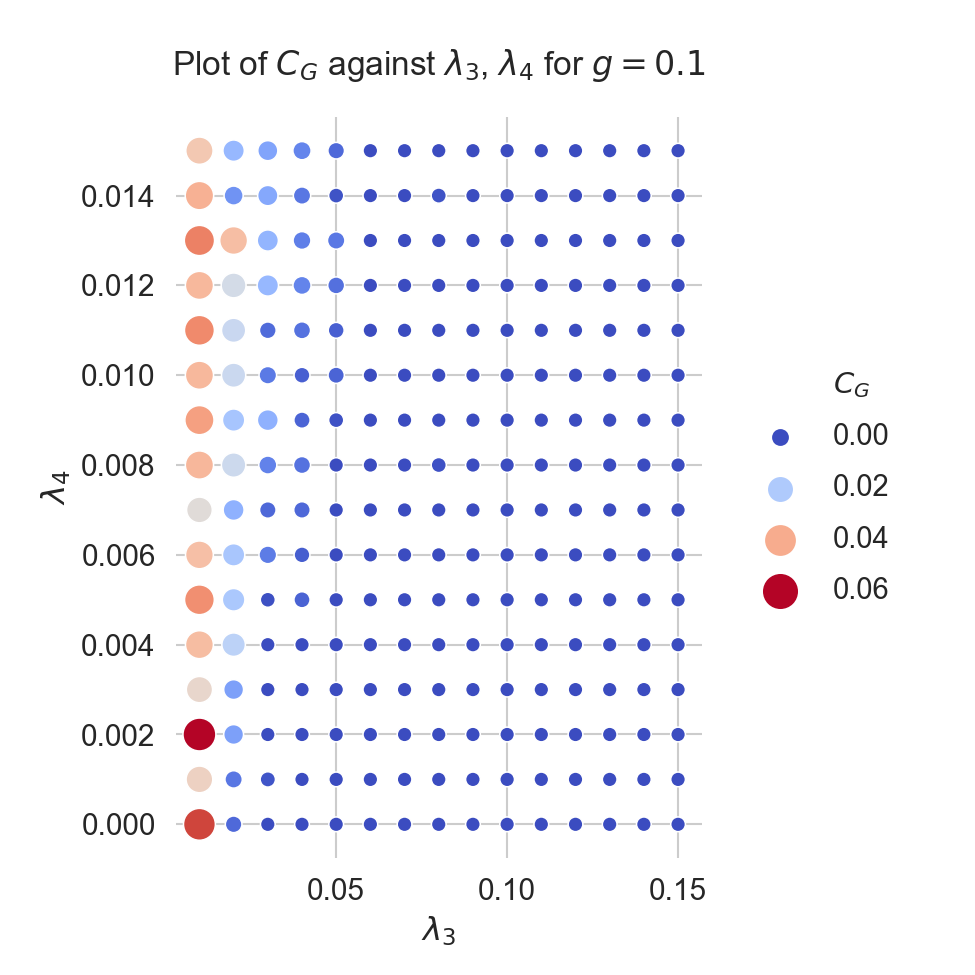}
		\label{fig:DCCg10}
	\end{subfigure}%
	~ 
	\begin{subfigure}[t]{0.3\textwidth}
		\centering
		\includegraphics[height=0.2\textheight]{DGMLamSpaceCC_g0_11}
		\label{fig:DCCg11}
	\end{subfigure}%
	~ 
	\begin{subfigure}[t]{0.3\textwidth}
		\centering
		\includegraphics[ height=0.2\textheight]{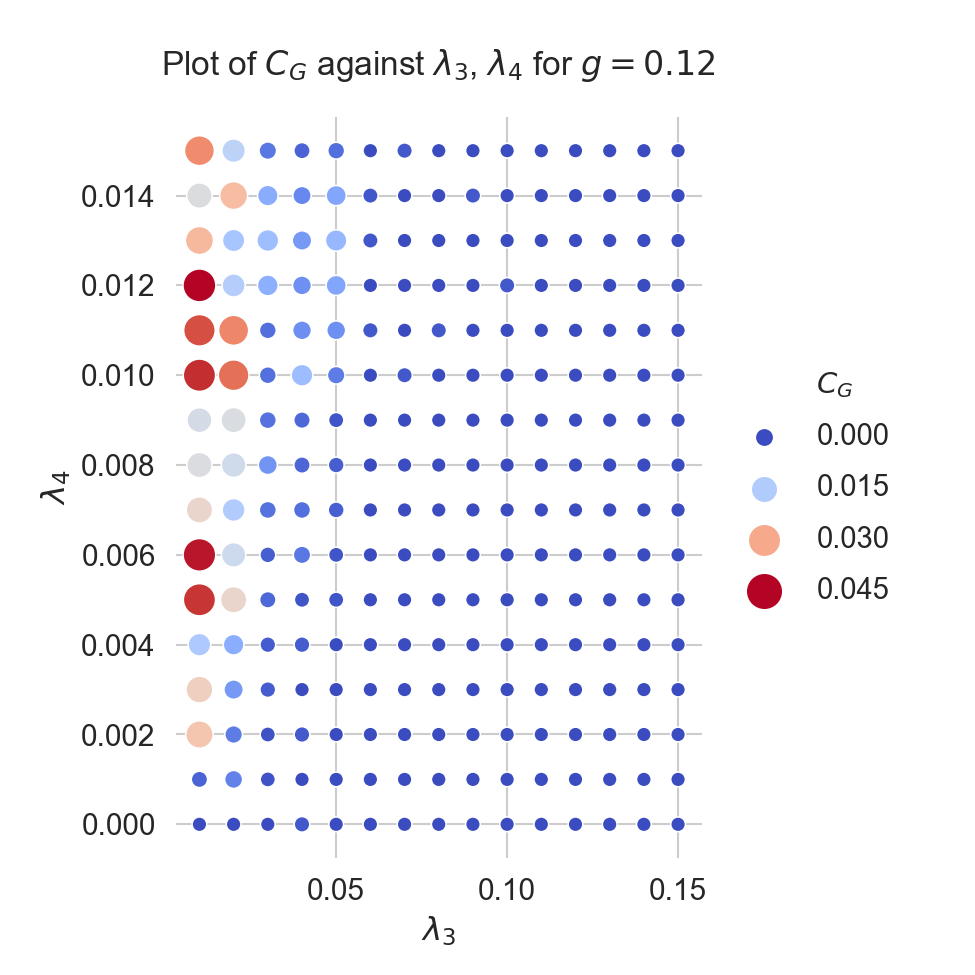}
		\label{fig:DCCg12}
	\end{subfigure}
	~ 
	\begin{subfigure}[t]{0.3\textwidth}
		\centering
		\includegraphics[height=0.2\textheight]{DGMLamSpaceCC_g0_07}
		\label{fig:DCCg07}	
	\end{subfigure}
	~
	\begin{subfigure}[t]{0.3\textwidth}
		\centering
		\includegraphics[height=0.2\textheight]{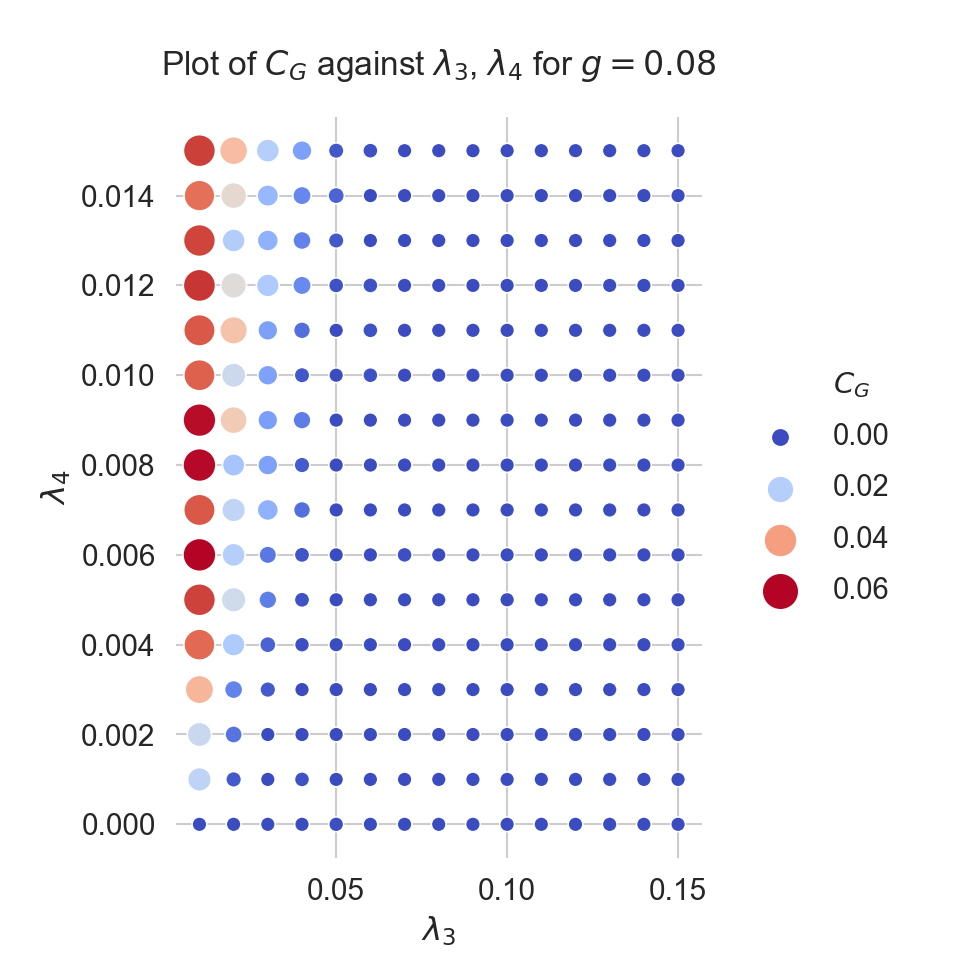}
		\label{fig:DCCg08}
	\end{subfigure}%
	~ 
	\begin{subfigure}[t]{0.3\textwidth}
		\centering
		\includegraphics[height=0.2\textheight]{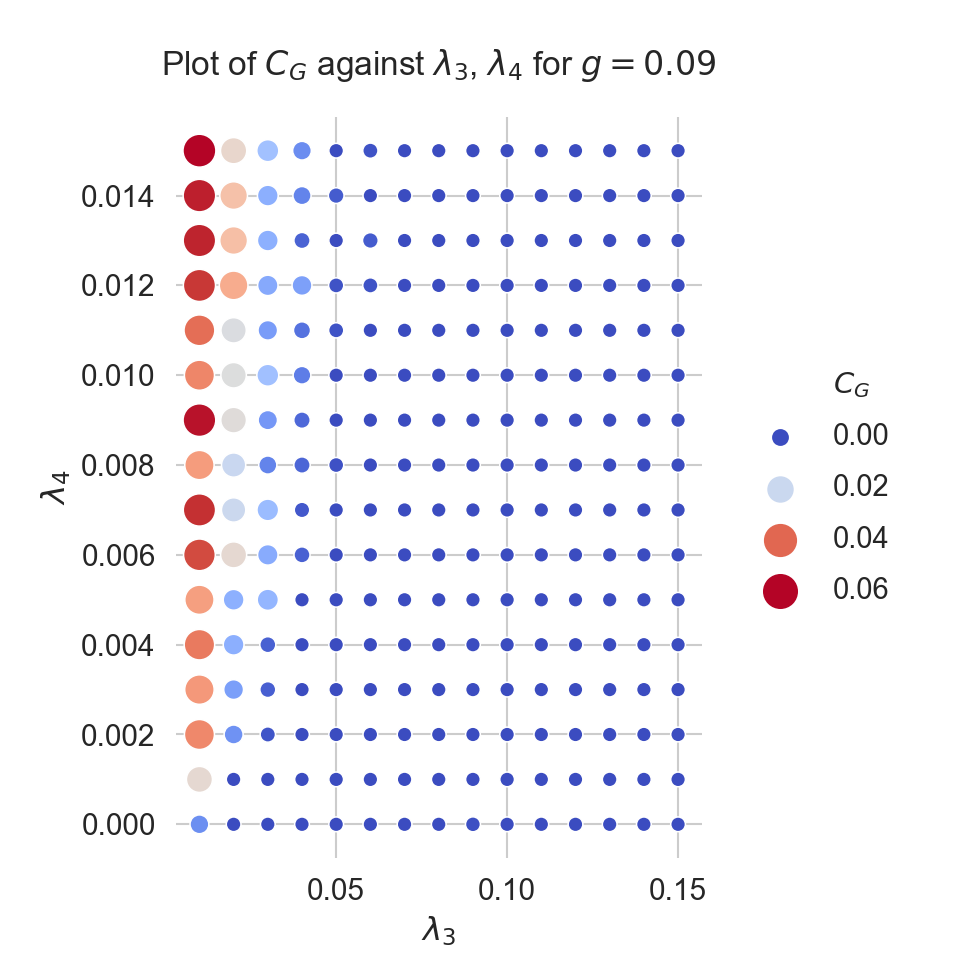}
		\label{fig:DCCg09}
	\end{subfigure}
	~ 
	\begin{subfigure}[t]{0.3\textwidth}
		\centering
		\includegraphics[height=0.2\textheight]{DGMLamSpaceCC_g0_04}
		\label{fig:DCCg04}	
	\end{subfigure}
	~
	\begin{subfigure}[t]{0.3\textwidth}
		\centering
		\includegraphics[height=0.2\textheight]{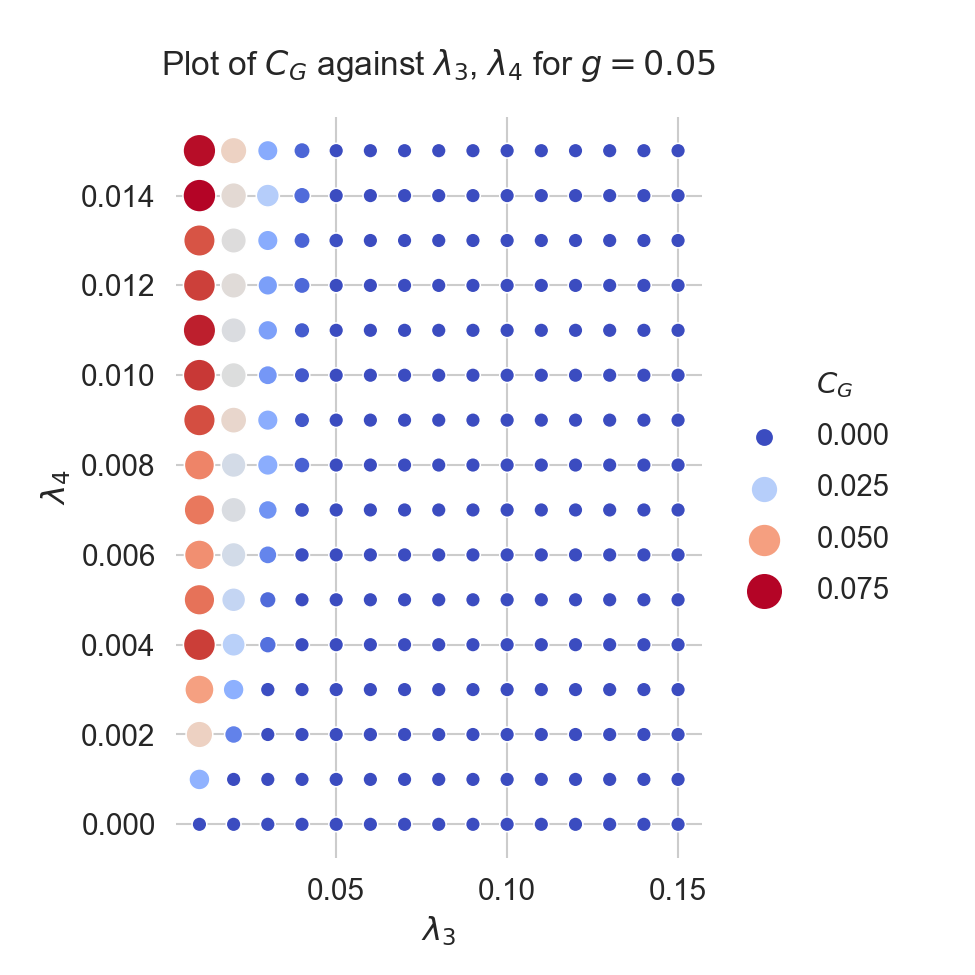}
		\label{fig:DCCg05}
	\end{subfigure}%
	~ 
	\begin{subfigure}[t]{0.3\textwidth}
		\centering
		\includegraphics[height=0.2\textheight]{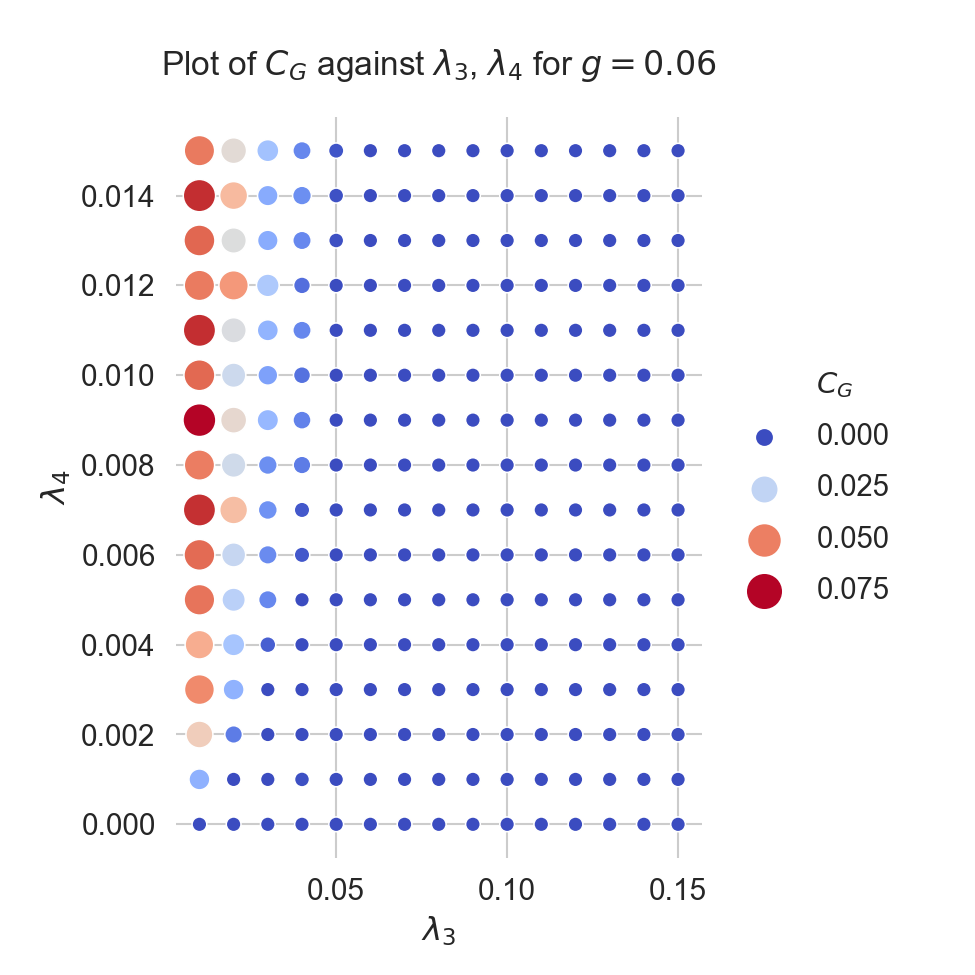}
		\label{fig:DCCg06}
	\end{subfigure}
	~ 
	\begin{subfigure}[t]{0.3\textwidth}
		\centering
		\includegraphics[height=0.2\textheight]{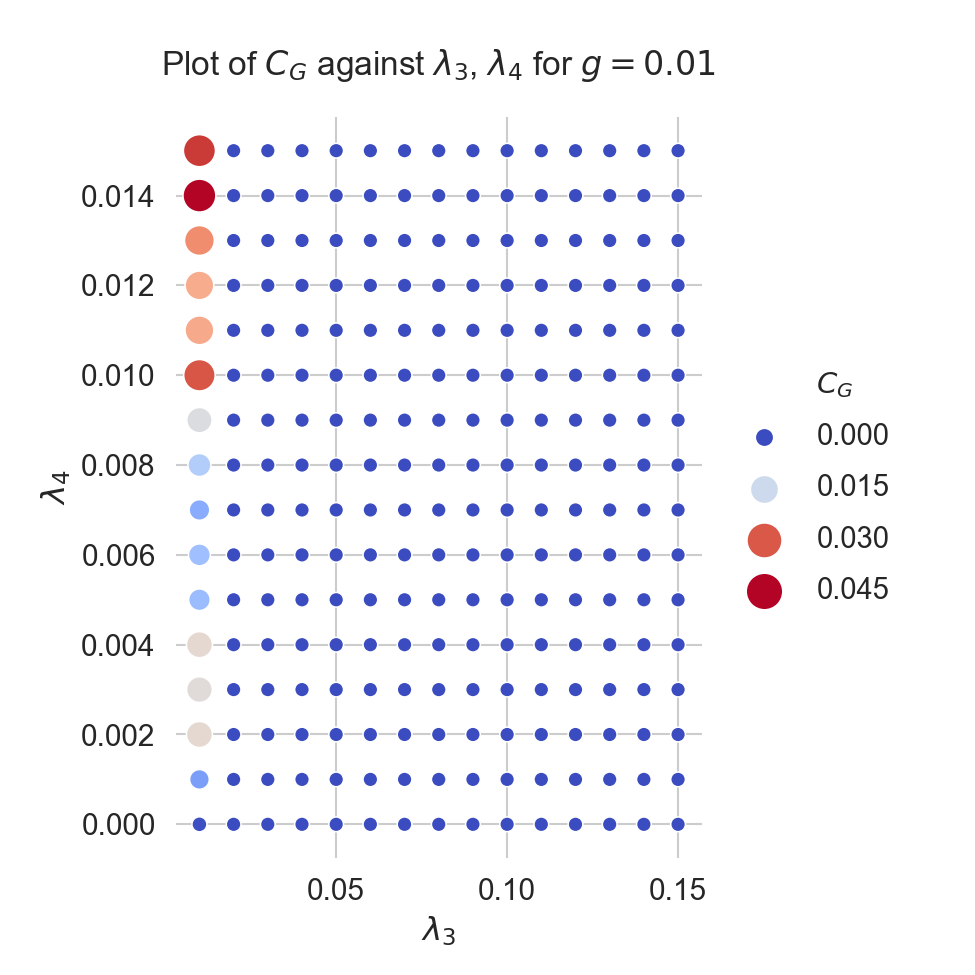}
		\label{fig:DCCg01}	
	\end{subfigure}
	~
	\begin{subfigure}[t]{0.3\textwidth}
		\centering
		\includegraphics[height=0.2\textheight]{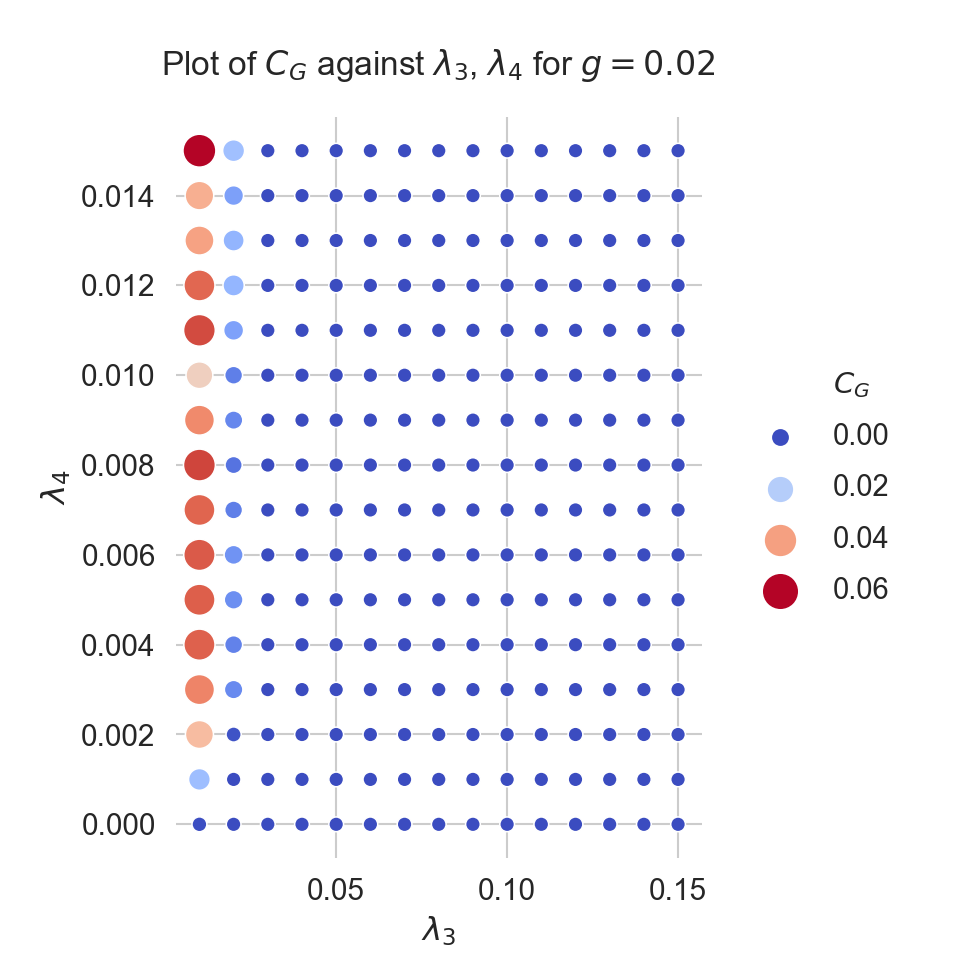}
		\label{fig:DCCg02}
	\end{subfigure}%
	~ 
	\begin{subfigure}[t]{0.3\textwidth}
		\centering
		\includegraphics[height=0.2\textheight]{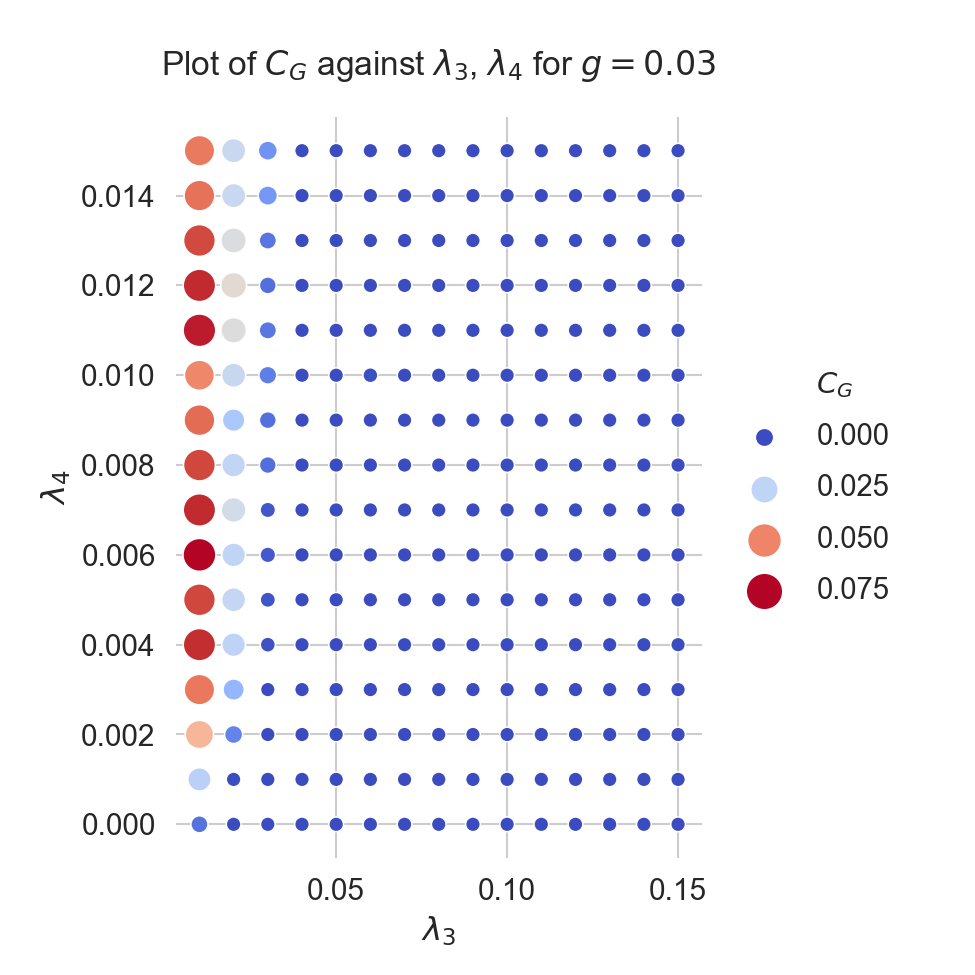}
		\label{fig:DCCg03}
	\end{subfigure}%
	\caption{For a mesh of $N=50$ the clustering coefficient is measured computed from the ground state of Eq. \eqref{eqn:pdgm_total}, across a range of values of $\lambda_3$,$\lambda_4$ from $0.0$ to $0.15$. Clustering coefficient decreases as $\lambda_3$,$\lambda_4$  increases, but $\lambda_4=0$ is most effective. }
	\label{fig:pdgm_c3}
\end{figure*}

%
%
\begin{figure*}[ht]
	\centering
	\begin{subfigure}[t]{0.3\textwidth}
		\centering
		\includegraphics[height=0.2\textheight]{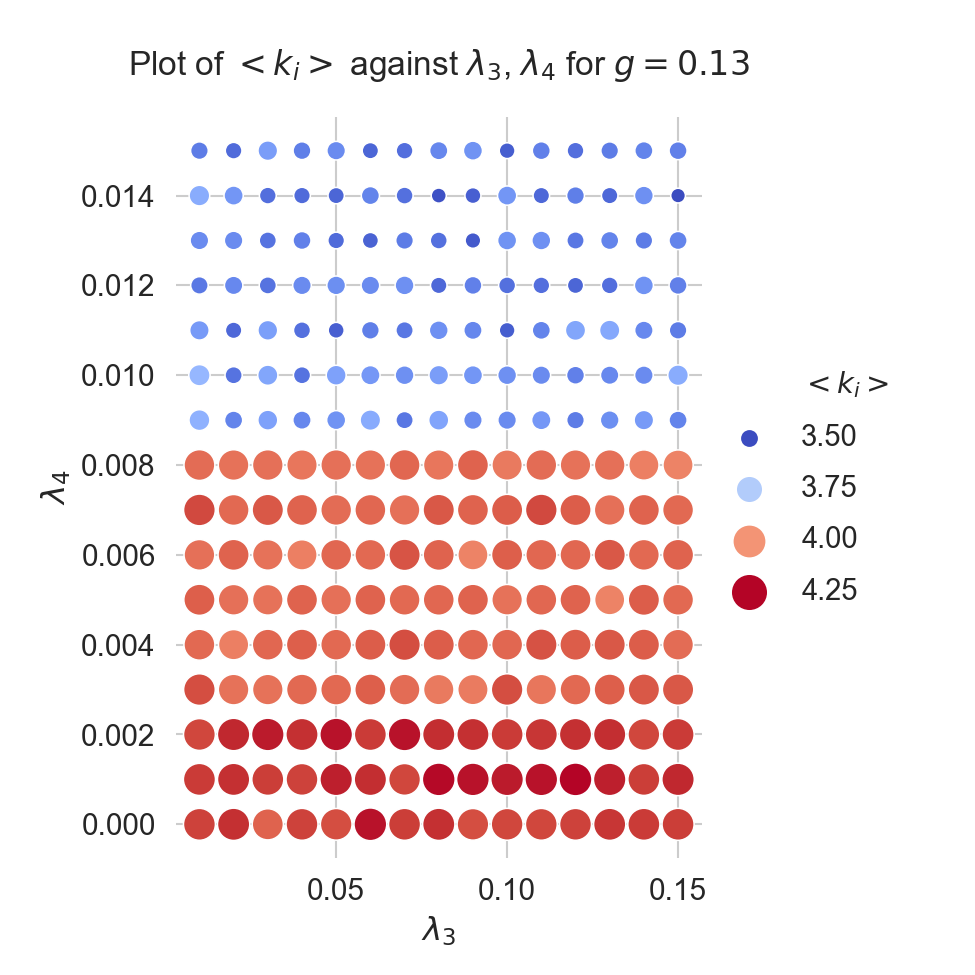}
		\label{fig:DADg13}
	\end{subfigure}%
	~ 
	\begin{subfigure}[t]{0.3\textwidth}
		\centering
		\includegraphics[height=0.2\textheight]{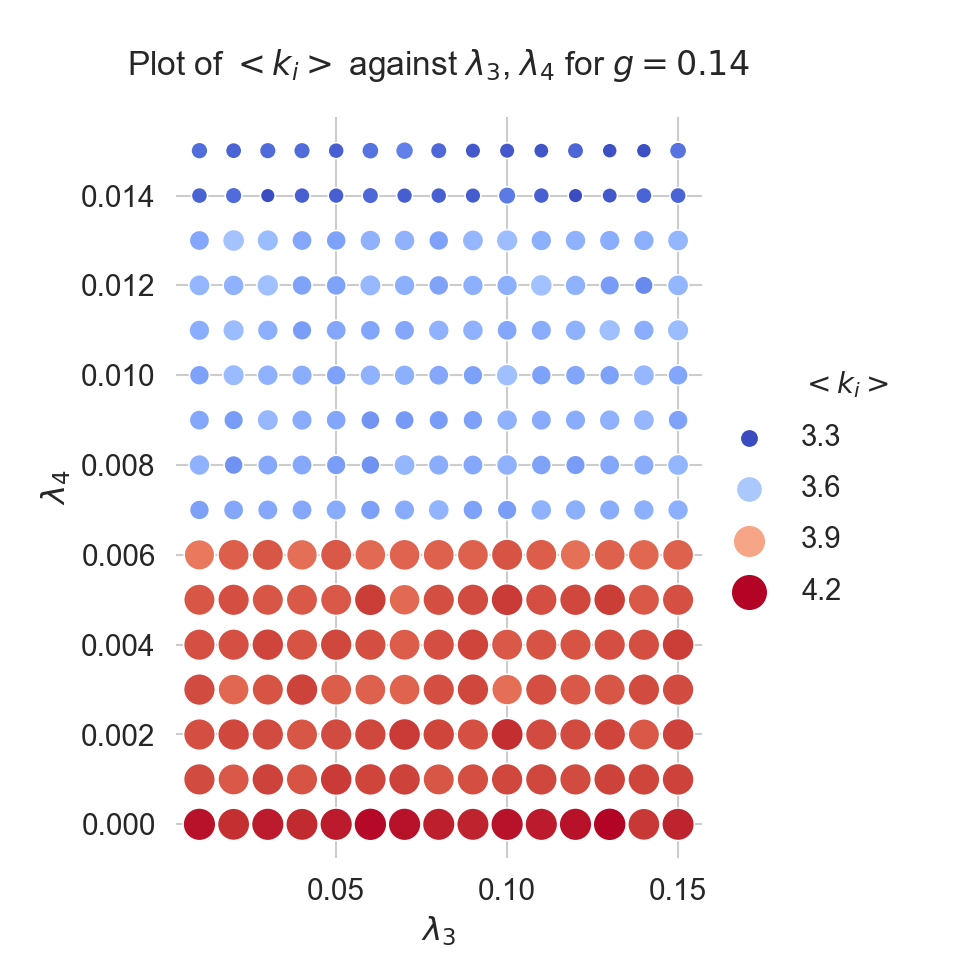}
		\label{fig:DADg14}
	\end{subfigure}
	~ 
	\begin{subfigure}[t]{0.3\textwidth}
		\centering
		\includegraphics[height=0.2\textheight]{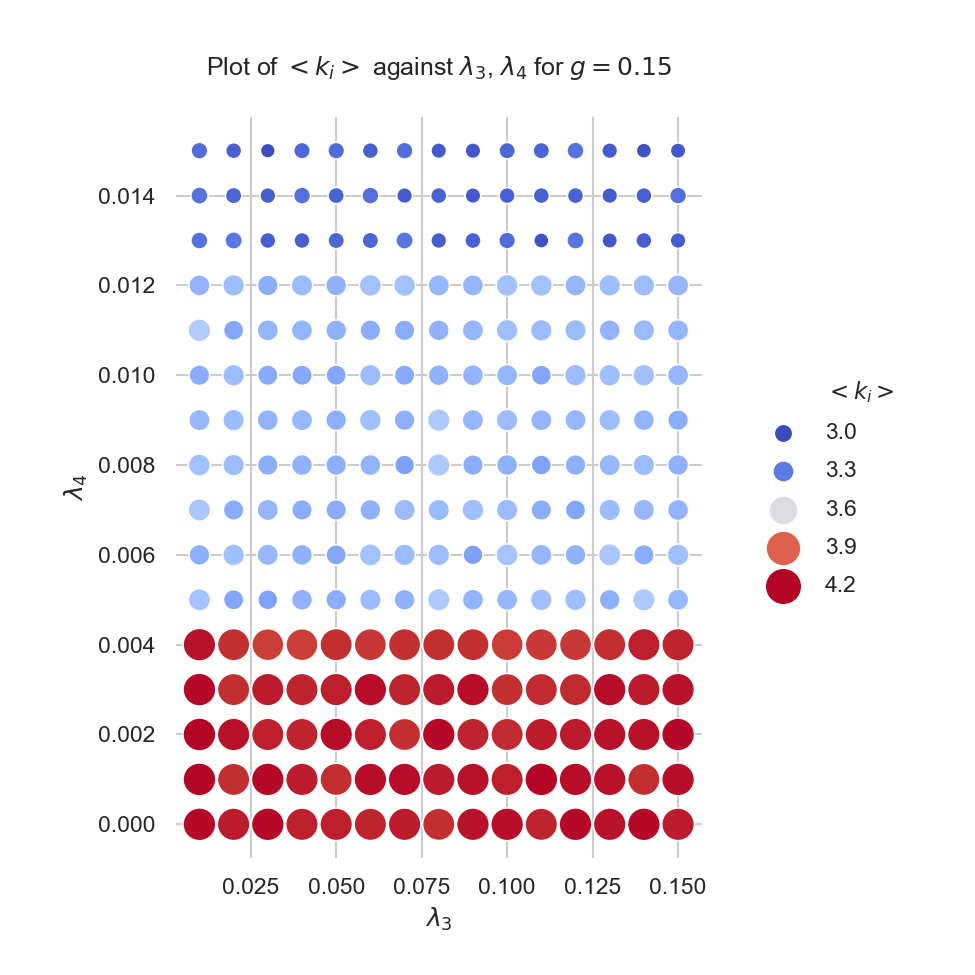}
		\label{fig:DADg15}	
	\end{subfigure}
	~
	\begin{subfigure}[t]{0.3\textwidth}
		\centering
		\includegraphics[height=0.2\textheight]{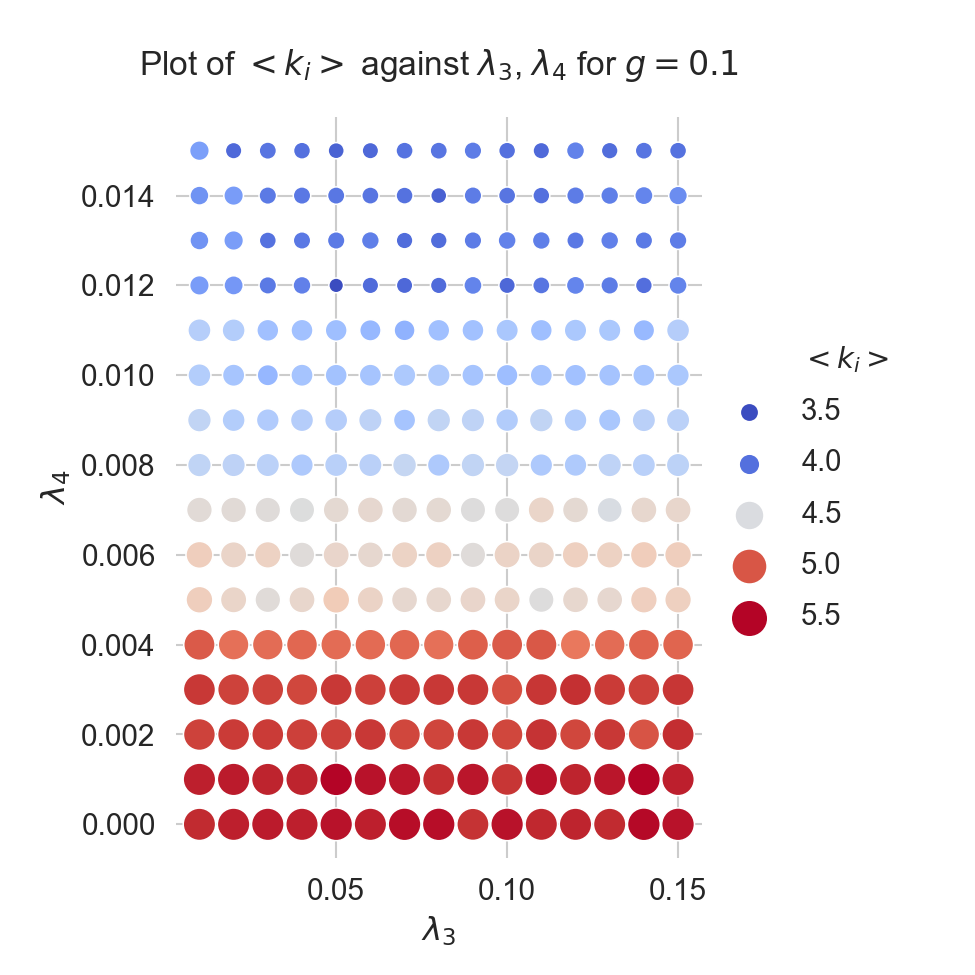}
		\label{fig:DADg10}
	\end{subfigure}%
	~ 
	\begin{subfigure}[t]{0.3\textwidth}
		\centering
		\includegraphics[height=0.2\textheight]{DGMLamSpaceDeg_g0_11}
		\label{fig:DADg11}
	\end{subfigure}%
	~ 
	\begin{subfigure}[t]{0.3\textwidth}
		\centering
		\includegraphics[ height=0.2\textheight]{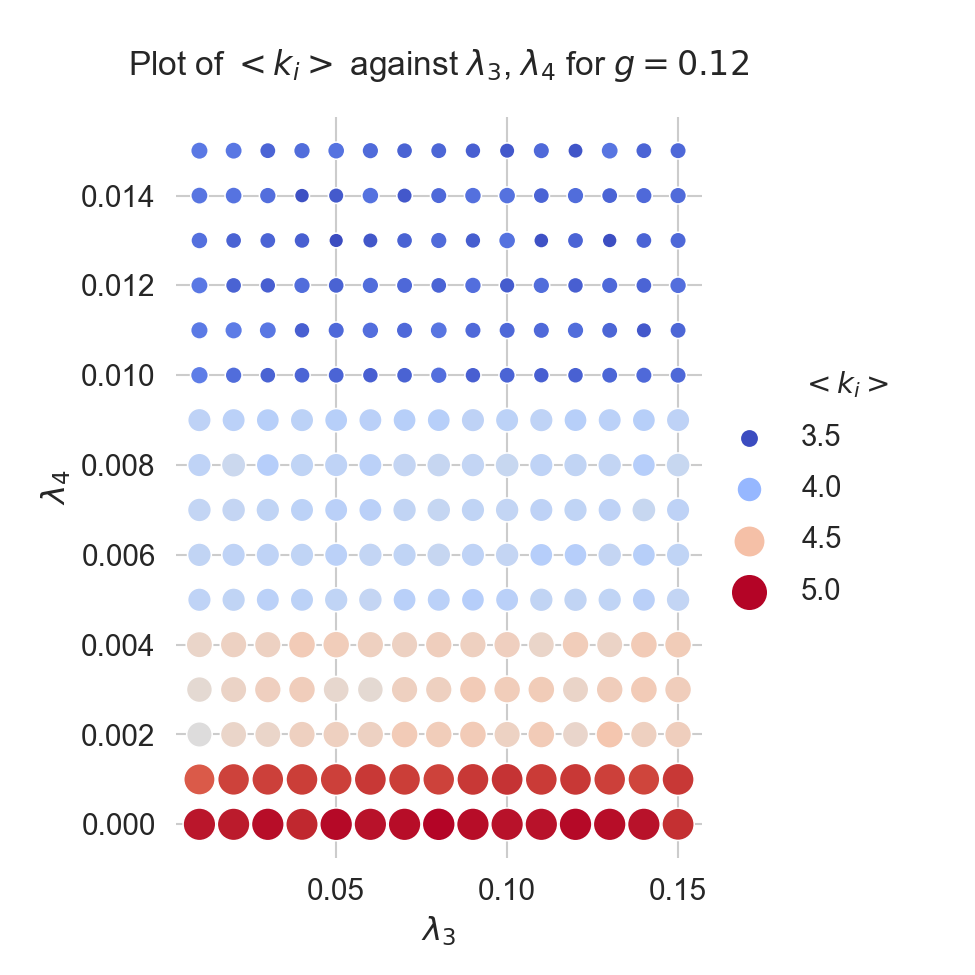}
		\label{fig:DADg12}
	\end{subfigure}
	~ 
	\begin{subfigure}[t]{0.3\textwidth}
		\centering
		\includegraphics[height=0.2\textheight]{DGMLamSpaceDeg_g0_07}
		\label{fig:DADg07}	
	\end{subfigure}
	~
	\begin{subfigure}[t]{0.3\textwidth}
		\centering
		\includegraphics[height=0.2\textheight]{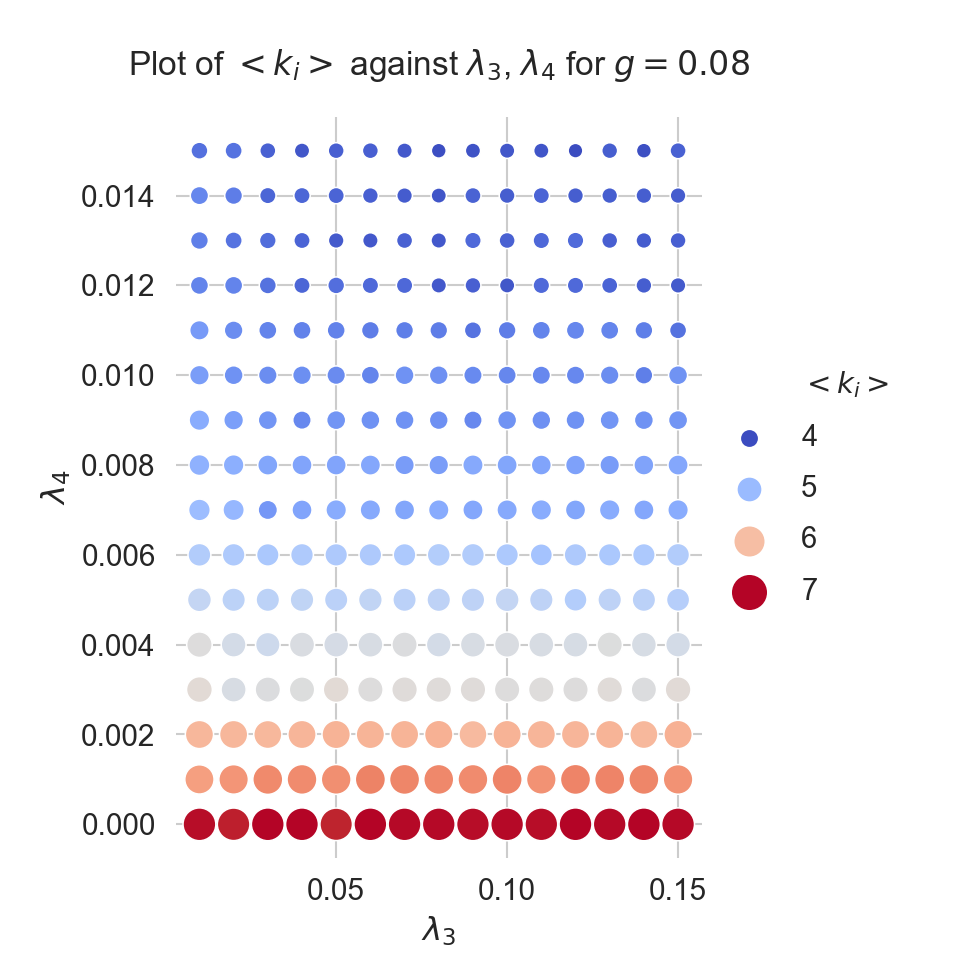}
		\label{fig:DADg08}
	\end{subfigure}%
	~ 
	\begin{subfigure}[t]{0.3\textwidth}
		\centering
		\includegraphics[height=0.2\textheight]{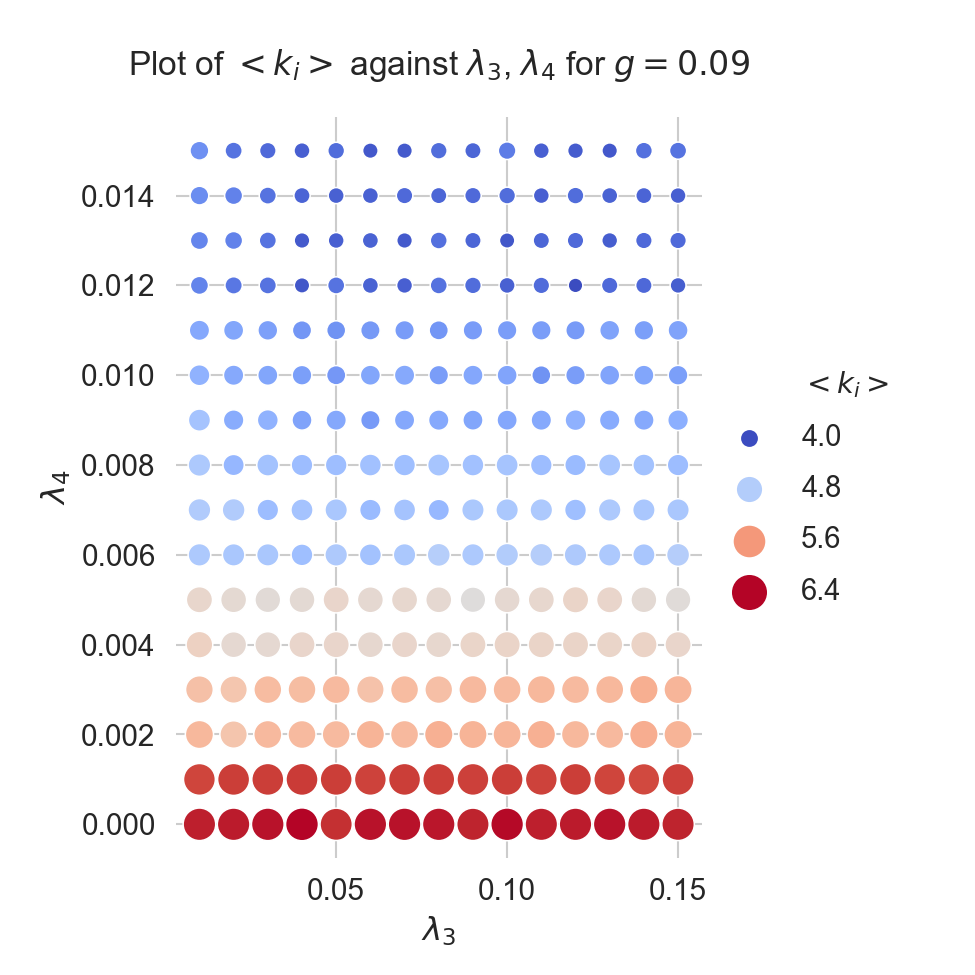}
		\label{fig:DADg09}
	\end{subfigure}
	~ 
	\begin{subfigure}[t]{0.3\textwidth}
		\centering
		\includegraphics[height=0.2\textheight]{DGMLamSpaceDeg_g0_04}
		\label{fig:DADg04}	
	\end{subfigure}
	~
	\begin{subfigure}[t]{0.3\textwidth}
		\centering
		\includegraphics[height=0.2\textheight]{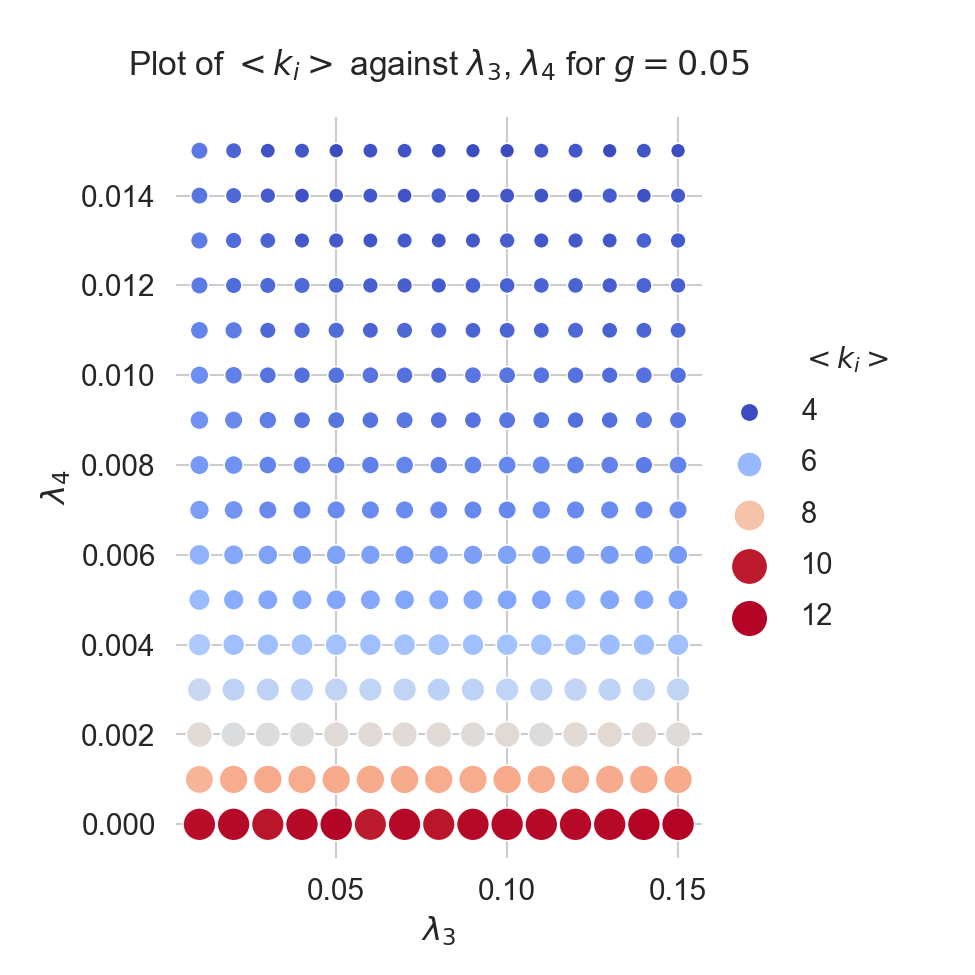}
		\label{fig:DADg05}
	\end{subfigure}%
	~ 
	\begin{subfigure}[t]{0.3\textwidth}
		\centering
		\includegraphics[height=0.2\textheight]{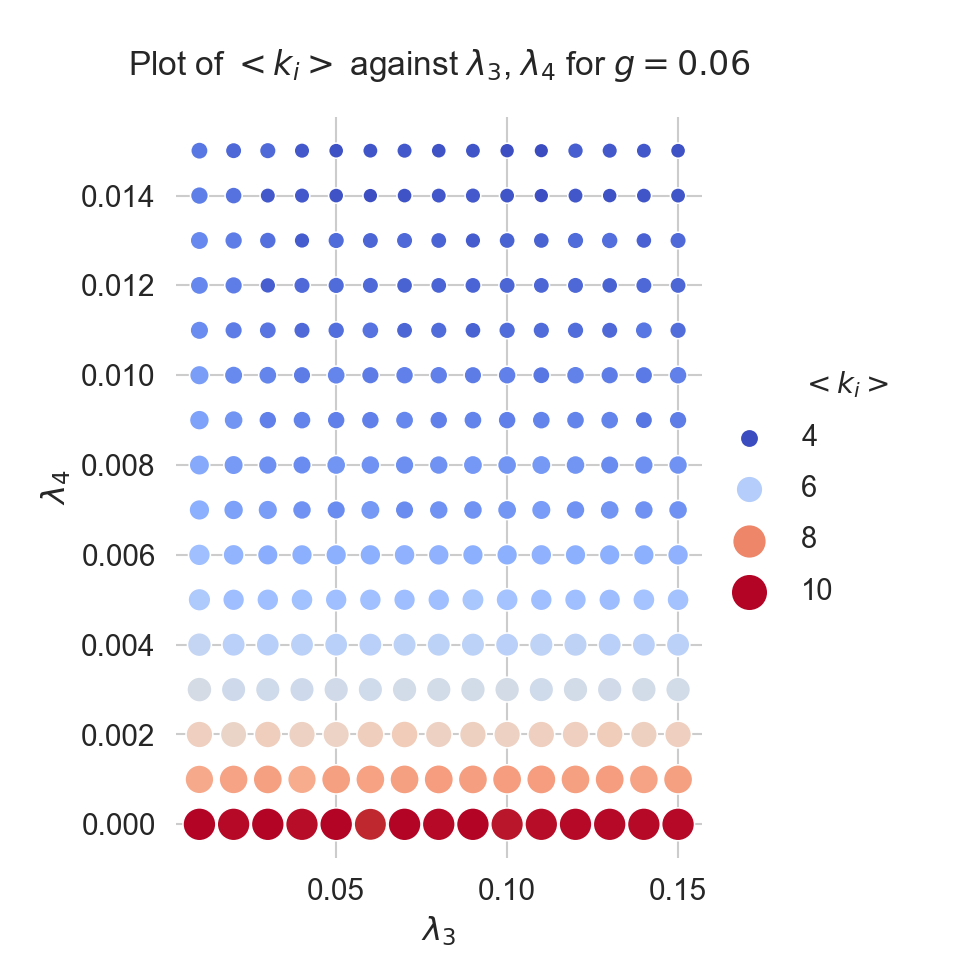}
		\label{fig:DADg06}
	\end{subfigure}
	~ 
	\begin{subfigure}[t]{0.3\textwidth}
		\centering
		\includegraphics[height=0.2\textheight]{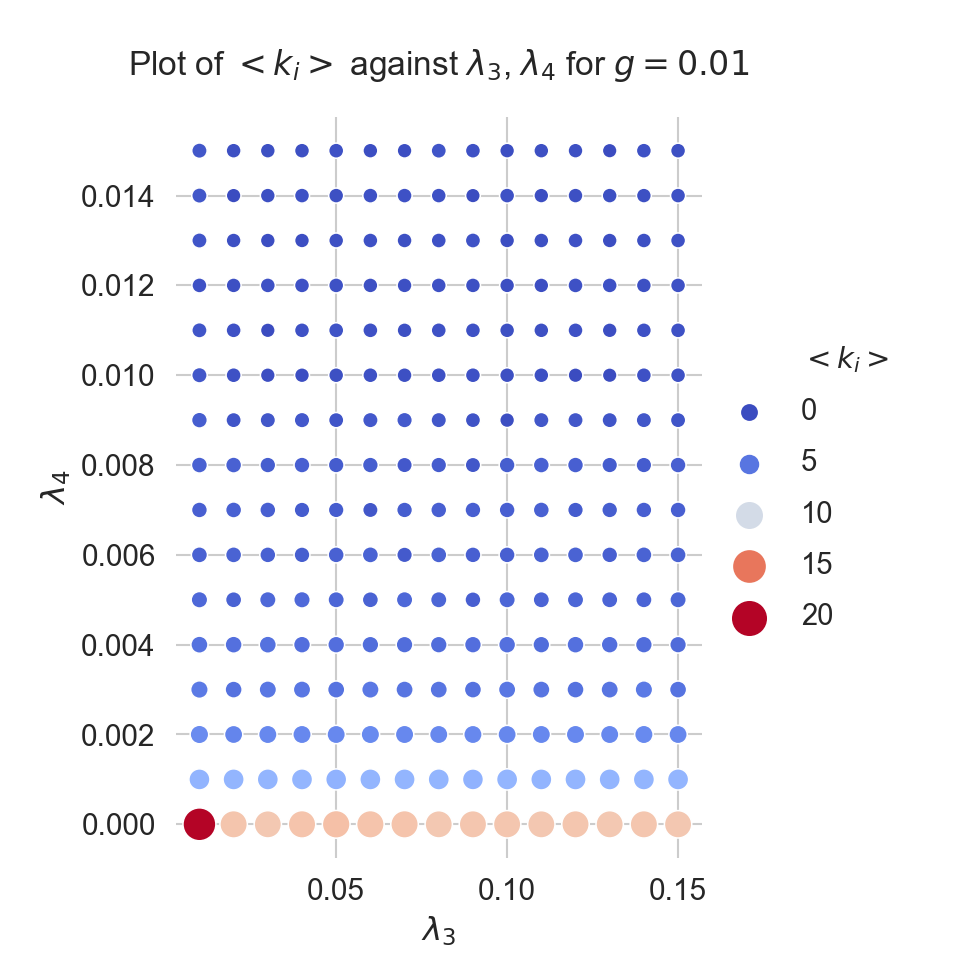}
		\label{fig:DADg01}	
	\end{subfigure}
	~
	\begin{subfigure}[t]{0.3\textwidth}
		\centering
		\includegraphics[height=0.2\textheight]{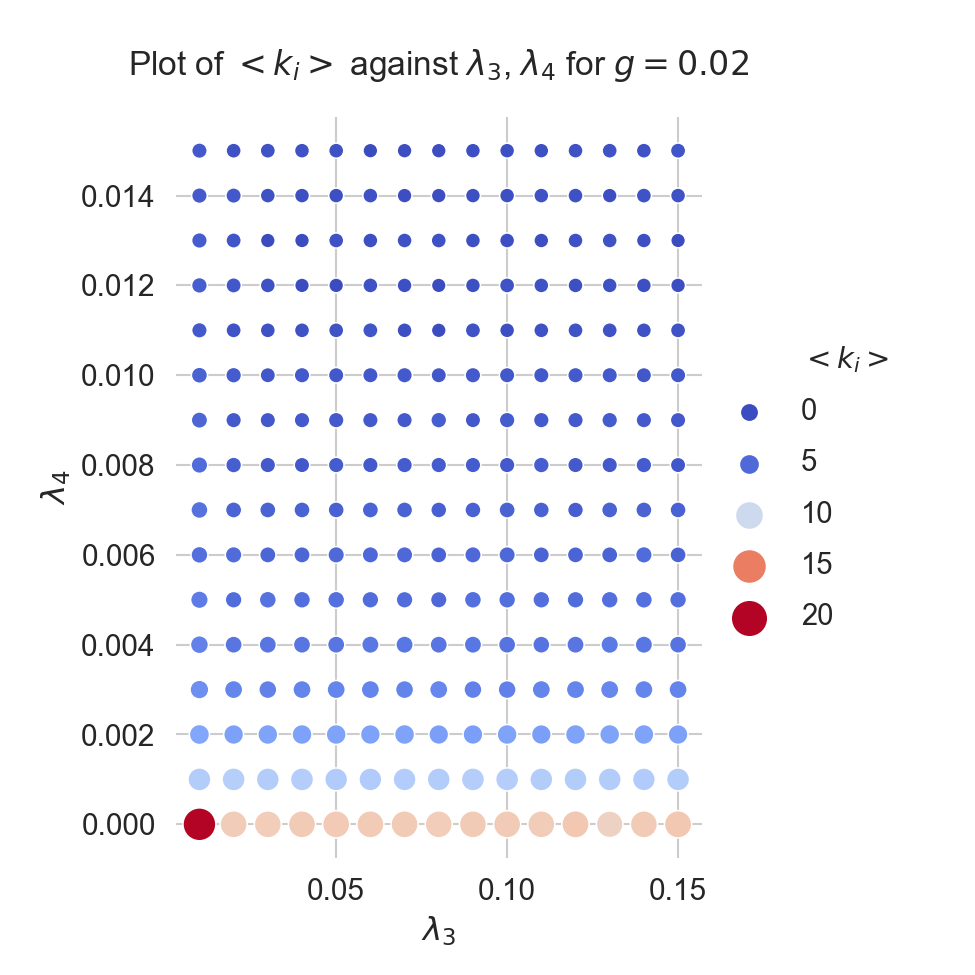}
		\label{fig:DADg02}
	\end{subfigure}%
	~ 
	\begin{subfigure}[t]{0.3\textwidth}
		\centering
		\includegraphics[height=0.2\textheight]{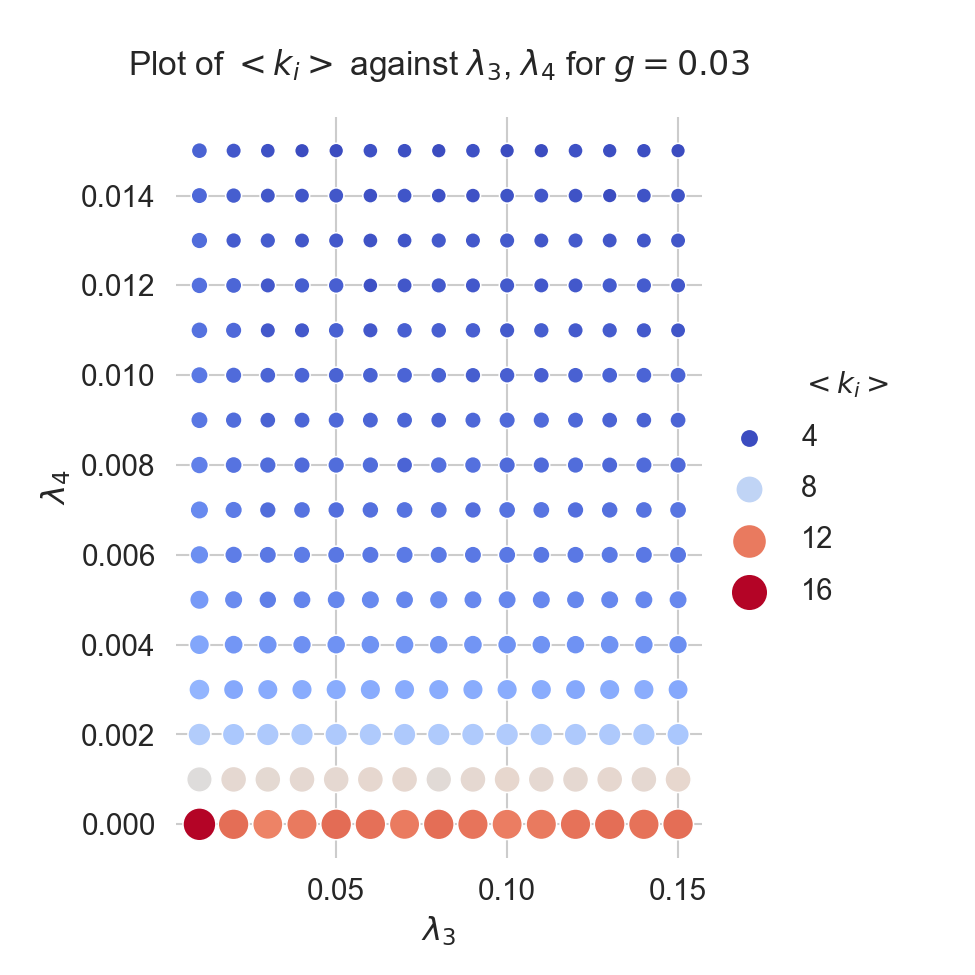}
		\label{fig:DADg03}
	\end{subfigure}%
	\caption{For a mesh of $N=50$ the average node degree is measured computed from the ground state of Eq. \eqref{eqn:pdgm_total}, across a range of values of $\lambda_3$,$\lambda_4$ from $0.0$ to $0.15$. Average degree decreases as $\lambda_4$  increases, but is constant for fixed $\lambda_4$.}
	\label{fig:pdgm_dad}
\end{figure*}


\clearpage
\bibliographystyle{epj}
\bibliography{EPJMeshMatter}
\end{document}